\documentclass{emulateapjd}				

\slugcomment{Version of \today}                                                
\shortauthors{Harrington et al.}                                            
\shorttitle{$\alpha$ Vir Spectral Variability}

\begin{document}

\title{Line-profile variability from tidal flows in Alpha Virginis (Spica)}

\author{David Harrington}
\affil{Institute for Astronomy, University of Hawaii, {2680 Woodlawn Drive}, Honolulu, HI, 96822}
\email{dmh@ifa.hawaii.edu}

\author{Gloria Koenigsberger\altaffilmark{1}}
\affil{Instituto de Ciencias Fisicas, 
       Universidad Nacional Aut\'{o}noma de M\'{e}xico,
       {Adpo. Postal 48-3}, 
       Cuernavaca, Morelos
       62251 M\'exico}
       
\altaffiltext{1}{Miembro del Cuerpo de Tutores,
                 Instituto de Astronom\'{\i}a,
                 Universidad Nacional Aut\'onoma de M\'exico,
                 Apdo. Postal 70-264
                 D.F., 04510  M\'exico.}
\email{gloria@fis.unam.mx}

\author{Edmundo Moreno}
\affil{Instituto de Astronom\'{\i}a,
       Universidad Nacional Aut\'onoma de M\'exico,
       Apdo. Postal 70-264
       D.F., 04510  M\'exico.}
\email{edmundo@astroscu.unam.mx}

\and 

\author{Jeffrey Kuhn}
\affil{Institute for Astronomy, University of Hawaii, {2680 Woodlawn Drive}, Honolulu, HI, 96822}
\email{kuhn@ifa.hawaii.edu}

\begin{abstract}

	We present the results of high precision, high resolution (R$\sim$68000) optical observations of the short-period (4d) eccentric  binary system Alpha Virginis (Spica) showing the photospheric line-profile variability that in this system can be attributed to non-radial pulsations driven by tidal effects. Although scant in orbital phase coverage, the data provide S/N$>$2000 line profiles at full spectral resolution in the wavelength range $\Delta\lambda$4000--8500{\AA}, allowing a detailed study of the night-to-night variability as well as changes that occur on $\sim$2 hr timescale. Using an {\it ab initio} theoretical calculation, we show that the line-profile variability can arise as a natural consequence of surface flows  that are induced  by the tidal interaction.

\end{abstract}


\section{Introduction}

\begin{figure*} [!h, !t, !b]
\begin{center}
\includegraphics[width=0.35\linewidth, angle=90]{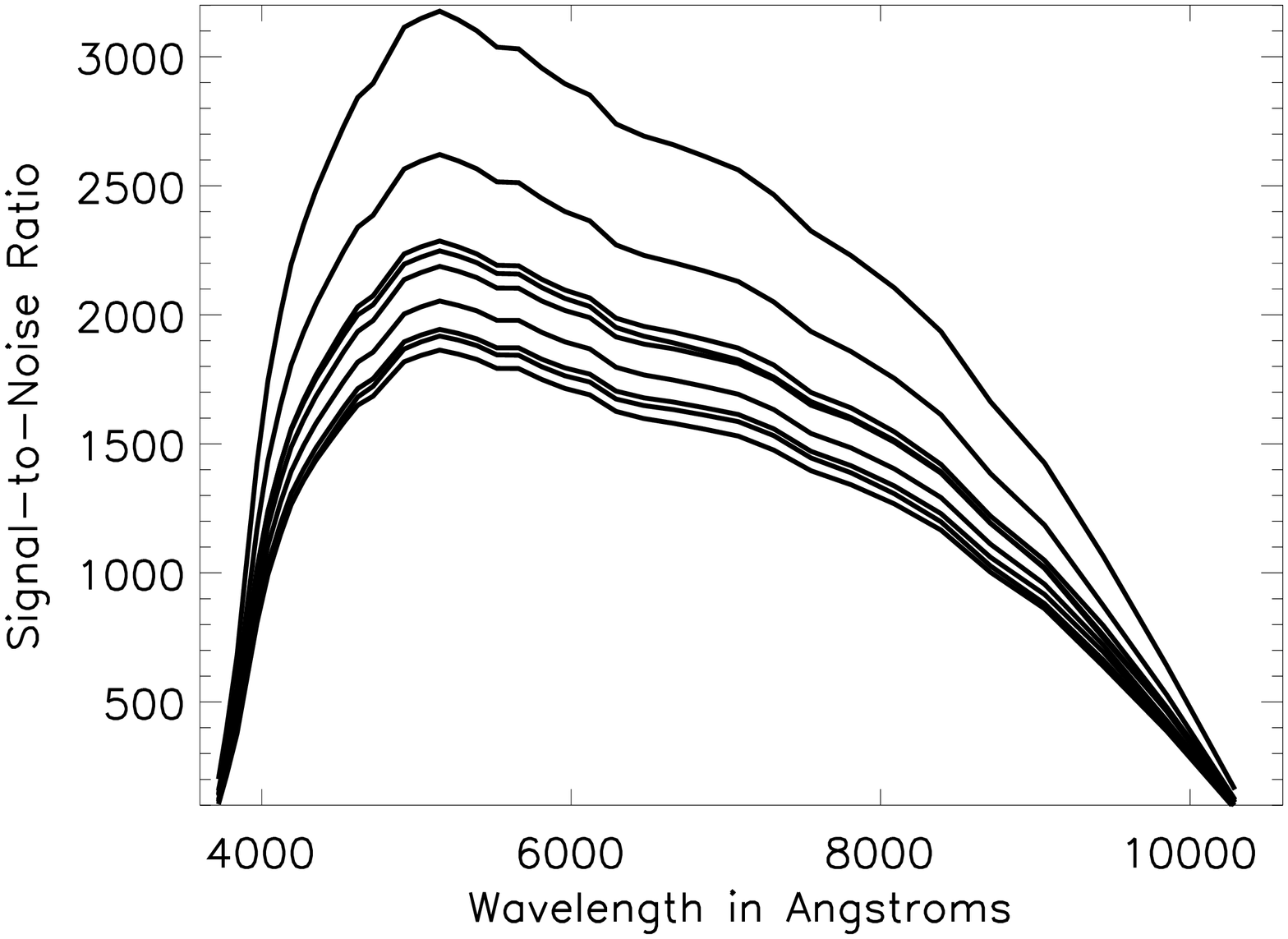}  
\includegraphics[width=0.35\linewidth, angle=90]{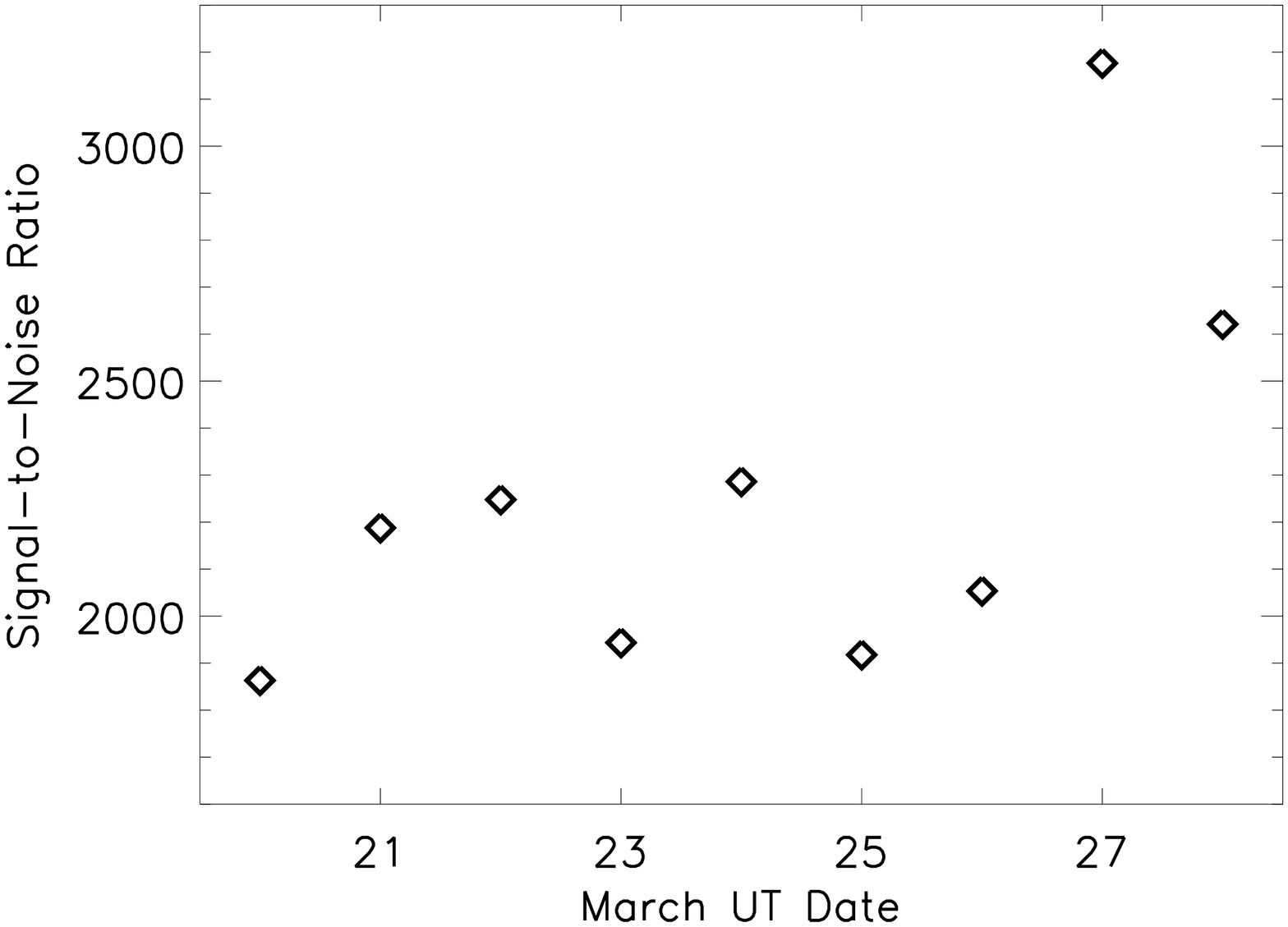}  
\caption{The top panel shows the signal-to-noise measurements output by the Libre-Esprit package as a function of wavelength after averaging all the individual polarimetric exposures. The bottom panel shows the night-to-night differences in the signal-to-noise for order 44 (the 18th) at 5150{\AA}. \label{signoi}}
\end{center}
\end{figure*}

$\alpha$ Virginis (Spica, HD 116658)  is a double-lined spectroscopic binary consisting of two early B-type stars in a short-period ($\sim$4 d) eccentric  orbit.  The  B1 III-IV primary was discovered to be a $\beta$ Cephei-type star in early observations, with a pulsation period of 0.1738 d in both the light curve (Shobbrook et al. 1969) and radial velocity (RV) variations (Smak 1970). Although the  light variations became undetectable a short time after they were first detected (Lomb 1978; Sterken et al. 1986), the line-profile variability  persists (Smith 1985a, 1985b; Riddle 2000).  Its most striking characteristic is the presence of discrete  absorption and emission features travelling from the blue towards the red wing of the absorption line. 

     The first detection of this type of periodic line-profile variability was made by Walker et al. (1979) in the rotationally-broadened O9.5V star $\zeta$ Ophiuchi.  In their seminal study, Vogt \& Penrod (1983) showed that this variability could be explained by non-radial pulsations (NRP). Subsequent observations show a prevalence of this type of line-profile variability  (c.f. Baade 1984; Gies \& Kullavanijaya 1988; Reid et al. 1993; Fullerton et al. 1996; Rivinius et al. 2001; Uytterhoeven et al. 2001). The numerous models based on the NRP mechanism are reviewed by Townsend (1997a, 1997b), some of which incorporate the effects of rapid rotation. A general feature of these models is to assume a prescription for the surface velocity field of the pulsating star, from which the perturbed photospheric absorption line-profiles are computed.  

	A very different approach was used by Moreno et al. (2005), which calculates the surface velocity field from first principles. This calculated (but not assumed) field is then projected along the line-of-sight to the observer to produce photospheric absorption line profiles.  The one basic approximation in this model is to assume that only the external layer of the star oscillates in response to the forces in the system, while the interior region is assumed to rotate rigidly. This one-layer approximation is essentially equivalent to the assumption that the surface layer behavior is primarily controlled by the binary companion perturbations. 

	The advantages of this method are threefold: 1) it  makes no {\it a priori} assumption regarding the mathematical formulation of the tidal flow structure since the velocity field {\boldmath $v$} is derived from first principles; 2) the method is not limited to slow stellar rotation rates nor to small orbital eccentricities; and 3) it is computationally inexpensive. It is not clear to what extent the response of the inner layers may affect the line-profile calculation because the amplitudes of induced oscillations decrease sharply in deeper layers (Dolgivon \& Smel'Chakova 1992). In this paper we confront the results obtained from the Moreno et al. (2005) method with very high-quality observational data of Spica. We show that the azimuthal velocity components of the tidal perturbations (the ``tidal flows") play a dominant role in determining the line-profile variability as produced using this theoretical framework.

	The structure of this paper is as follows: In Sections 2 and 3 we describe the observations and the characteristics of Spica's spectrum and variability; in Section 4 we present the results of the model calculations of Spica's surface velocity field and line-profile variability; the conclusions are presented in Section 5; and details of the data reduction process and a discussion of the local line profile effects and size of the perturbations on the stellar surface are given in the Appendices.

\begin{table}[!h,!t,!b]
\begin{center}
\begin{small}
\caption{ESPaDOnS Observations \label{espobs}}
\vskip 0.1cm
\begin{tabular}{lcccccc}
\hline
\hline
{\bf Date \& } 	& {\bf N}	&{\bf S/N}		&{\bf S/N}		&{\bf S/N}   	& {\bf JD}		& {\bf Orbital}	 \\ 
{\bf Time UT} 	&		&{\bf 4270\AA}	&{\bf 4720\AA}  &{\bf 6660\AA}	& 			&{\bf Phase} 	\\
\hline
\hline
20  12:32:28 	& 8		& 1358		& 1685		& 1580   		&  46.0225  	&  0.3776\\
21  14:23:35 	& 8	 	& 1592	  	& 1978		& 1868		&  47.1000  	&  0.6523\\
22  12:44:32 	& 8	 	& 1660	 	& 2039		& 1892		&  48.0309  	&  0.8838\\
23  13:17:38 	& 8	 	& 1401	 	& 1753		& 1662		&  49.0539  	&  0.1387 	\\
24  14:21:52 	& 8	 	& 1670	 	& 2074		& 1933		&  50.0985  	&  0.3987 	\\
25  14:45:23 	& 8	 	& 1360	 	& 1723		& 1633		&  51.1148  	&  0.6516 	\\
26  09:34:45 	& 8	 	& 1493	 	& 1856		& 1747		&  51.8991	&  0.8472 	\\
27  11:30:58 	&16	 	& 2349	 	& 2897		& 2659		&  52.9798  	&  0.1167 	\\
28  11:16:02 	&16           & 1933	 	& 2386		& 2202		&  53.9695  	&  0.3623 	\\
\hline
\hline
\end{tabular}
\end{small}
\end{center}
This Table shows the UT day and time of the March 2008 observations, the number of exposures summed, the flux based estimate of the signal-to-noise ratio output by the Libre-Esprit package for 4270, 4720 and 6660\AA, the Julian date - 2454500 and the orbital phase of the binary at the time of observation. 
\end{table}

\section{Observations and data reduction}

	The observations were performed in queue mode at the Canada France Hawaii 3.6m telescope (CFHT) with the ESPaDOnS spectropolarimeter. ESPaDOnS is a fiber-fed cross-dispersed echelle covering 3700{\AA} to 10480{\AA} in a single exposure with 40 spectral orders on the EEV1 detector with a nominal spectral resolution of R=68,000 in polarimetric mode. The instrument has a dedicated reduction package, Libre-Esprit, that automatically processes the data. This script does typical flat fielding, bias subtraction, wavelength calibration using both calibration lamp and Fabry-Perot frames and optimal spectral extraction (see Donati et al. 1997). Our integration times were 6 seconds per exposure in polarimetric mode. This mode uses a Wollaston prism to produce two separate orthogonally polarized spectral orders on the detector. A polarimetric sequence consists of two groups of 4 individual exposures at changing wave-plate angles. Each group of four exposures is processed by Libre-Esprit into average intensity and polarization spectra. The doubling of the spectral orders when combined with the intrinsically broadened spatial point-spread-function caused by the image slicer gives a peak signal-to-noise ratio above 1500 in individual spectral pixels without saturation. Thus, a polarimetric sequence consisting of 8 exposures (16 individual spectra) spatially broadened by the image-slicer gives very high precision at full spectral resolution.

	On each night during 2008 March 20--March 26 we obtained one full polarimetric sequence. On March 27 and 28 two complete sequences were obtained. Table \ref{espobs} lists the date and time of the exposures, the number of images combined to produce the spectrum, the signal to noise of for various lines in this combined spectrum, the Julian date and finally the corresponding orbital phase. The orbital phase was computed using P and T$_0$ listed in Table \ref{spicaprop} from Herbison-Evans et al. (1971). A phase of zero corresponds to periastron passage.

	ESPaDOnS is optimized for a broad wavelength coverage, having a high efficiency from roughly 4000-8000{\AA}. Libre-Esprit automatically outputs an estimate of the signal-to-noise for each spectral order based on the average flux in each order. Figure \ref{signoi} shows the wavelength and temporal dependence of the signal-to-noise ratio for our data set. The peak signal-to-noise occurs in order 17 around 5150{\AA}. The ratio is above half the peak value over the 4000-8500{\AA} range. Table \ref{espobs} shows that the precision obtained for all the lines presented in this paper are similar. The ratio only changes by 35\% from 4270 to 6660{\AA}.

\begin{figure} [!h, !t, !b]
\begin{center}
\includegraphics[width=0.9\linewidth, height=1.0\linewidth, angle=270]{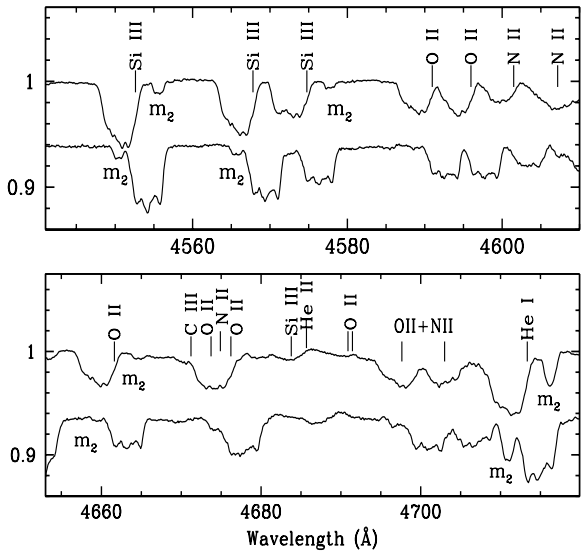}
\caption{\label{specpart} Amplified portions of the March 26 (top) and 28 (bottom) spectra showing the relative strengths and locations of the absorptions from $m_1$ and $m_2$, as well as the principal identifications of each line. The tick marks coincide with the laboratory wavelength of the identification. \label{portionfig}}.
\end{center}
\end{figure}
	
	All cross-dispersed echelle spectrographs have multiple spectral orders with wavelength-dependent order overlap. Typically shorter wavelength orders overlap significantly while longer wavelength orders leave order gaps. The spectral orders must be individually continuum-normalized before being spectrally merged to account for the wavelength dependent sensitivity. The Libre-Esprit package produces both continuum-normalized and un-normalized spectra. In this package, each individual spectral order is rectified using a high-order polynomial response function to the continuum regions. However, this polynomial fit was not accurate enough for this work. The continuum normalization was done on each un-normalized individual exposure after orders were spectrally merged. This merging and continuum mis-fitting can be problematic for accurate spectroscopic analysis. The analysis presented in this paper avoids lines falling near order boundaries and in regions where the order overlap is incomplete. This continuum fitting provides sufficient accuracy for this work (see appendix and ESPaDOnS instrument description on the web).
	
	We have created IDL scripts to deal with the non-uniform overlapping spectral regions by binning the intensity values into essentially constant wavelength intervals of $\Delta\lambda=$0.1{\AA}. This wavelength interval is typically 4 times the individual spectral-pixel size. These scripts result in a monotonic and nearly-regular wavelength coverage as well as increased signal-to-noise per wavelength bin resulting from the 2-12 pixel averaging. The properties of the ESPaDOnS spectral coverage and a detailed description of the effects of this script are contained in the appendix. As an example, Figure \ref{specpart} shows a plot of the March 20 merged and rectified spectrum where the very high signal-to-noise profiles can be seen.

\begin{table}[!h,!t,!b]
\begin{center}
\begin{small}
\caption{Spica Parameters \label{spicaprop}}
\vskip 0.1cm
\begin{tabular}{lcc}
\hline
\hline
{\bf Parameter  }       		& {\bf Set ``HE"}       		&{\bf Set ``Aufden"}      \\
\hline
\hline
$m_1$ (M$_\odot$)		& 10.9$\pm0.9^1$          	& 10.25$\pm0.68^4$       \\
$m_2$ (M$_\odot$)		&  6.8$\pm0.7^1$          	&6.97$\pm0.46^{4,a}$                 \\
R$_1$ (R$_\odot$)		& 8.1$\pm0.5^1$           	&7.40$\pm0.57^{4,a}$                 \\        
R$_2$ (R$_\odot$)		& \nodata                 		&3.64$\pm0.28^4$                 \\
P (days)                        	&4.014597$^1$             	&4.0145898                      \\
T$_0$ (JD)                      	& 2440678.09$^1$          	&2440678.09                     \\
e                               		&0.146$^1$                	&0.067$\pm0.014^3$      \\
$i$ (deg)                       	& 66$\pm2^1$               	& 54$\pm6^4$                   \\        
$\omega$(deg) at $T_0$ 	& 138 $\pm15$           	&140 $\pm10^4$                     \\
Aps. Period (yrs)               	& 124$\pm11^1$           	&135$\pm15^4$                       \\
$v_1 sin i$ (km/s)              	& 161$\pm2^2$            	&161$\pm2^2$            \\
$v_{rot1}$ (km/s)               	& 176$\pm$5              	& 199$\pm$5                     \\         
$v_2 sin i$ (km/s)              	&  70$\pm5^2$           	& 70$\pm5^2$                    \\
$v_{rot2}$ (km/s)               	&  77$\pm$6              		&  87$\pm$6                     \\
$\beta_0(m1)$                   	& 1.3                   		& 1.88$\pm$0.19                 \\        
$\beta_0(m2)$                   	& \nodata                       	& 1.67$\pm$0.5                  \\           
\hline
\hline
\end{tabular}
\end{small}
\end{center}
This Table lists the parameters of the Spica binary system from Herbison-Evans et al. 1971 (H-E 1971) and Aufdenberg et al. 2009 (Aufden 2009) Reference$^1$ is Herbison-Evans et al. 1971, reference$^2$ is Smith, 1985a, reference$^3$ is Riddle, R.L. 2000; reference$^4$ is Aufdenberg et al. 2009; note$^a$ these are the radii at the poles.
\end{table}

\section{Spectrum and Line-profile variability}

	The  spectrum of Spica is very similar to that described by Riddle (2000).  It contains two sets of photospheric absorptions lines, each typical of an early B-type spectrum. The components arising in the primary star, $m_1$ are significantly stronger than the lines arising in the secondary component, $m_2$.  Figure \ref{portionfig} displays a fragment of the spectrum at two orbital phases close to elongations, where the two sets of lines are clearly resolved.  The $m_2$ components are indicated in this Figure, along with the identifications of the strongest absorption lines.

	The most noteworthy features of Spica's spectra are the following:  1) The width of  $m_1$'s lines is consistent with the value of $ v sin i=$161 km/s reported by Smith (1985), while that of $m_2$ is, indeed, significantly smaller. 2) We are able to clearly identify He II 4685.7. Although its blue wing is blended with Si III 4683.80, the relatively small shift  ($-$20 km/s) of the centroid of the feature with respect to $m_1$'s reference frame indicates that He II is the dominant contributor to this blend. If both Si III and He II contributed equally, the expected shift would be $-$64 km/s.  We also confirm the absence of He II 4541.6, thus concurring with Riddle (2000) and Aufdenberg et al. (2009) that $m_1$ is a  B0.5 star.  3) All weak lines present in $m_1$'s spectrum display the discrete narrow features, while lines having larger opacities, such as the H-Balmer series and the He I lines do not. We do not measure any significant difference in the equivalent widths of $m_2$ lines at the two opposite elongation phases illustrated in Figure \ref{portionfig}.

\begin{figure} [!h, !t, !b]
\begin{center}
\includegraphics[width=0.95\linewidth, height=0.85\linewidth]{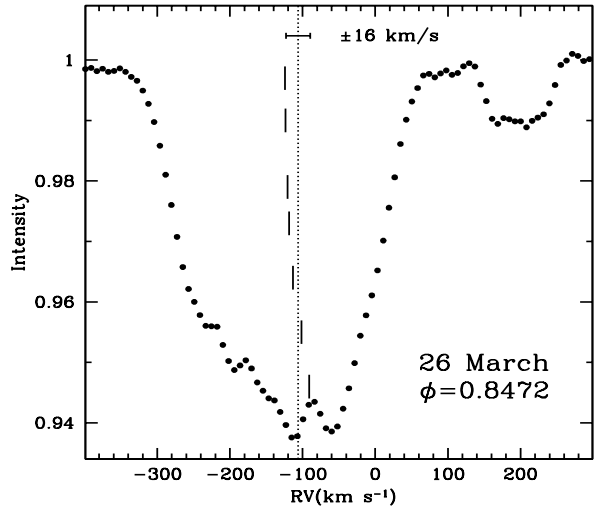}
\caption{Line centroids measured at different depths on the Si III 4552.62 line. The observation date and orbital phase are listed in the Figure. Each vertical dash corresponds to the centroid measured at that depth. The vertical dotted line is the average of all centroids shown in this Figure. The horizontal line shows the range of these calculated centroids. Any individual measurement can differ from the average by up to 16 km/s. This illustrates the source of uncertainty in the RV measurements. Note that the spectrum of the secondary is clearly separated yet the intrinsic line profile variability of the primary produces a very large systematic uncertainty. \label{centroid_levels} }
\end{center}
\end{figure}

\subsection{Radial Velocities}

	The first step in the line-profile analysis consists of transforming the wavelength scale to a velocity scale in the frame of reference of  $m_1$, which requires knowledge of its orbital motion. There are currently at least two determinations of the stellar and orbital parameters of Spica, the first from Herbison-Evans et al. (1971), the second from Aufdenberg et al (2009).  The parameters we used are listed in Table \ref{spicaprop} in two columns, the first derived from the results from Herbison-Evans et al. (labeled ``H-E 1971" and the second derived from those of Aufdenberg et al. (labeled ``Aufden 2009"). 

For the analysis presented in this paper, we chose the lines of Si III 4552.62 and He II 4685.7 (+Si III) for the following reasons:  1) they  are relatively isolated from other neighboring lines; 2) they are  relatively free from contamination by other atomic transitions; and 3) the contribution from $m_2$ is clearly visible in Si III, but absent from He II. We also note that the equivalent widths of these Si III lines are insensitive to effective temperature.

\begin{table}[!h,!t,!b]
\begin{center}
\begin{footnotesize}
\caption{Radial Velocities \label{rvtable}}
\vskip 0.1cm
\begin{tabular}{cccrrrrrrr}
\hline
\hline
{\bf Day}	& {\bf JD}	& {\bf Phase}	& {\bf Si III}	& {\bf } 	& {\bf He II} & {\bf} & {\bf He I} & {\bf m$_2$} & {\bf Na I} \\
{\bf Mar}	& 		& 			& {\bf RV}		& {\bf err}	& {\bf RV}   & {\bf err}     & {\bf RV}   & {\bf EW}       & {\bf ISM} \\
\hline
\hline
20 &  546.022  &  0.377 	& 94   	& 3    	&  64   	& 4   	&-115 	& 0.086 	& -7		\\ 
21 &  547.100  &  0.652 	& -77  	& 14   	& -91   	& 1   	& --  		& --    	& -8		\\ 
22 &  548.031  &  0.878 	& -87  	& 14   	& -118  	& 10	& 179 	& 0.11  	& -8		\\ 
23 &  549.054  &  0.132 	& 73   	& 6    	& 44    	& 3   	&-135 	& 0.12  	& -9		\\ 
24 &  550.098  &  0.393 	& 84   	& 3    	& 43    	& 2   	& --  		& --    	& -10	\\ 
25 &  551.115  &  0.646 	& -79  	& 9    	& -96   	& 1   	& --  		&  --   	& -8		\\ 
26 &  551.899  &  0.841 	& -106 	& 16   	& -141  	& 8   	& 193 	& 0.10  	& -10	\\ 
27 &  552.980  &  0.110 	& 68   	& 3    	&  39   	& 4   	& --  		& --    	& -9		\\ 
28 &  553.970  &  0.357 	& 103  	& 2    	&  71   	& 2   	&-139 	& 0.11  	& -8		\\ 
\hline
\hline
\end{tabular}
\end{footnotesize}
\end{center}
This Table shows the radial velocities measured for each observation with respect to laboratory wavelengths. The columns show the March 2008 observation date, Julian date (JD-2454000), orbital phase of observation, radial velocity (RV) and uncertainty (err) of the primary star for Si III (4552.2{\AA}) and He II (4685.7{\AA}) lines, the secondary companion He I (6678.15{\AA}) radial velocity and equivalent width in angstroms and finally the Na I D ISM line radial velocity.
\end{table}

The strong line-profile variability in Spica requires that the measurement of spectral lines be performed with great caution.  We exemplify the problem in Figure \ref{centroid_levels} with the Si III 4552.62{\AA} line observed on 26 March. The location of the line centroid is indicated for measurements performed at different intensity levels; as the measured intensity level approaches the continuum, the centroid of the line becomes increasingly more negative. The RV measured near the minimum of the line differs by 32 km/s from the RV measured at the continuum level. This particular example is very illustrative because the contribution from $m_2$, clearly visible in the plot, cannot contribute in any manner to the shape of $m_1$'s profile.  Thus, it is very difficult at this stage to decide what portion of the line profile most reliably represents orbital motion and this situation introduces a high degree of uncertainty in the determination of the stellar and orbital parameters.

\begin{figure} [!h, !t, !b]
\begin{center}
\includegraphics[width=0.95\linewidth, height=0.5\linewidth]{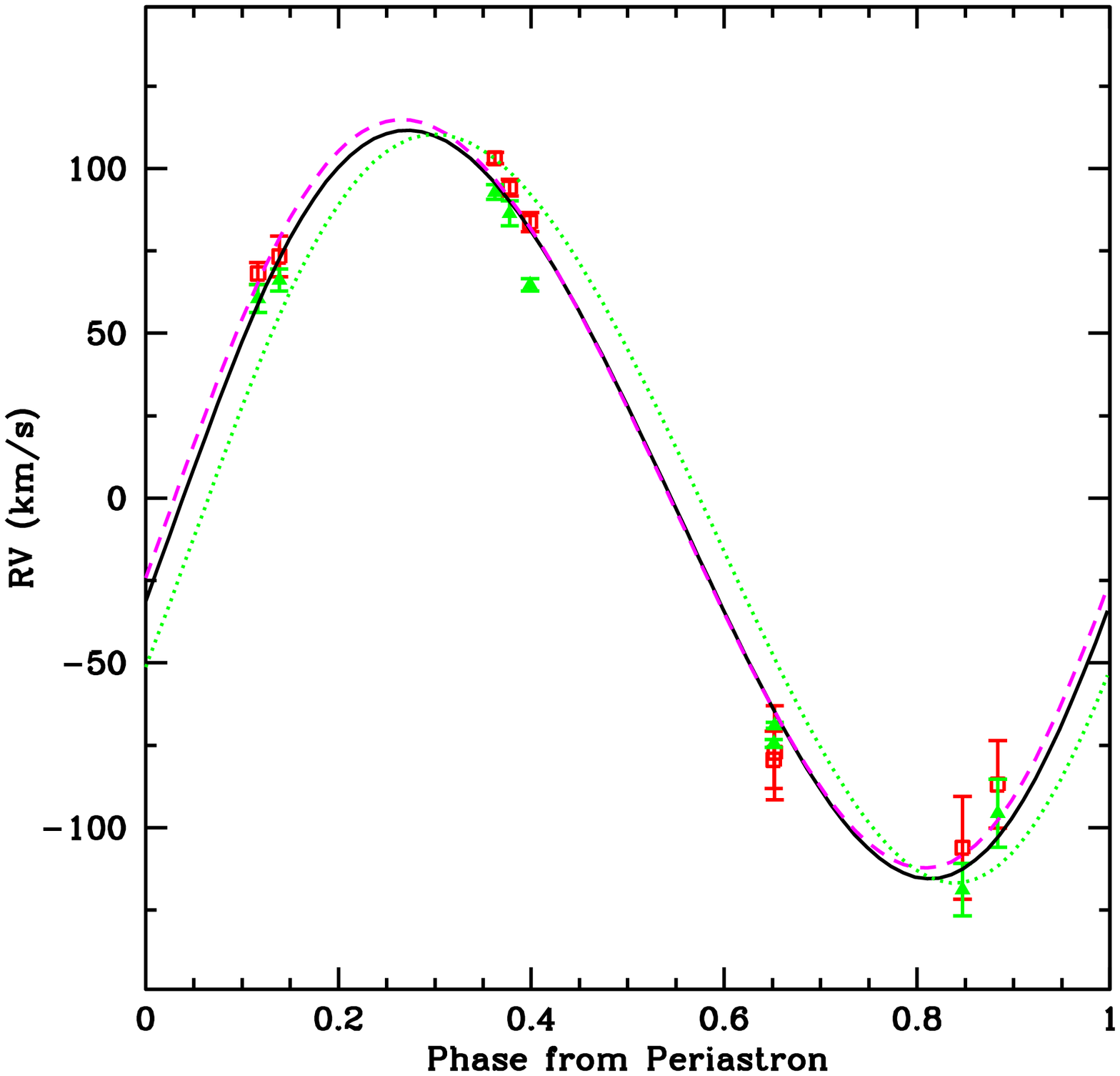}  
\includegraphics[width=0.95\linewidth, height=0.5\linewidth]{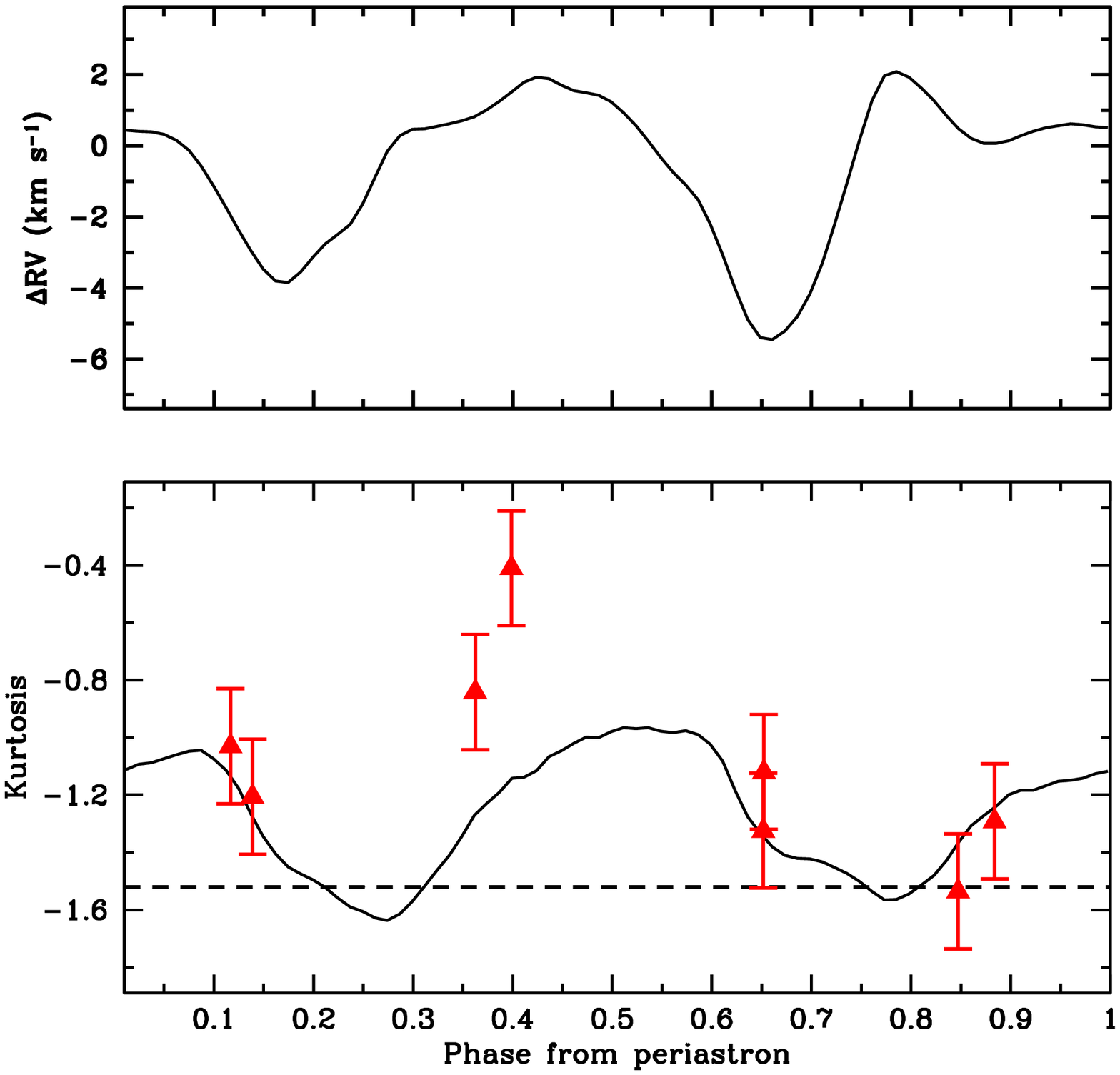}
\caption{Top: Radial velocity measurements of Si III 4552.62 (open squares)  and He II 4685.7 (shifted by $+$22 km/s; filled-in triangles) for the primary component of Spica plotted as a function of orbital phase measured from periastron. The curves correspond to the predicted RV velocities  according to the Aufdenberg et al. (2009) parameters and adopting $\omega=$255$^\circ$ (solid),  $\omega=$257$^\circ$ (dash )and $\omega=$248$^\circ$  (dots). Middle: predicted difference between the centroid of a tidally-perturbed  and the unperturbed line profiles. Bottom: kurtosis of the theoretical line-profiles. The kurtosis of the observed line profiles (triangles) measured at an intensity level that avoids the contribution from $m_2$ displays a similar trend. \label{rv}}
\end{center}
\end{figure}

Given this uncertainty, we measured Si III and He II 4685.7{\AA} at as many intensity levels as possible, without reaching levels where $m_2$'s absorption is evident. The average value of these measurements is listed in Table \ref{rvtable}, together with an uncertainty that corresponds to 1/2 of the total range in RV values measured for the line. The resulting RVs are plotted in Figure \ref{rv}.   Also plotted in this Figure are the RV curves corresponding to the Aufdenberg et al. (2009) parameters (solid) for three different values of $\omega$, the argument of periastron, which defines the location of line of apsides with respect to the plane of the sky.  Aufdenberg et al. (2009) give $\omega=$140$^\circ\pm$10$^\circ$ at T$_0=$ JD 2440678.09, as well as the apsidal period P$_{aps}=$135$\pm$15 years.  Thus, for the current observations, $\omega_{2008}$ lies between 248$^\circ$ and 257$^\circ$.  The measured radial velocities indicate that  $\omega_{2008}=$255$^\circ$, as illustrated in Figure \ref{rv}. Since this lies within the range of acceptable values, we adopt $\omega_{2008}=$255$^\circ$ for this study.

\begin{figure} [!h, !t, !b]
\begin{center}
\includegraphics[width=0.95\linewidth, height=0.95\linewidth]{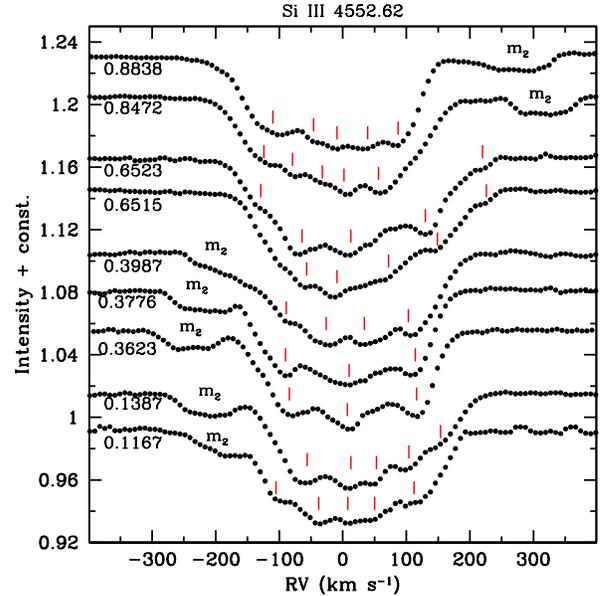}  
\caption{Montage of  the observed Si III line profiles on a velocity scale centered on 4552.62 {\AA}, and corrected for the orbital motion of the primary using RV values listed Table \ref{rvtable}. The spectra are vertically arranged by orbital phase. The tick marks show the troughs of traveling bumps and the secondary's component is marked with $m_2$. \label{mont1}}
\end{center}
\end{figure}

\begin{figure*} [!h, !t, !b]
\begin{center}
\includegraphics[width=0.45\linewidth]{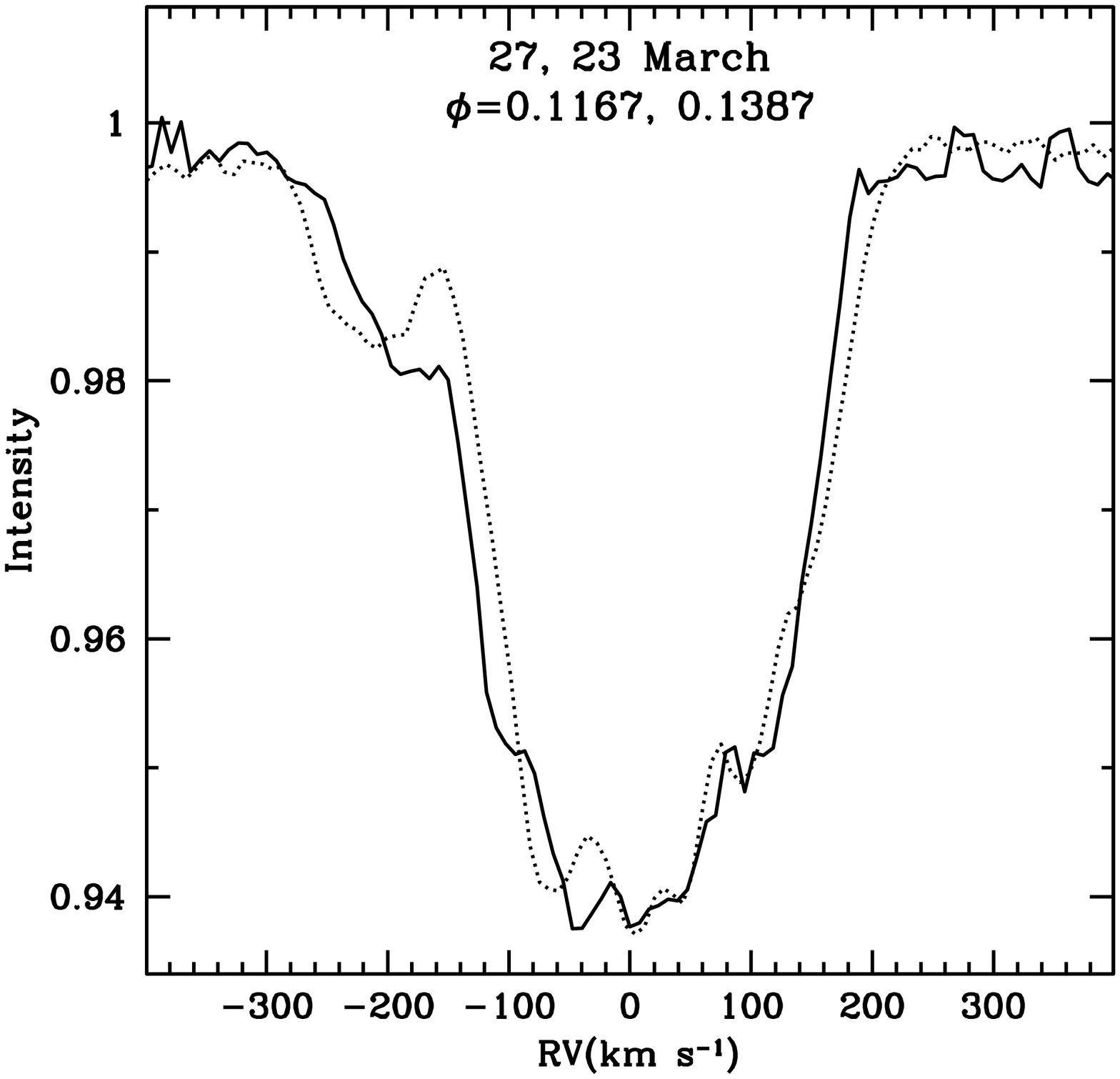}
\includegraphics[width=0.45\linewidth]{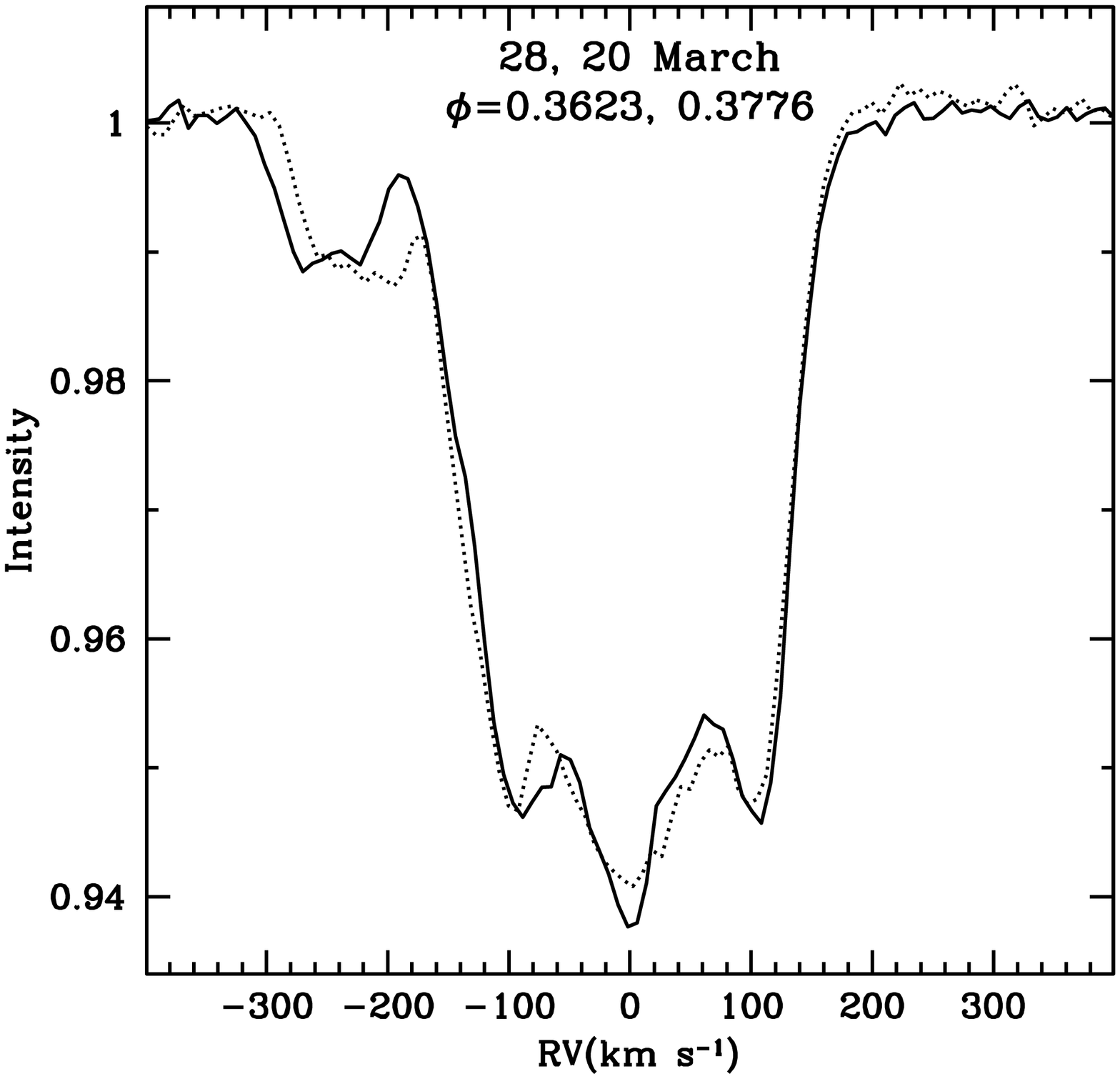}
\includegraphics[width=0.45\linewidth]{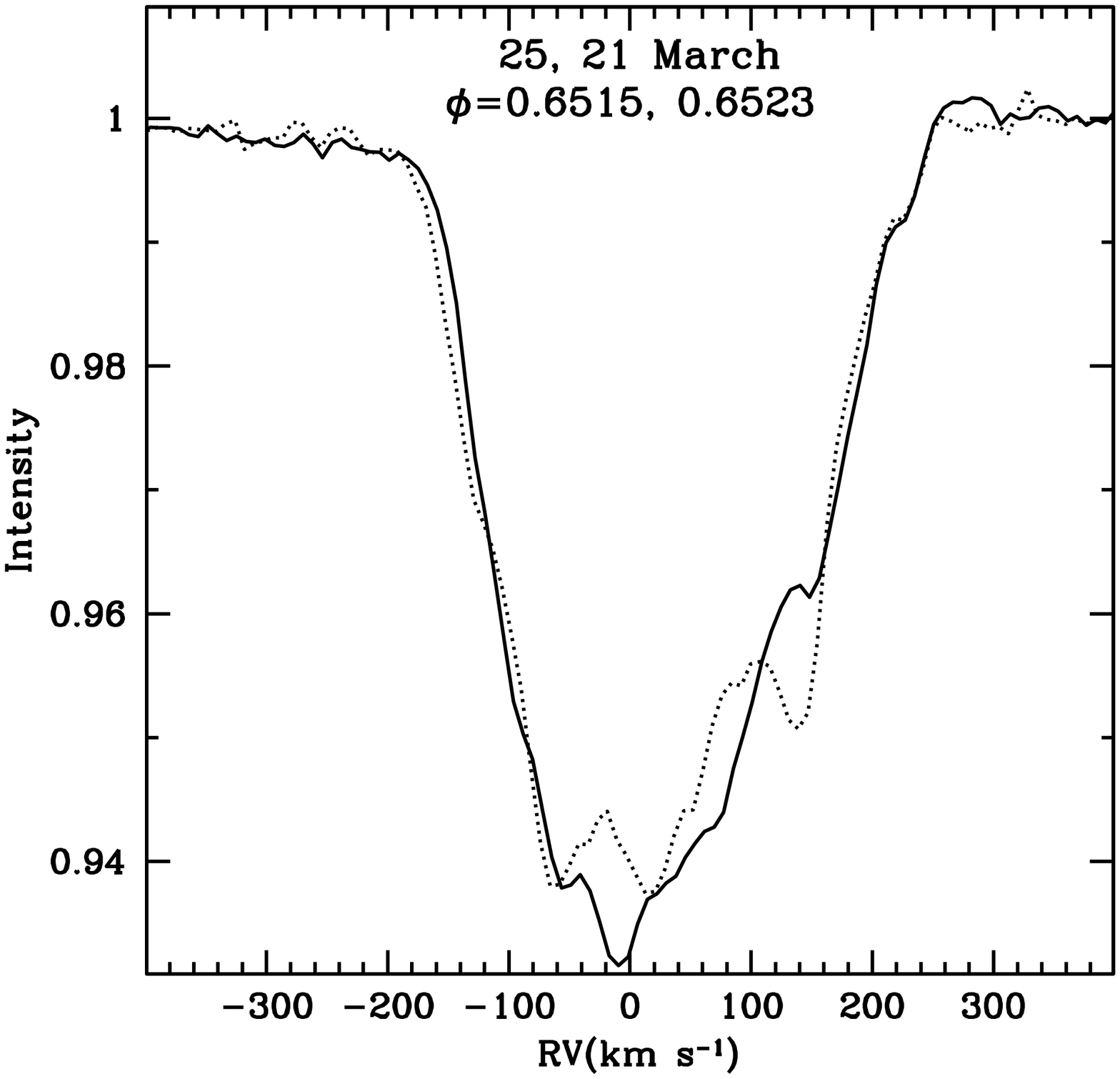}
\includegraphics[width=0.45\linewidth]{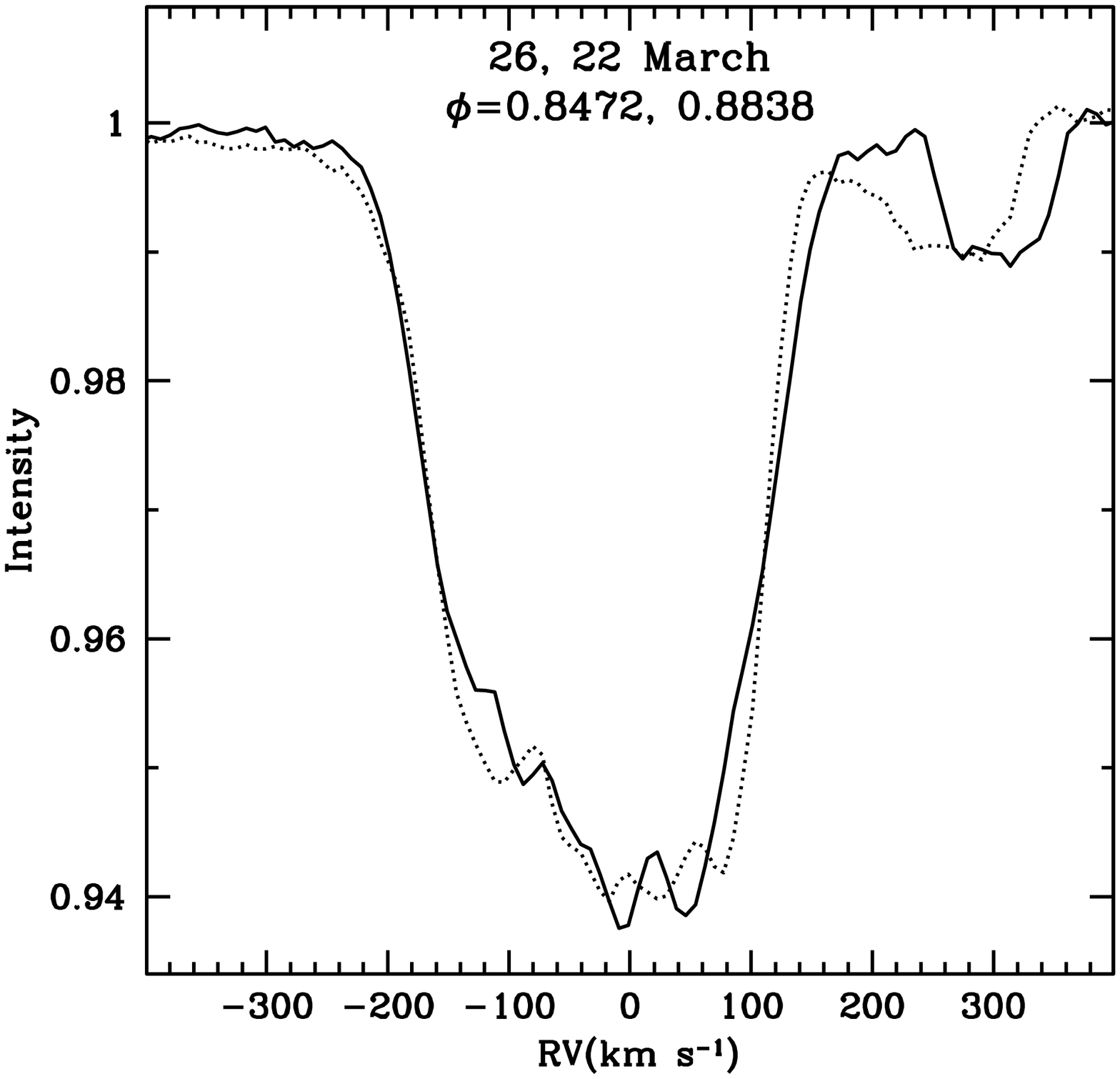}
\caption{Comparison of observed Si III 4552{\AA} line profiles, illustrating the changes that occur over  $\Delta\phi\sim$0.02--0.03 (2-3 hours). Each panel shows two observations at similar phases, though separated by several days. The velocity scale is shifted for each profile according to the RV values listed in Table \ref{rvtable}. The general shape of the profile is very consistently reproduced after one or two full orbits but the small scale structure of the bumps is sometimes remarkably consistent (top right) and sometimes highly variable (bottom left). The dates and orbital phases for each observation are noted in each panel with the solid curve being the first date/phase. \label{compare_same_phase}}
\end{center}
\end{figure*}

\subsection{Line-profile variability}

Line profile variability in Spica was first analyzed in detail by Smith (1985a, 1985b) in a closely-spaced set of data obtained over the complete orbital cycle. He described the variability in  $m_1$ in terms of the presence of ``bumps" that deform the shape of the photospheric absorption line profile from that which it would have if its surface were rotating as a rigid body.  The individual ``bumps" were found to travel across the line-profile, from blue to red, at rates of $\sim$12--17 km/s per hour.  The analysis of the pattern of moving bumps seen in the SiIII lines at 4552, 4567, 4574 \AA\, led him to conclude that they could be accounted for by four nonradial pulsation modes. Two of the modes had periods of $\sim$1/12P and 1/2P, the latter being associated to ``the spectroscopic equivalent of the ellipsoidal light variability" which is due to the tidal distortion.

\begin{table*}[!h,!t,!b]
\begin{center}
\begin{normalsize}
\caption{Summary of model parameters \label{tablemodpar}}
\vskip 0.1cm
\begin{tabular}{lrrrrrrrrr}
\hline   
\hline
{\bf Model} & {\bf m1} & {\bf m2} & {\bf R1} & {\bf i} & {\bf $\beta_0$} & {\bf $\nu$} & {\bf e} & {\bf $\omega$} & {\bf dR/R1}  \\
\hline
\hline
All sets   &10.25--11.8& 6.1--6.97&6.84--8.1&54--66 &1.29--2.07 &  0.009--0.058 & 0.016--0.146 & 209--258 & 0.01--0.087 \\
C          	&10.90      & 6.80     & 8.10    &66     & 1.30      & 0.028        & 0.146         & 248      & 0.05    \\
D          	&10.90      & 6.80     & 8.10    &66     & 1.30      & 0.028        & 0.146         & 248      & 0.06    \\
E          	&10.90      & 6.80     & 8.10    &66     & 1.30      & 0.028        & 0.146         & 248      & 0.07    \\
G          	&10.90      & 6.80     & 8.10    &66     & 1.30      & 0.018        & 0.146         & 248      & 0.05    \\
GGD		&10.25      & 6.97     & 6.84    &54     & 2.07      & 0.028        & 0.067         & 255      & 0.07    \\
GGD3       &10.25      & 6.97     & 6.84    &54     & 2.07      & 0.028        & 0.067         & 209      & 0.07    \\
GGD4       &10.25      & 6.97     & 6.84    &54     & 2.07      & 0.018        & 0.067         & 255      & 0.07    \\
GGD6       &10.25      & 6.97     & 6.84    &54     & 2.07      & 0.018        & 0.067         & 255      & 0.07    \\
GG2        	&10.90      & 6.80     & 8.10    &66     & 1.30      & 0.018        & 0.146         & 258      & 0.07    \\
H          	&10.90      & 6.80     & 8.10    &66     & 1.30      & 0.010        & 0.146         & 248      & 0.07    \\
\hline
\end{tabular}
\end{normalsize}
\end{center}
For the models in this Table, the polytropic index was 1.5. However, we have run models with an index of 1.5, 1.67 and 3.
\end{table*}

Our limited data set is unable to fully address the short-timescale variability since we only have a single spectroscopic observation set of 8 exposures taken over less than 5 minutes on most nights. The two nights with 16 exposures (27th and 28th) only cover 10 minutes. However, we do have two pairs of nights (March 27$^{th}$ \& 23$^{rd}$, 25$^{th}$ \& 21$^{st}$) where there are small orbital phase differences between observations corresponding to a few hours of stellar rotation. These pairs allow the characterization of changes in the ``bump" structure over short time-scales, though with one complete orbit actually in between the observations. The night-to-night variability can be easily addressed with such high quality data.

Figure \ref{mont1} illustrates the Si III 4552.62{\AA} line in the 9 different orbital phases covered by our observations, plotted on a velocity scale after applying the correction for orbital motion as listed in Table \ref{rvtable}. The variability in this line is typical of that seen in all other weak lines. Note that the absorption from $m_2$ can be clearly followed in these spectra. In the first five profiles ($\phi=$0.1--0.4, from bottom to top), the absorption line originating in the secondary component, $m_2$, is observed to move towards maximum negative velocity and then return towards the rest-frame velocity. In the last two profiles ($\phi=$0.8472, 0.8838), $m_2$'s contribution may be seen near its maximum positive velocities. In the two profiles at $\phi=$0.652  the contributions from $m_1$ and $m_2$ overlap.

We have marked in Figure \ref{mont1} the location of the discrete absorptions that are resolved in our data and whose radial velocity variations could be measured.  In general, at least 5 such troughs may be identified, although at times some of these may blend together leading to the appearance of only 3 troughs (for example, at orbital phases 0.3623, 0.3776 and 0.3987).  If we assume that the troughs retain their identity between two consecutive orbital phases \footnote{Although it is important to keep in mind that the two consecutive orbital phases in our data correspond to different orbital cycles}, then we find that between $\phi=$0.1167 and 0.1387, the rate by which they travel from blue to red is $\Delta V_{bump}/\Delta t=$22 km/s per hour; and between $\phi=$0.8472 and 0.8838, $\Delta V_{bump}/\Delta t=$8 km/s per hour. For $\phi=$0.35--0.39 one of the bumps appears to remain stationary, the second moves slightly "red-ward", and  the third bump seems to split into at least two absorption-like features.

Thus, the traveling bumps described by Smith (1985b) are clearly visible in Figure \ref{mont1}.  However, only the troughs for $\phi\sim$0.1 in Figure \ref{mont1} travel at an average displacement speed similar to the $\sim$17 km/s per hour speed that may be inferred from Smith (1984a; Figures 1--4).  At the other orbital phases covered by our data the velocity is significantly slower.  These speeds, of course, are the result of the actual motions on the stellar surface combined with the projection of the stellar rotation velocity along the line-of-sight to the observer.  Further insight into the phenomenon requires the use of a theoretical model.

The stability on $\sim$2 hour timescales may be fully appreciated in Figure \ref{compare_same_phase}.  In three of the four comparisons of Figure \ref{compare_same_phase} there is excellent agreement and the identity of individual bumps is almost entirely preserved. This is despite the fact that the observations are separated by a full orbital cycle and that diverse variability sources could possibly be present.  The most striking difference appears in the third comparison, corresponding to the line profiles from March 21st and 25th obtained at nearly the same orbital phase, 0.6515 and 0.6523. These observations have the closest phase overlap of all four shown in the Figure and yet they present the most significant line profile differences.  This indicates that the variability is not fully orbital-phase locked.

Figure \ref{compare_same_phase} also provides a view of the manner in which the absorption line wings rise from the core  to the continuum level.  There is a marked difference between the shape of the line-wings at different orbital phases.  For example, at  orbital phases $\sim$0.3--0.4, the red wing gives the profile  a more ``boxy" shape than at other phases,   changes that may be quantified using the {\it kurtosis}, which describes the degree to which a distribution is centrally peaked or flat.  The bottom panel of Figure \ref{rv} shows the phase-dependent behavior of the kurtosis, confirming the difference that is apparent from Figure \ref{compare_same_phase} in the line shapes. The kurtosis was computed over a wavelength range  that avoids the contribution from $m_2$, and the error bars correspond to an uncertainty  $\pm$0.2 \AA\ on the definition of each line wing.

In summary: Two categories of line-profile variability are present in our data: 1) there are traveling bumps that retain their identity over timescales of at least a day, though some may change significantly; and 2) there is a varying shape of the line-wings which go from being very extended to ``box-like" on a nightly timescale. We will show below that both of these characteristics are a consequence of surface flows that can be produced by the tidal interaction between the two components.

\section{The tidal flow model}

A system is in synchronous rotation when the orbital angular velocity $\Omega$ equals the angular velocity of rotation $\omega_0$.  In eccentric orbits, the degree of synchronicity varies with orbital phase.  We use periastron passage as the reference point for defining the synchronicity parameter, $\beta_{per}$\footnote{$\beta_{per}=\omega_0/\Omega_{per}=$ 0.02 P $\frac{v_{rot}(1-e)^{3/2}}{R_1(1+e)^{1/2}}$ The orbital eccentricity is denoted $e$, the rotation velocity $v_{rot}$ is given in km/s, the orbital period P is given in days, and the stellar equilibrium radius R$_1$ is given in solar units.}. When at any orbital phase $\beta\neq$1, non-radial oscillations are excited, driven by the tidal interactions. We refer to the azimuthal components of the forced oscillations as ``tidal flows".

In the following paragraphs we summarize the general characteristics of a simple model developed to gain some insight into the effects that the tidal flows introduce in the photospheric line profiles.  A more detailed description is provided in Moreno \& Koenigsberger (1999) and Moreno et al. (2005).

Consider  a binary system in which the light of the primary star with mass $m_1$ and radius $R_1$ dominates the shape and intensity of a photospheric spectral line. The main body of the star, below the surface layer of thickness $\Delta R$ is assumed to have a constant spin angular velocity $\omega_0$ with respect to an inertial frame. The binary companion has mass $m_2$ and an instantaneous orbital angular velocity $\Omega$ in its (in general) elliptic orbit around $m_1$.  The axis of stellar rotation is assumed to be perpendicular to the orbital plane.  We define $\beta_0=\omega_0/\Omega_0$, where $\Omega_0$ is the orbital angular velocity at periastron. $\beta_0$ specifies whether the stellar rotation and the orbital periods are synchronized ($\beta_0=1$) or whether the rotation is super- or sub-synchronous ($\beta_0>1$ or $\beta_0<1$, respectively).  If $\beta_0\neq$1.000, then tidal flows on the stellar surface are induced and non-radial oscillations are excited.  Our model is based on the fact that the surface layer is the one most strongly affected by the external gravitational potential of the companion star (see, for example, Dolginov \& Smel'Chakova 1992; Eggleton et al. 1998) and we therefore compute the response of this layer to all the forces in the system, while the rest of the star below it is treated as a rigidly-rotating body.  We find {\it a posteriori}, that the dominant contribution to the photospheric line-profile variability is produced by the azimuthal component of the velocity field and not the radial component. In our model, the azimuthal flow structure is governed by the tidal forcing and appears to be very weakly dependent on the radial component of the velocity field.

A non-uniform Lagrangian grid is constructed over the stellar surface by  specifying the number of surface elements along the equatorial belt and  the number of latitudes  for the computation.  The number of surface elements for each latitude is  defined so that in the unperturbed star all elements have similar sizes in the longitudinal coordinate.  For each surface element  in the grid, an equation of motion is specified in which the following forces are included:  the gravitational field of $m_1$  and $m_2$, gas pressure, Coriolis force, centrifugal force and the viscous shear forces associated with the interface of the thin surface layer with the inner stellar body and the interface between two adjacent surface elements.  The surface layer may be treated either in the polytropic or in the isothermal approximation. The cases described below correspond to calculations with polytropic indices of 1.5.  Cases run with a different polytropic index or in the isothermal approximation yield similar global characteristics.  The simultaneous solution of the equations of motion for all surface elements yields values of the radial,$v_r$, and azimuthal, $v_\varphi$, velocity fields over the stellar surface.  Azimuth angle, $\varphi$, is measured from the line joining the centers of $m_1$ and $m_2$, in the non-inertial reference frame centered on $m_1$, and rotating with the instantaneous orbital angular velocity $\Omega$.  Polar angle, $\theta$, is measured from the pole of $m_1$ to its equator.

The model is not restricted to slow rotation nor to small excentricities. The only {\it a priori} assumptions built into the calculation are: 1) only the  thin outer layer responds to the tidal perturbation;  2) the stellar rotation axis is perpendicular to the orbital plane; and 3) the response of the outer layer is either  polytropic or isothermal. It must be kept in mind, however that because of the single-layer approximation, we compute only the motion of the center of mass of each surface element, thus ignoring the details of the inner motions within these elements, and thus the buoyancy force is neglected.

\subsection{Input parameters}

The input parameters that determine the physical characteristics of the tidal interaction effects are: the stellar masses, m$_1$, m$_2$, the orbital period, P$_{orb}$, the eccentricity  of the orbit, e, the primary star radius, R$_1$, the  synchronicity parameter, $\beta_0$, the relative depth of the surface layer, $\Delta$R/R$_1$, and the kinematical viscosity of the stellar material, $\nu$.  In addition, in order to  compute the projection of the surface velocity field along the line of sight to the observer, the longitude of periastron, $\omega$, and  the inclination  of the  orbital plane, $i$, are required.  Finally, the properties of the spectral lines  that need to be specified are the strength of the absorption line $a_0$ and  the line-broadening parameter, $k$.  The latter is associated with the microturbulence velocity and for the calculations presented here, we use values in the range 10--15  km/s.

	We performed model calculations for both sets of parameters listed in Table \ref{spicaprop}, as well as for intermediate values.  The range in parameters that was explored is listed in the first line of Table \ref{tablemodpar}.  However, since the Aufdenberg et al. (2009) parameter set yields a closer match to our observational RVs  (Figure \ref{rv}), they are adopted for the remainder of this paper.  The two free parameters that remain to be discussed are  the depth of the surface layer and the kinematical viscosity. The constraint imposed by the basic assumption of a thin outer layer is here adopted to mean that the depth of the oscillating layer, $\Delta$R/R$_1<$0.1. We computed models with $\Delta$R/R$_1=$0.05--0.08 to determine the dependence of the results on this parameter.

	The kinematic viscosity, $\nu$ determines the efficiency of angular momentum transfer between the surface layer and the rigidly-rotating inner body of the star, and it also couples the radial motion of each  surface element with that of its neighboring elements. In order to model  many of the interesting binary systems, we generally require a value of $\nu$ in the range 10$^{13}$--10$^{16}$ cm$^2$ s$^{-1}$ to limit the velocity amplitudes.  When the amplitudes grow excessively, neighboring surface elements may become detached, or they partially overlap, causing the computation to halt.  Our choice of $\nu$ is thus determined by the smallest value that allows the computation to proceed for as many orbital cycles as initially specified.  The need for a large $\nu$ may be related to the fact that the oscillations are restricted to a single surface layer and that there is no built-in mechanism for energy to be transported out of the layer. On the other hand, high values of the viscosity are not unusual in astrophysical plasma calculations. For example, for a so-called alpha-viscosity disk such that $\nu= \alpha c H$, where $c$ is the sound speed and $H$ is the disk scale height, using a typical value $\alpha = $0.01 for T Tauri disks (Hartmann et al. 1998) and a typical disk structure (D'Alessio et al. 1999) one derives viscosities precisely in the same range 10$^{13}$--10$^{16}$ cm$^2$ s$^{-1}$.

\begin{figure} [!h, !t, !b]
\begin{center}
\includegraphics[width=0.95\linewidth, height=0.7\linewidth]{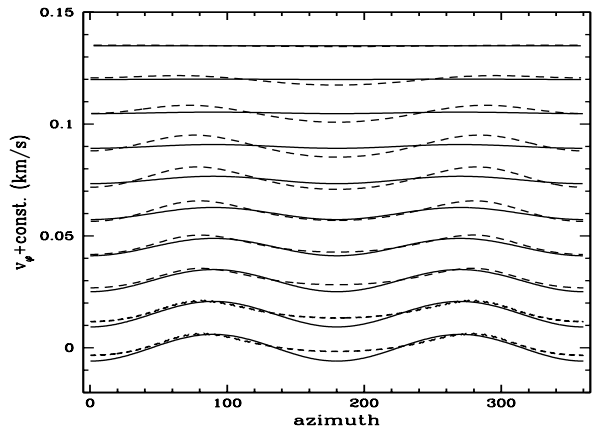}
\caption{The azimuthal velocity field obtained from Scharlemann's (1981) analytical formulation (continuous curves) for a binary system with m$_1 =$ 10.9 M$_\odot$, m$_2 =$ 6.8 M$_\odot$, R$_1 =$ 8.1 R$_\odot$, $\beta =$ 1.001 for latitudes from bottom to top 0$^\circ$  (equator), 9$^\circ$, 17$^\circ$, 26$^\circ$, 35$^\circ$, 44$^\circ$, 52$^\circ$, 61$^\circ$, 70$^\circ$, 93$^\circ$. The results for the same binary parameters obtained from our one-layer calculation are illustrated with the dash curves. \label{scharlcomp}}
\end{center}
\end{figure}

	One final remark concerns the synchronization parameter $\beta_0$.  In an eccentric orbit, $\beta_0$ changes its value as a function of orbital phase, being smallest at periastron (where the orbital motion is fastest) and largest at apastron.  The values of $\beta_0$ that we quote to identify a particular model calculation refer to its value at periastron passage, although its changes over the orbital cycle are included in the numerical calculations. From an observational standpoint, it is not straightforward to determine $\beta_0$ since it requires knowledge of the  rotation rate of the star in the layers underlying the visible portion of the stellar atmosphere.  As will be shown below, the effects produced by the tidal interaction modify the shape of the photospheric absorption-line profiles.  Hence, although the rotational velocity of the star  is in principle a quantity that is derived from observations, it is not clear whether the value derived from the line profiles corresponds to the actual rotation speed of the inner stellar regions.  The uncertainty in $ v sin i$ translates into an uncertainty in $\beta_0$.

	As the code is a time-marching algorithm, we let the simulations run for at least $\sim$45000 time steps (roughly corresponding to 30 orbital cycles), long enough to get rid of the perturbations generated by the initial conditions. Once these initial perturbations disappear, the solution is  stable over many more orbital cycles for all model parameters shown here.

\subsection{The azimuthal velocity field}

	We use the term ``tidal flows" to refer to the large-scale azimuthal velocity patterns that are induced by the gravitational field of the binary companion.  An intuitively simple way to grasp the essence of the idea of tidal flows is given by the description of Tassoul (1987): ``Assume that at some initial instant the primary is rotating rigidly with a constant supersynchronous angular velocity.  If there were no companion, the system would remain forever axisymmetric.  Because of the presence of the secondary, however,  a fluid particle moving on the free surface of the primary will experience small accelerations and decelerations in the azimuthal direction." In the simplest of cases, the tidal flow pattern may be characterized in terms of four large contiguous zones over which the acceleration alternates between positive and negative values.  This defines four quadrants over the stellar surface, the first of which always lies near the sub-binary longitude, and in which the acceleration is negative for supersynchronous rotation.  Figure \ref{scharlcomp} is an example of such velocity perturbations obtained from the analytical solution given by Scharlemann (1981) for a binary system with m$_1=$10.9 M$_\odot$, m$_2=$6.8 M$_\odot$, R$_1=$8.1 R$_\odot$, and P$=$4.01d.  The Figure shows a plot of the azimuthal velocity {\it vs.} azimuth angle for different latitudes, starting near the equator.  Because his formulation is valid only for circular orbits, near-synchronous rotation and inviscid fluids, we used $\beta=$1.001 and $\nu=$0.04 R$_\odot^2$/day, the minimum allowed viscosity by the code for this case. The dashed curves in this Figure illustrate the results from the calculations of our code for a $\Delta$R/R$_1=$0.025 surface layer and  the same orbital and stellar parameters. The one-layer velocity perturbations are very similar in shape and in amplitude to those of Scharleman's model at latitudes lower than  $\sim$44$^\circ$. At intermediate latitudes ($\sim$44$^\circ$--70$^\circ$) the shape is still the same, although the one-layer amplitudes are systematically larger. 

\begin{figure} [!h, !t, !b]
\begin{center}
\includegraphics[width=0.95\linewidth, height=0.7\linewidth]{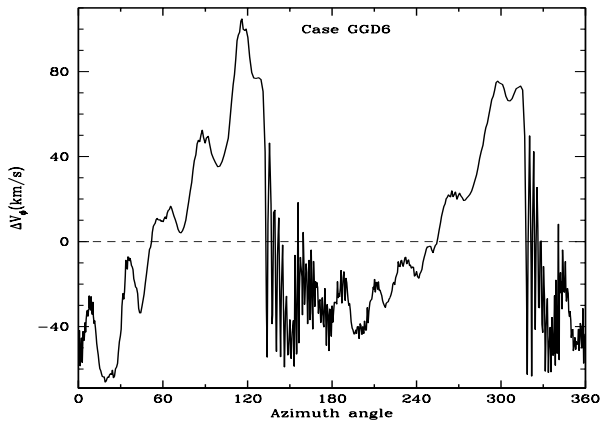}
\caption{An example of the azimuthal velocity perturbations at the equator from one of our eccentric binary models (CaseGGD6). These azimuthal perturbations do show the typical two-lobe tidal morphology but are significantly more complex than those shown in the previous Figure due to the larger eccentricity and rotation rates. The usual two-lobe tidal structure is always reproduced but  with significant smaller-scale structure induced by the non-zero eccentricity and larger stellar rotation rates. \label{propaz_plots}}
\end{center}
\end{figure}

	The velocity field becomes significantly more complex when eccentricity and faster stellar rotation rates are introduced, as illustrated in Figure \ref{propaz_plots}. There we show the azimuthal velocity field for the GGD6 one-layer model calculations for one of the parameter sets similar to those of Spica. Note that the pattern still consists of two maxima and minima, but their shape can no longer be described as simply sinusoidal and, in addition, a great deal of substructure is apparent.  Furthermore, because the orbit is eccentric, the pattern changes from one orbital phase to another. The maxima appear around azimuth angle $\sim$90--150$^\circ$ and  $\sim$270--330$^\circ$,  and correspond to material that is flowing in the direction of the stellar rotation.  The minima correspond to surface material that, with respect to the rigidly-rotating inner region, is flowing in the opposite direction; with respect to the external observer, this material is flowing slower than its corresponding rotation speed. 

\begin{figure*} [!h, !t, !b]
\begin{center}
\includegraphics[width=0.30\linewidth]{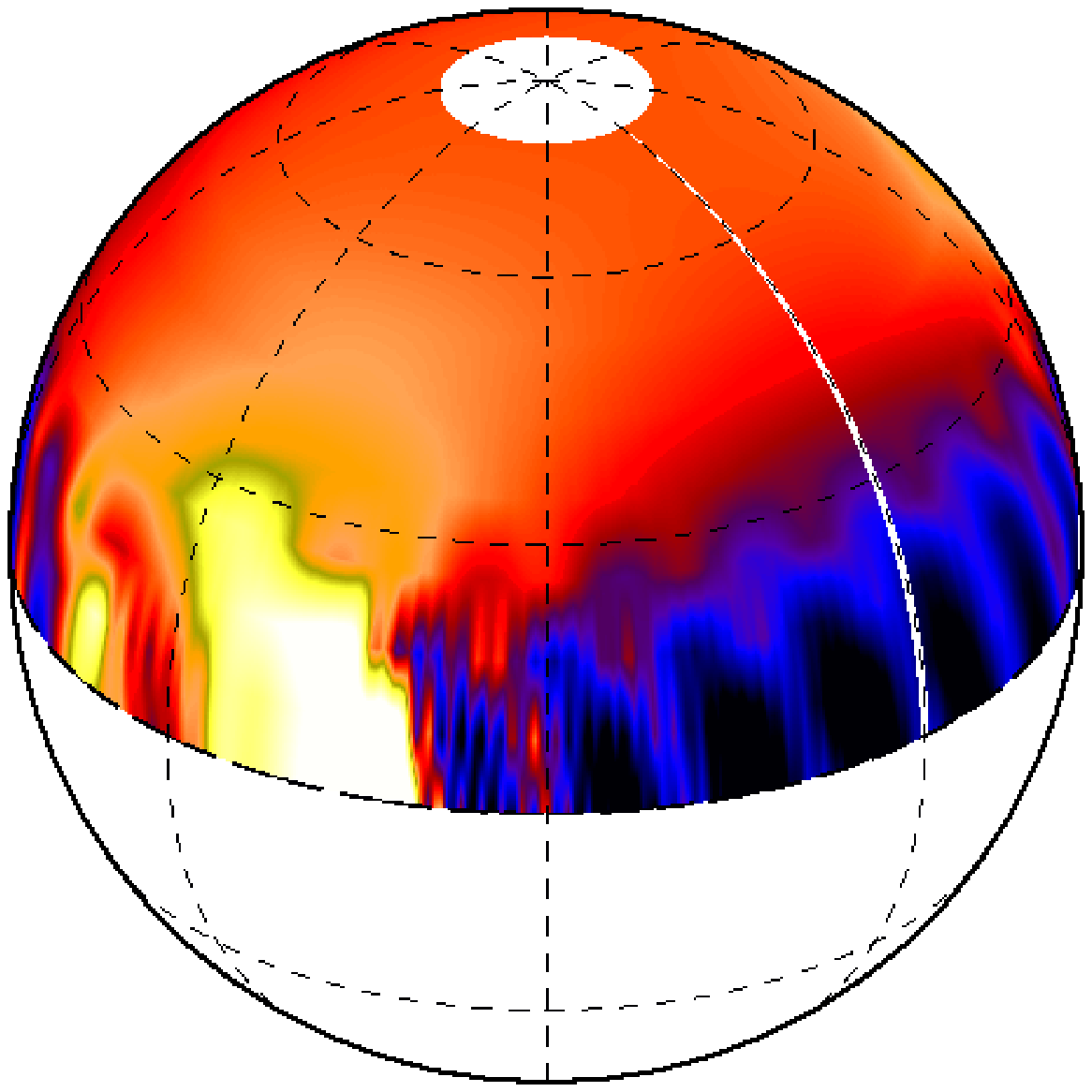}
\includegraphics[width=0.30\linewidth]{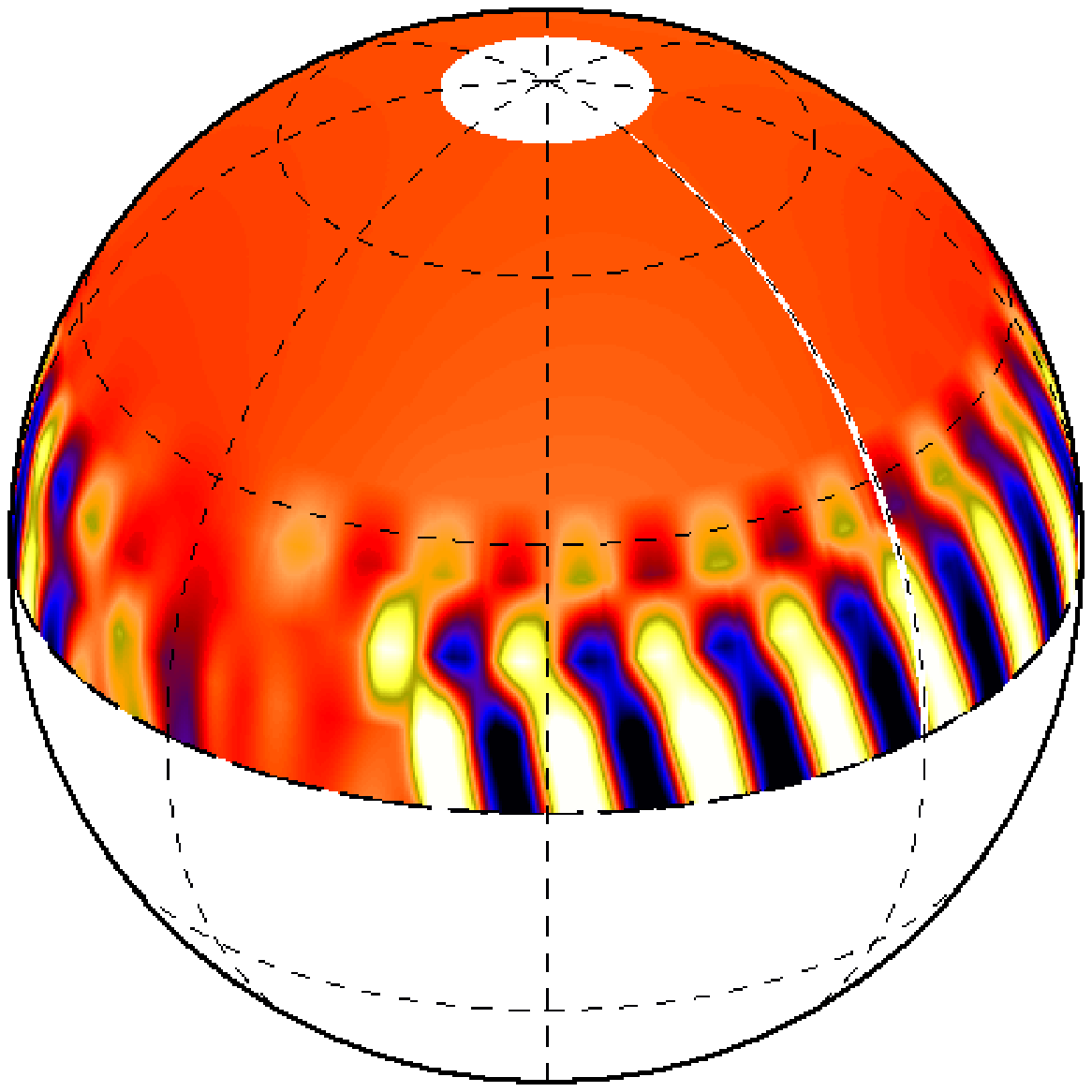}
\includegraphics[width=0.30\linewidth]{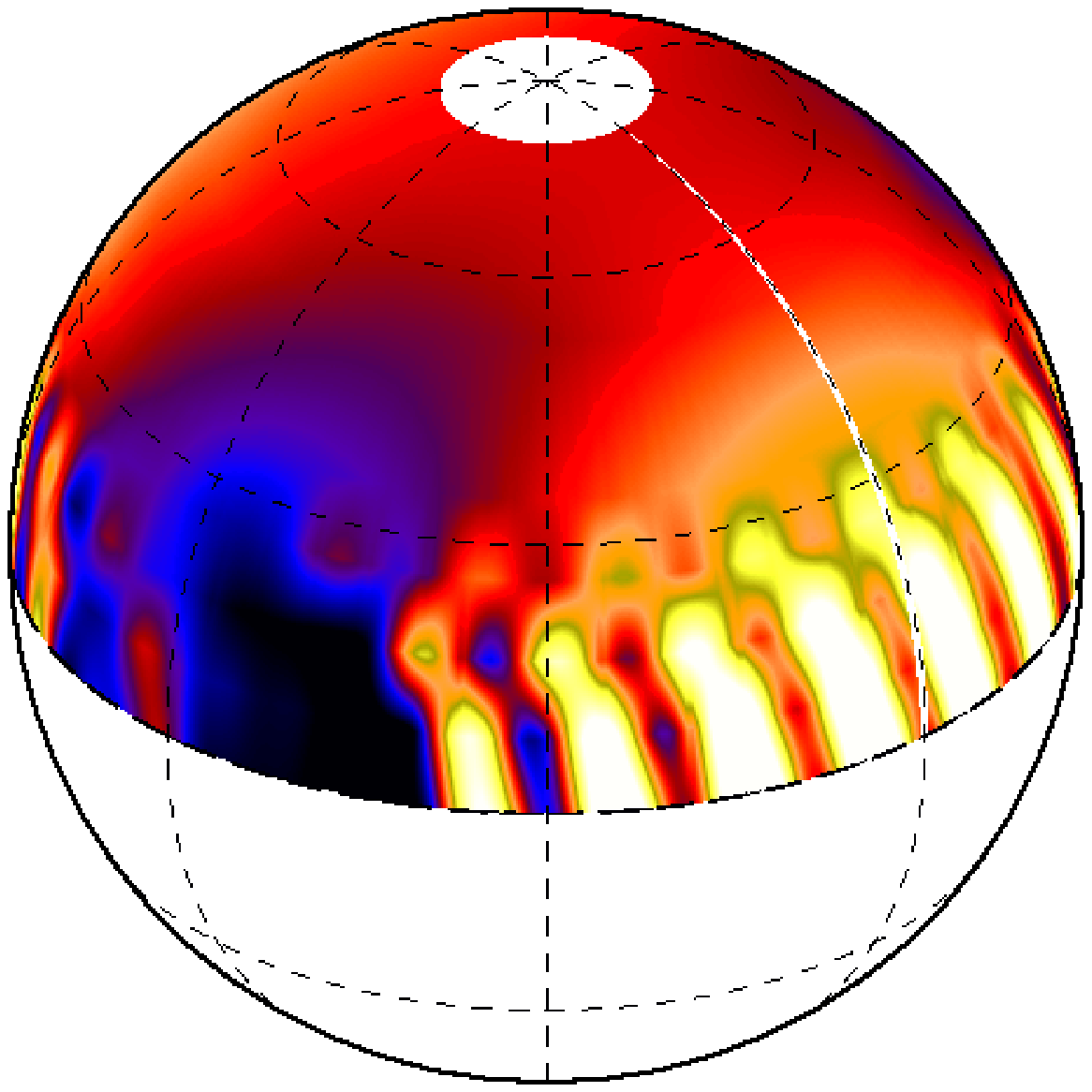}
\includegraphics[width=0.30\linewidth]{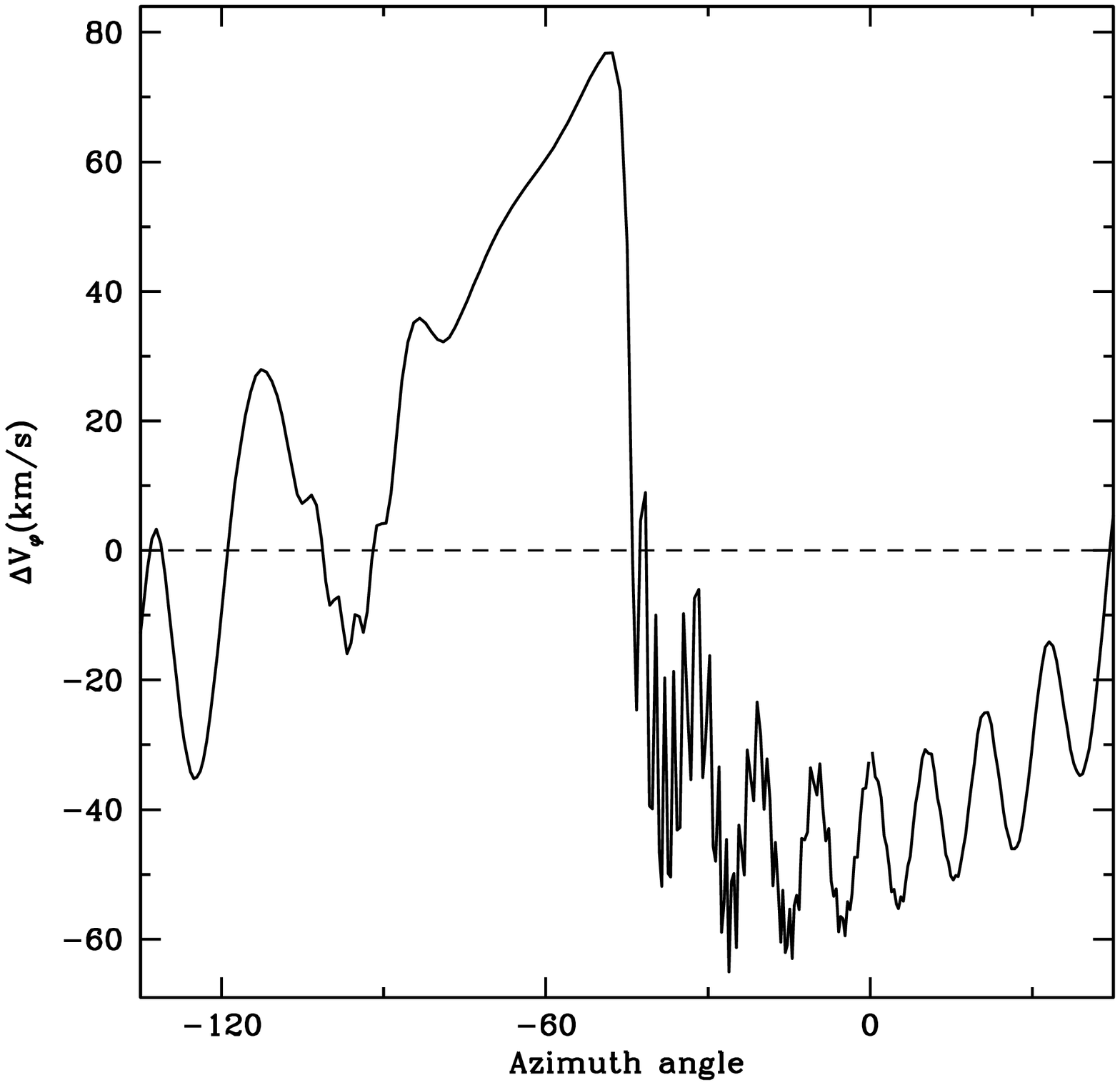}
\includegraphics[width=0.30\linewidth]{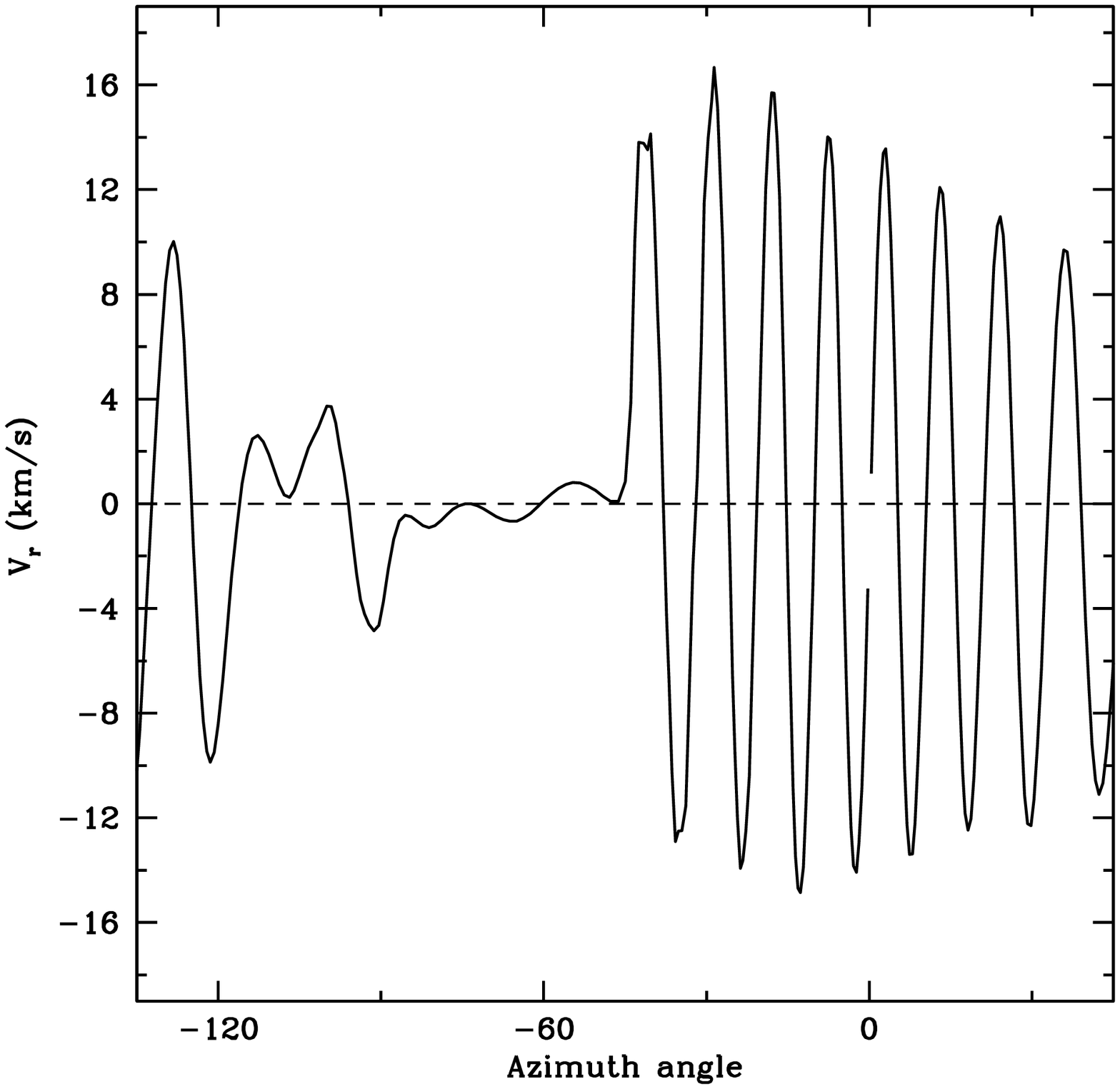}
\includegraphics[width=0.30\linewidth]{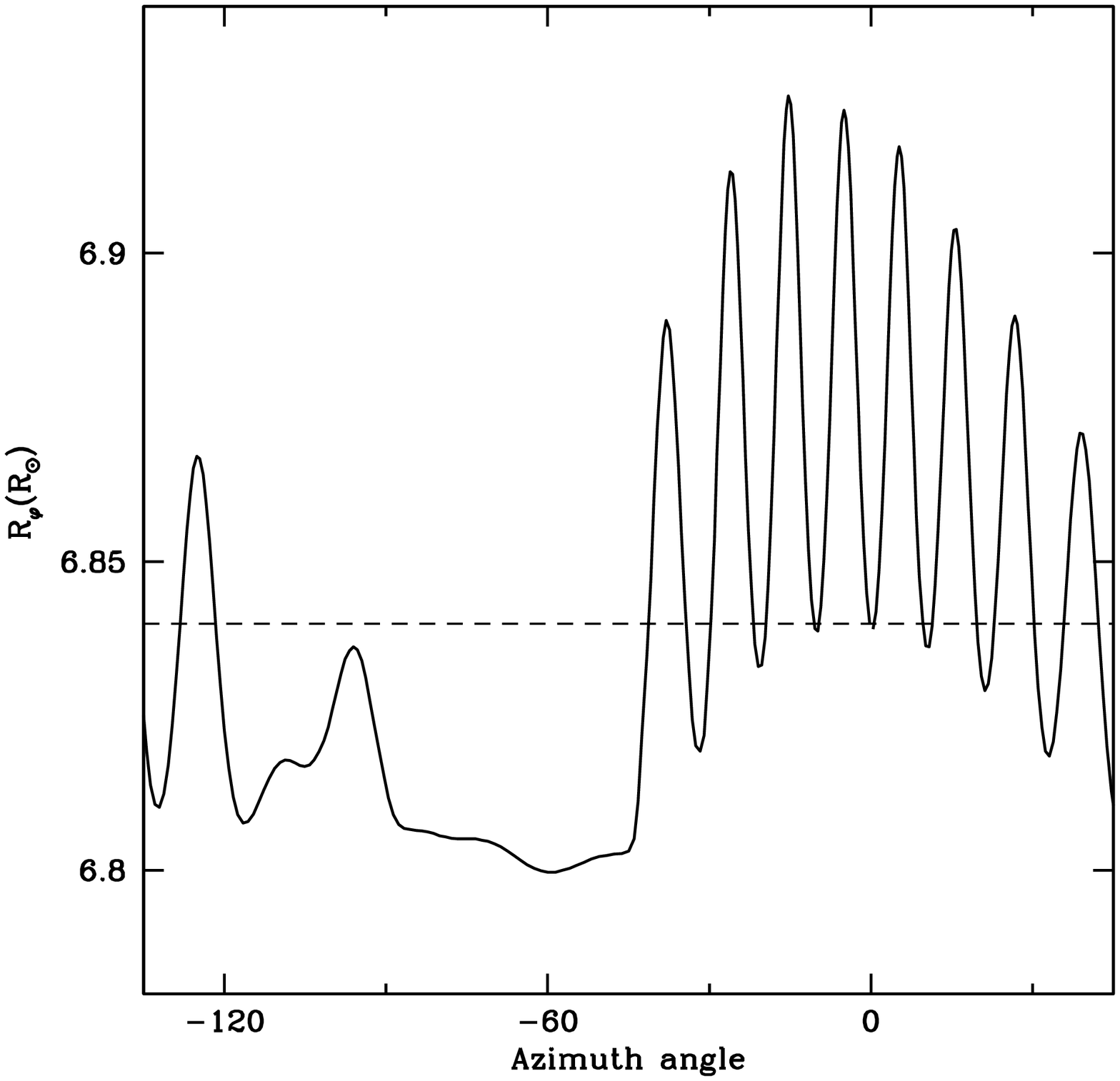}
\caption{Azimuthal (left) and radial (middle) velocity perturbations, and radial extent of the surface (right). Maps are color-coded such that white corresponds to maximum  perturbation and black to minimum. For $\Delta v_\varphi$,  maximum (positive) is in the direction of stellar rotation; for $v_r$ and $R_\varphi$, maximum correponds to expansion.  The accompanying plots provide the values for the equatorial belt and are centered the same as the maps are at  $\varphi=$270$^\circ$.  The displayed model is GGD at orbital phase 0.11.\label{GGDmaps}}
\end{center}
\end{figure*}

	These ``tidal flows" are represented for our model run GGD in Figure \ref{GGDmaps} (top left) with light and dark colors, corresponding to positive and negative perturbation speeds, respectively. This Figure also includes maps of the radial perturbations in both velocity, $v_r$ (middle), and in the stellar radius, $R_\varphi$ (right).

\begin{figure*} [!h, !t, !b]
\includegraphics[width=0.49\linewidth]{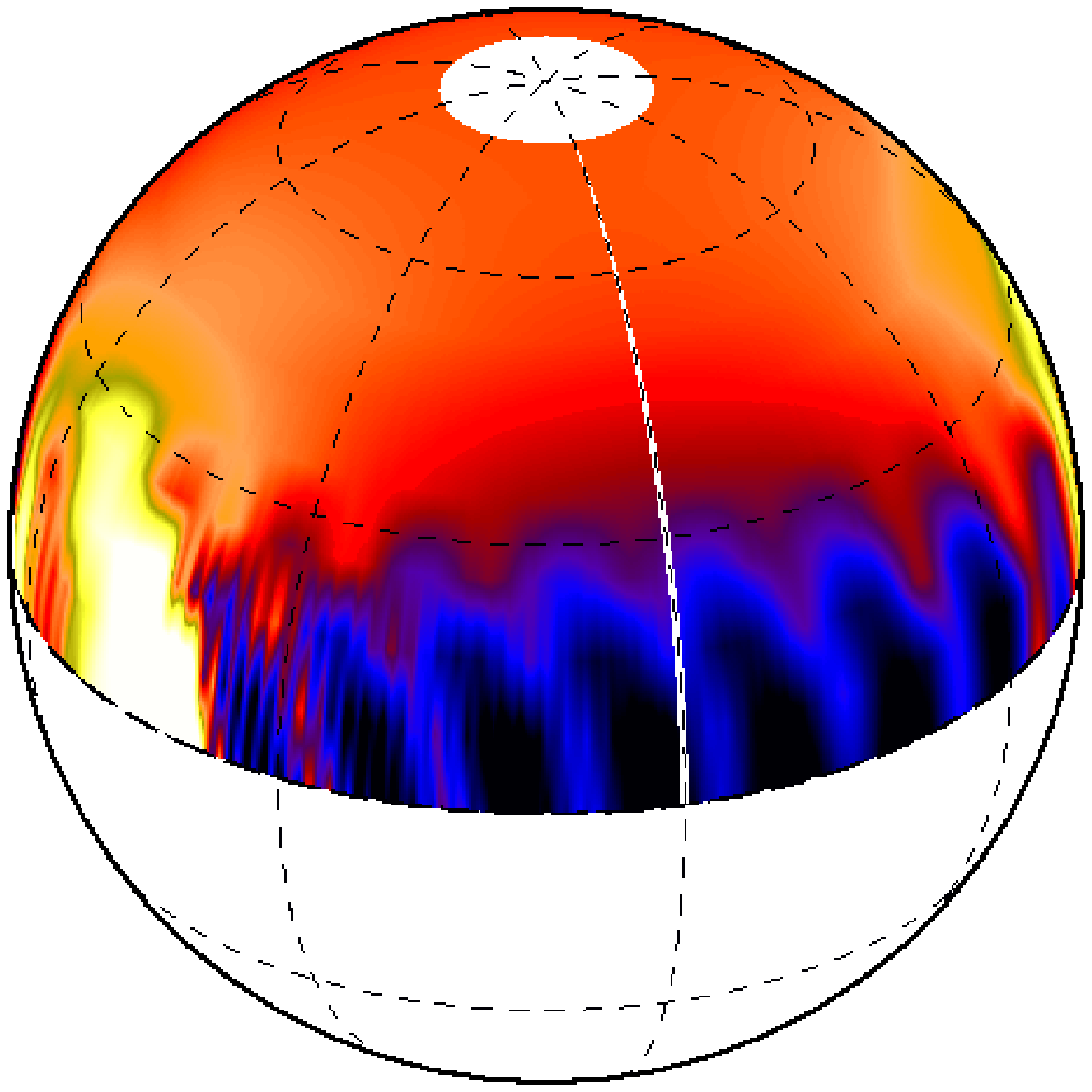}
\includegraphics[width=0.49\linewidth]{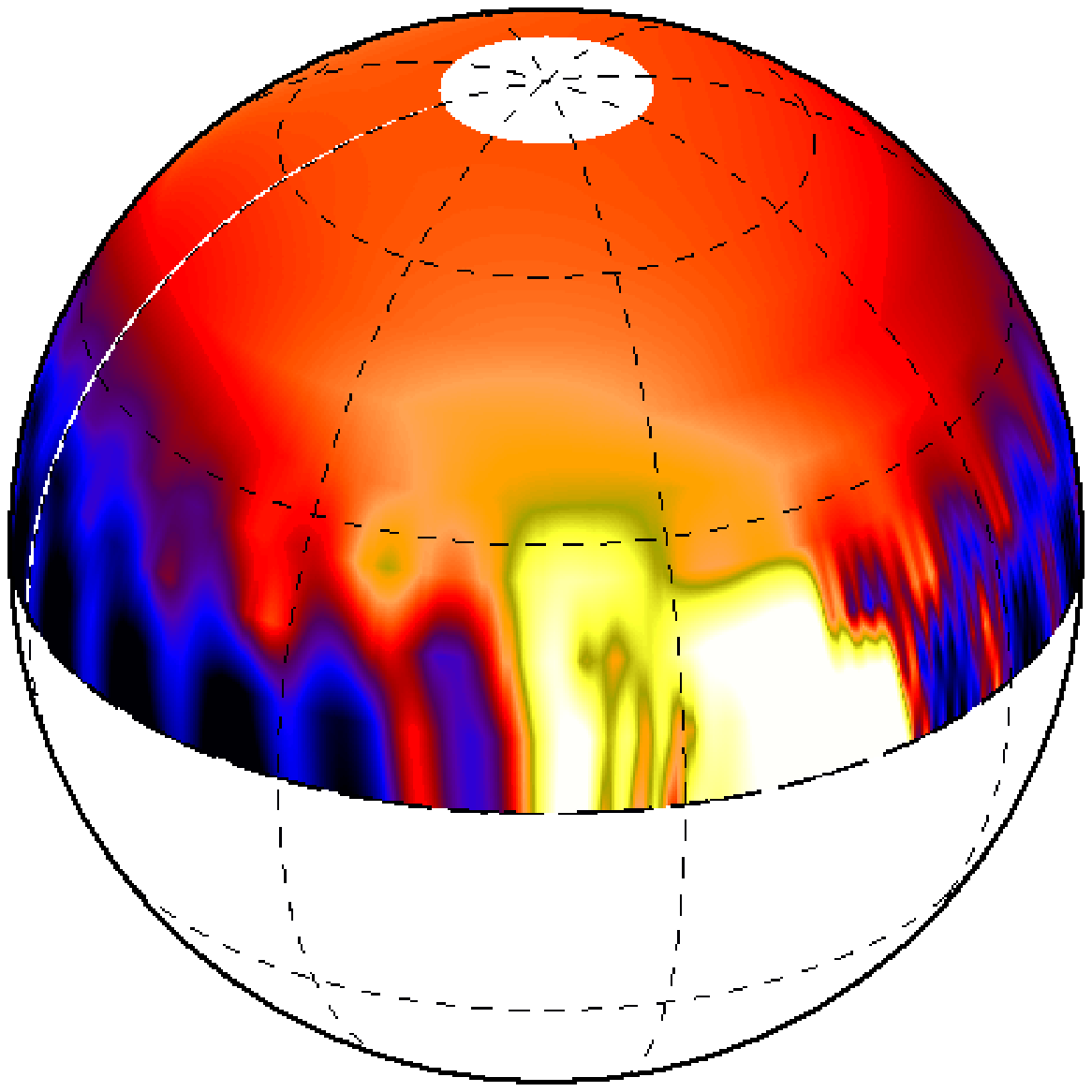} \\
\includegraphics[width=0.49\linewidth]{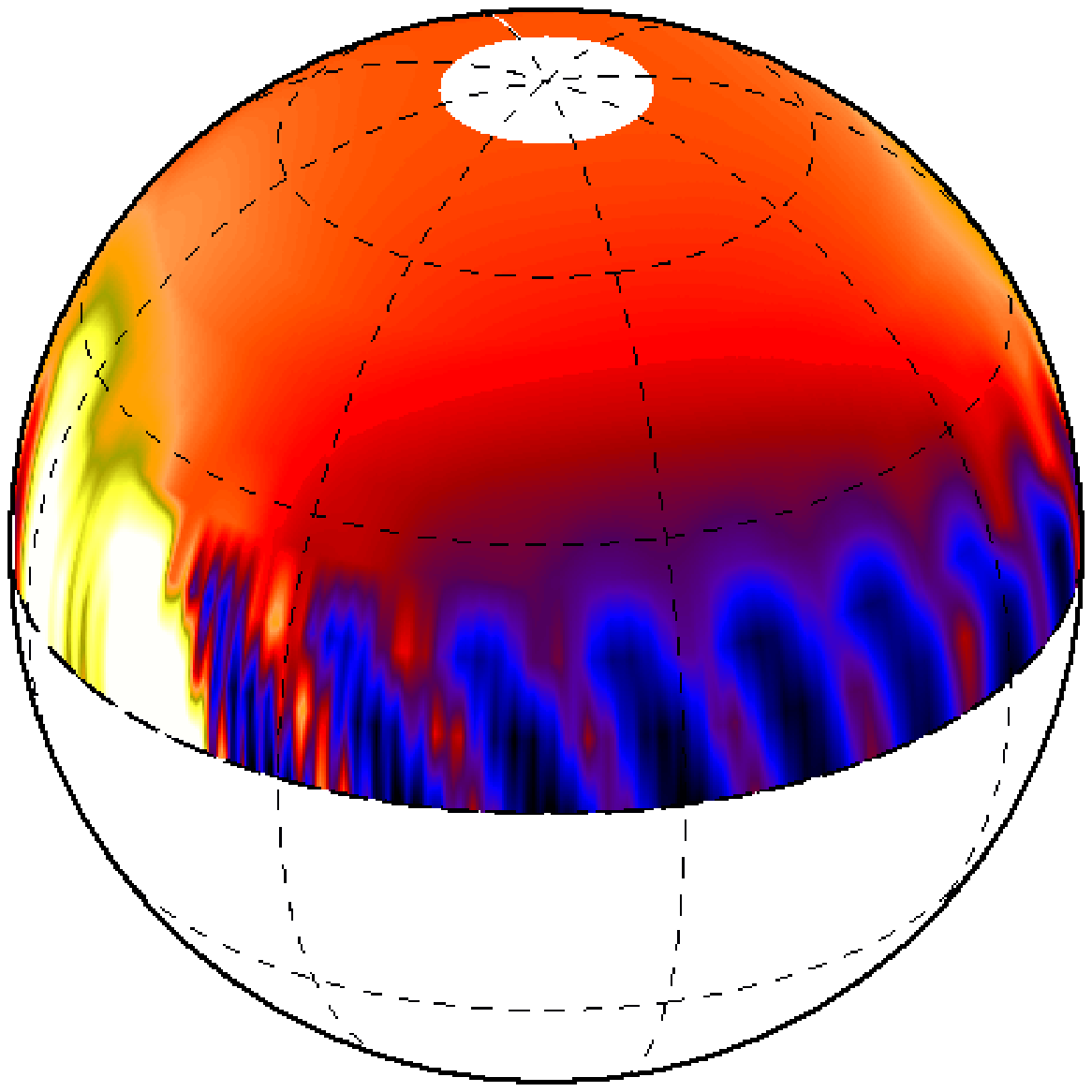}
\includegraphics[width=0.49\linewidth]{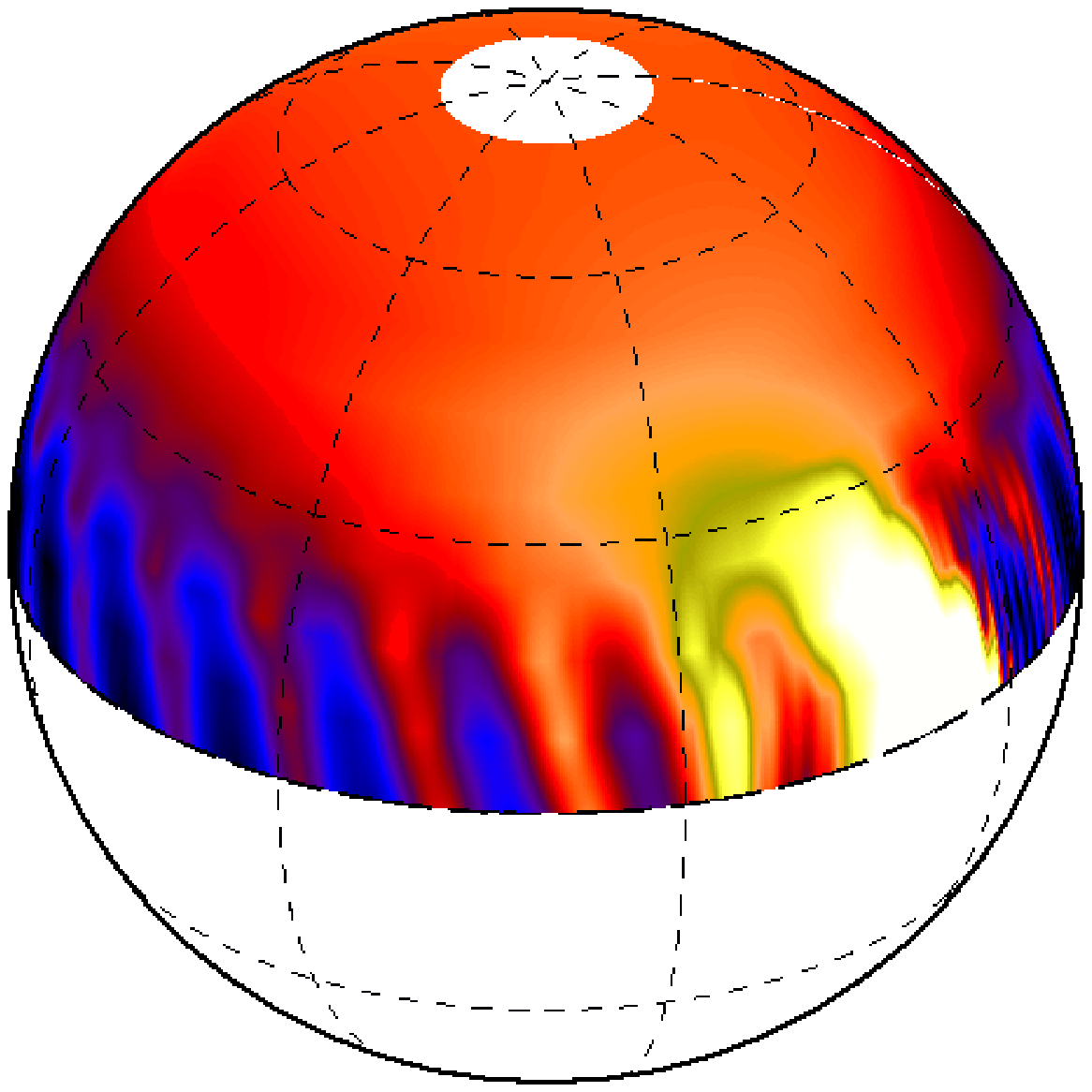}
\vspace{2mm}
\caption{The azimuthal velocity field from the GGD model calculation at four different orbital phases, oriented so as to illustrate the hemisphere that would be visible to an observer at these particular times; clockwise, starting at the top left map:  phase 0 (periastron), 0.25,  0.50 (apastron), and 0.75.  The rotation axis is tilted by 60$^\circ$ with respect to the line-of-sight to the observer.  The white broken line corresponds to the sub-binary longitude.  White(black) corresponds to maximum azimuthal perturbation in the direction of rotation (opposite to rotation).
 \label{GGDmaps_phases}}
\end{figure*}

	It is important to note that the  ``tidal bulge" is usually characterized as a region over which the maximum radial extent of the star is attained. Our model suggests that, although  on average the tidal bulge region is more extended, the detailed structure of the tidal bulge for an eccentric, super-synchronously rotating star such as Spica consists of small regions of alternating large and small radial extent.  These are accompanied by similar oscillations in the velocity perturbations. Hence, we conclude that even for a relatively small eccentricity, the ``dynamical tide" plays a major role and cannot be neglected. In Figure \ref{GGDmaps_phases} we illustrate the azimuthal velocity field evolving over an orbit on the star's hemisphere that would be visible to a distant observer. The velocity fields are shown for four different orbital phases: $\phi=$0 (periastron), 0.25, 0.5 (apastron) and 0.75, and assuming a longitude of periastron of $\omega=$255$^\circ$.
	
	A second  important point to note is  that the pattern of flow velocities is relatively fixed with respect to the coordinate axis that rotates with the binary system; it is {\em not} fixed with respect to the coordinate axis that is rotating with $m_1$.  Thus, this is conceptually very different from the idea of ``spots" that have relatively fixed locations on the stellar surface and rotate in and out of our field of view with the rotation angular velocity. Rather, the actual pattern changes with the orbital period; but given that there are two maxima and two minima per cycle, this  leads to variability on timescales 1/2 P.

\begin{figure} [!h, !t, !b]
\begin{center}
\includegraphics[width=0.95\linewidth, height=0.85\linewidth]{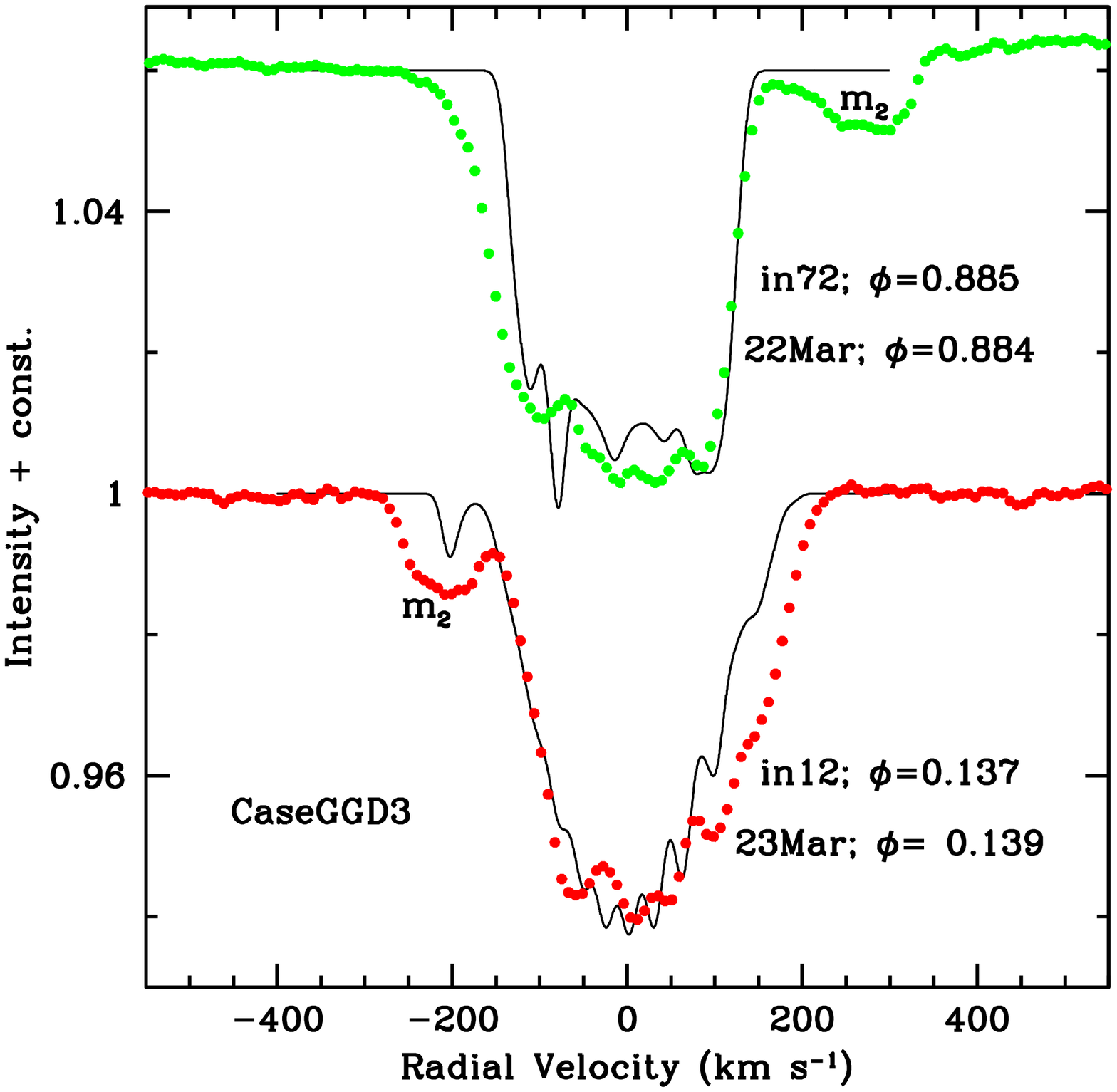}
\caption{Comparison of observations (dots) with theoretical profiles for orbital phases when the $m_2$ line profile is clearly separated from that of $m_1$.  The size of the dots corresponds to the uncertainty in the data. The theoretical profiles reproduce the general shape of the line-wings as well as the approximate number and location of the ``bumps".\label{montage6_GGD3}}
\end{center}
\end{figure}

\subsection{Line profile variability}

	Once the surface velocity perturbations have been computed, the projection of the velocity field along the line of sight to the observer provides the Doppler shift needed to produce the rotationally-broadened and tidally-perturbed line-profile that may be compared with observations.  The equations needed to perform the projection along the line-of-sight to the observer are described in Moreno et al. (2005).  The Doppler shifts derived from these projections are then applied to the  ``local" line profiles; i.e., the shape of the spectral line that is assumed to correspond to the emergent line radiation from each surface element. 

	Ideally, the local line profile should be derived from a model atmospheres computation.  In our current calculation, we adopt a Gaussian local line profile for the following reasons.  

	First, we have investigated the effect of various local line profile shapes on the shape of the total perturbed integrated profile and found the variation to be a very small. We used CMFGEN model atmosphere profiles (Hillier \& Miller 1998) for T$_{eff}$=25000K and log(g)=3.75 (kindly provided by John Hillier) as well as Gaussian and Voigt profile fits to these CMFGEN profiles. The CMFGEN model provides emergent flux (local line profiles) for a set of locations across the stellar disk ($\mu$). These model atmosphere profiles are then fit with Gaussian and Voigt functions at every $\mu$ so an accurate integrated line profile can be made using different local line profile shapes at every $\mu$. For the Voigt function fits, both parameters are varied to minimize the deviation between Voigt and CMFGEN profiles. Typical values for the damping parameter were 0.27 to 0.48. This illustrates how the variation in local line profile shape influences the total perturbed line profile and is detailed in the appendix. All local profiles give essentially identical rotationally-broadened line widths. The width of the total line profile is rotational-velocity dominated and the intrinsic line width and shape plays little role. The local line profile shape is somewhat more significant when computing the detailed shape of the discrete absorption-like features. A wider local profile will tend to smooth out these bumps, increasing their width and decreasing their amplitude, but the bump velocity and number of identifiable bumps remains unchanged for all the cases we examined. It is interesting to note that the CMFGEN model spectra produce slightly narrower local line profiles for the Si III line (smaller wings, shallower core) than do the Gaussian or Voigt profiles and would slightly increase the relative contrast of the bumps. We have investigated this effect using our computed tidal velocity fields as well as simulated velocity fields. These local line profile effects were found to be very small and can be easily neglected for the velocity fields and perturbed line profiles calculated here. See the Appendix for more details. 

	Secondly, it is important to note that given the dynamical conditions on the stellar surface, non-uniform temperature and microthermal speed distributions are most likely present. In addition, the shearing flows may lead to radial temperature gradients that differ from point to point over the stellar surface. Without a full 3D radiative transfer tratment, the impact of these effects on the shape of the local line profiles is difficult to predict and constitutes a source of systematic error. Thus, adopting local line profiles from a 1-dimensional model atmosphere is subject to a wide range of systematic errors and is as much of an approximation as is the use of a Gaussian local line profile.  

	Table \ref{tablemodpar} summarizes the range in input parameters that was explored in the nearly 200 model calculations  run for Spica as well as some specific examples.  A general feature of all the calculations is that the line-profile undergoes a strong change  in its shape on a day-to-day timescale, transitioning from a Gaussian-like profile  to a more ``boxy" shape. These changes may be associated with the ``equilibrium tide" component of the gravitatinal perturbation.  As pointed out by Smith (1985a), it is  the spectroscopic analogue of the ellipsoidal variability in the light curves and produces a  period of 1/2 the orbital period.  

	The change in the  line-profiles associated with this effect can be quantified using the kurtosis. The bottom panel in Figure \ref{rv} shows the variation of the kurtosis in the computed line profiles for the model Case GGD over the orbital cycle. As expected from the above discussion, there are two maxima and two minima in this curve, corresponding to the transitions into and out of the ``boxy" line-profile shape.  In order to compare the theoretical kurtosis with the observational data, we also measured the kurtosis in the Si III 4552.62{\AA} line of our 9 spectra.  Here, however, the wavelength interval measured was chosen to avoid the contribution from the binary companion. The trends in these measurements are plotted in Figure \ref{rv} as the filled-in trianges.  The error bars in the data correspond to the uncertainty introduced by increasing or decreasing the wavelength interval by 0.1{\AA}.  Thus, we conclude that our model adequately reproduces the variability trends in the observational line profiles showing that tidal flows are an important source of non-radial velocities.

	It is interesting to note that the line-profile variability introduces a systematic error in the measurement of the line centroid, as illustrated in the middle panel of Figure \ref{rv}. There we plot the difference between the centroid measured on the perturbed line-profile and the unperturbed profile at the same orbital phase. For the cases considered here, the largest error  occurs  in the orbital phase interval 0.6--0.7.

\begin{figure*} [!h, !t, !b]
\begin{center}
\includegraphics[width=0.45\linewidth]{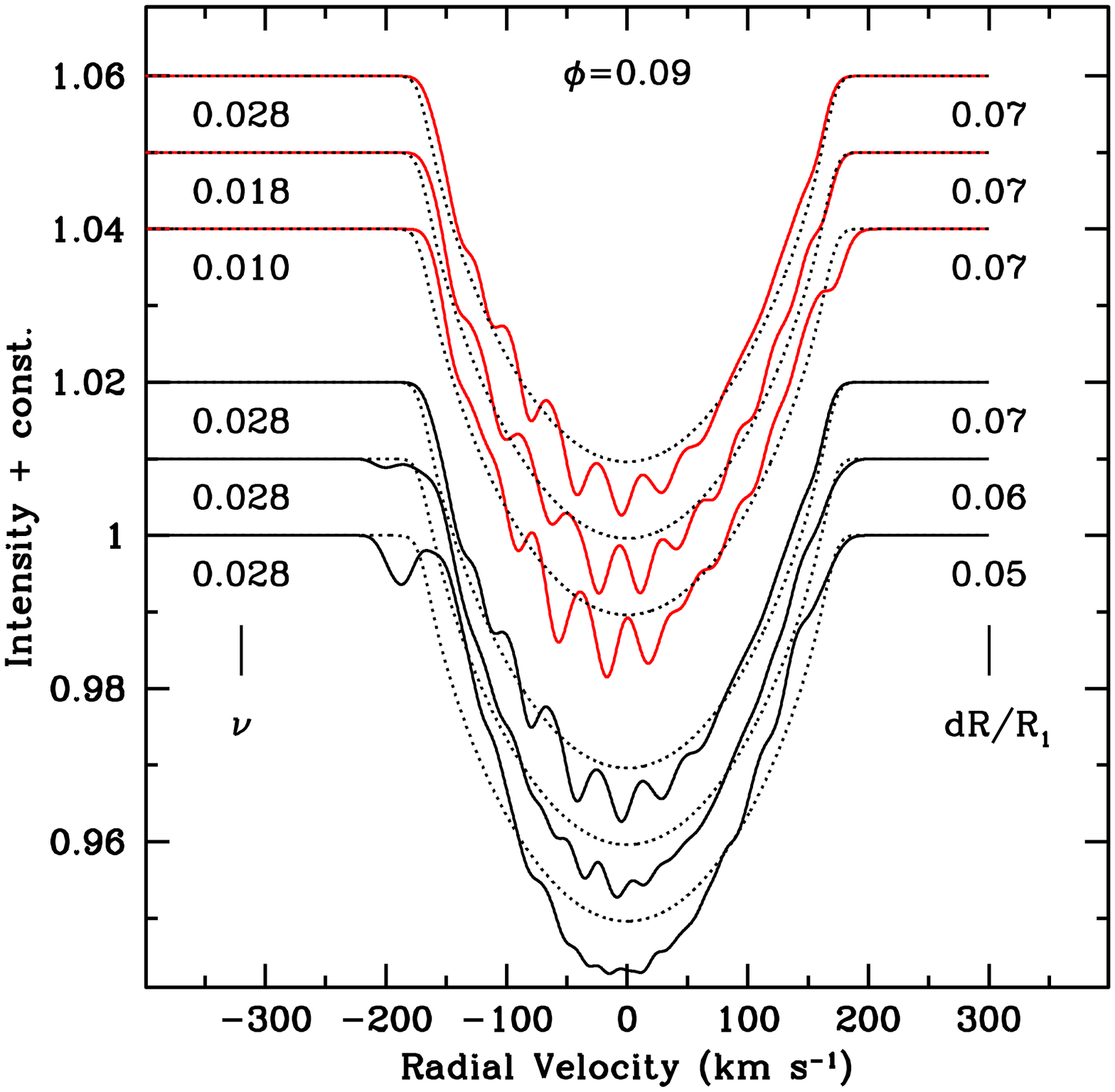}
\includegraphics[width=0.45\linewidth]{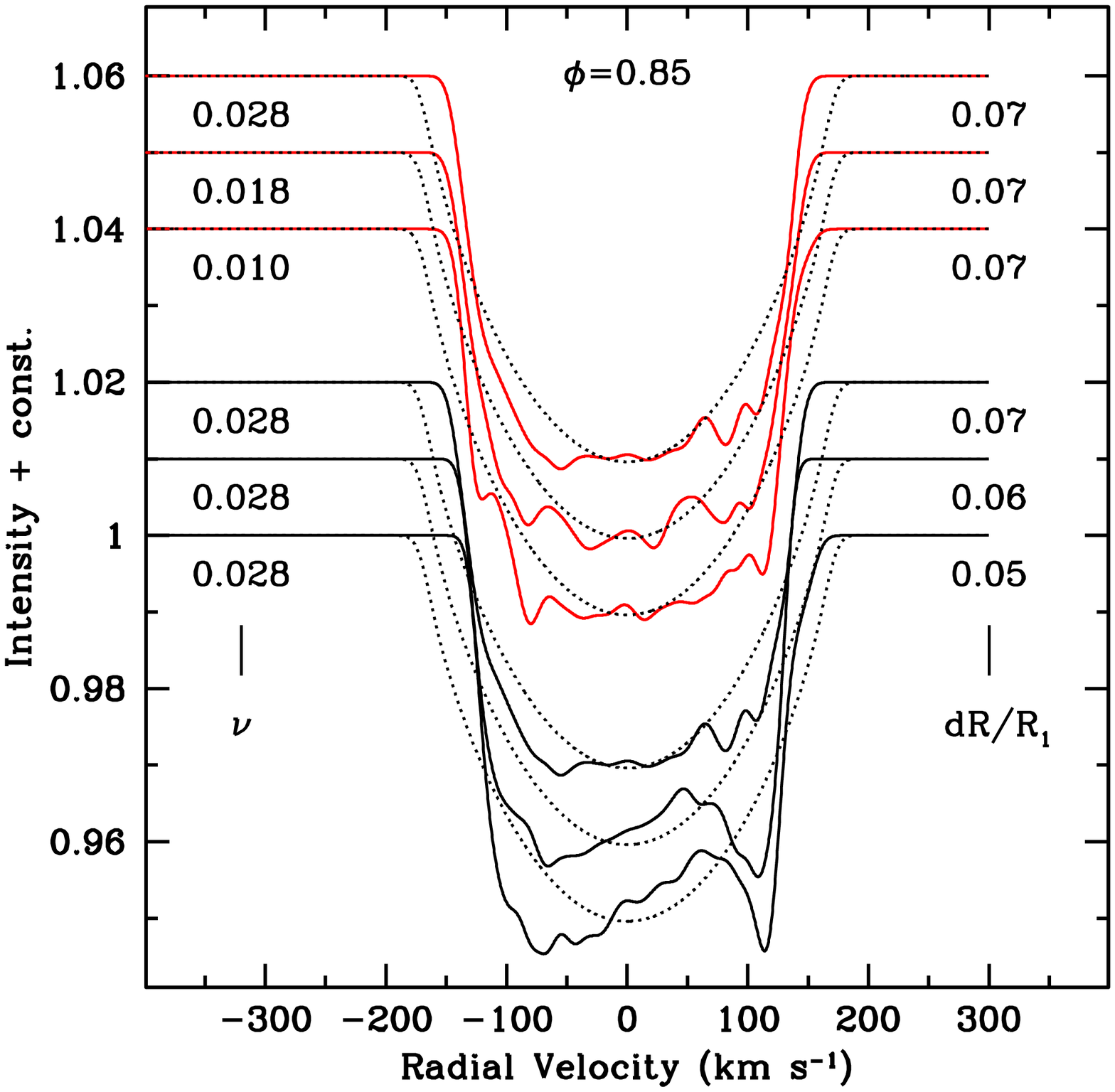}
\caption{Computed line profiles showing the effects of changing the depth of the oscillating layer, $dR/R_1$ and the viscosity, $\nu$.  Model runs C, D, E, G, GG2 and H were used to illustrate variation with layer depth and viscosity at two orbital phases - the left panel is $\phi=$0.09, the right is $\phi=$0.85. The dotted lines correspond to the unperturbed, rotationally-broadened line profile and the solid lines are perturbed. The layer depths and viscosities for each line are shown in each panel. The top group of line profiles have a fixed layer-depth with a varying viscosity while the bottom group varies layer depth. The general shape of the line is most sensitive to the depth of the oscillating layer, but the phase-dependent effects dominate over the relatively smaller differences introduced by the choice of this parameter. \label{montage5_plots}}
\end{center}
\end{figure*}

	A comparison of the observations with examples of the predicted line-profiles is shown in Figure \ref{montage6_GGD3}, illustrating that the model reproduces the orbital-phase dependent variations in both the general shape and  in the number and approximate location of the discrete absorptions.  It also reproduces the ``blue-to-red" motion of the discrete features as illustrated in the montage of all the theoretical line-profiles illustrated in the traveling bumps section  of the appendix. Also interesting to note is the presence in our computed profiles of the ``red spike" found by Smith (1985b)  and attributed to the horizontal motions. Finally, we note that since the profiles for all orbital phases are computed sequentially during the same run, the comparison of their time-dependent changes (from one orbital phase to the next) with that of the data constitutes yet another constraint.  Figure \ref{montage6_GGD3} illustrates that the model calculation is consistent with the data also in this respect.

\begin{figure} [!h, !t, !b]
\begin{center}
\includegraphics[width=0.95\linewidth, height=0.85\linewidth]{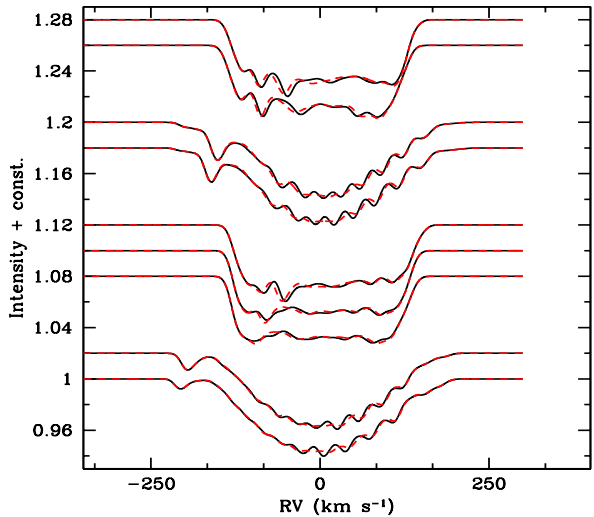}
\caption{Line profiles for the Spica GGD model computed with the total velocity field compared with the profiles computed with only the horizontal velocities, $\Delta v_\varphi$, showing that the radial component contributes very little to the line-profile variability.  Shown are orbital phases as in Figure \ref{mont1}. \label{azradcomp}}
\end{center}
\end{figure}

	The one important inconsistency between the model line-profile calculation shown in Figure \ref{montage6_GGD3} and the data is that we used $\omega=$209$^\circ$ for the imput parameter corresponding to the argument of periastron, instead of the value $\omega=$255$^\circ$ implied by the RV curve.  We wish to stress, however, that the theoretical profiles presented in Figure  \ref{montage6_GGD3} are not the product of a fitting process; these example profiles result directly from the tidal interaction computation using one of the sets of  stellar and orbital parameters that have been derived for the Spica system. Small modifications in some of the input parameters can have a significant impact on the details of the line-profiles. A more in-depth discussion of the effects on the line-profiles due to different  input parameters will  be presented elsewhere.  However, it is important here to illustrate the effects produced by the two free parameters of the model, $\nu$ and $dR/R_1$.  Figure \ref{montage5_plots} shows that the large-scale orbital phase-dependent variability is generally much stronger than the variations produced by changing  $\nu$ and $dR/R_1$.  Hence, once the stellar and the binary parameters are constrained, the number of free parameters that enter into our calculation are greatly reduced. Finally, Figure \ref{azradcomp} illustrates the perturbed line-profiles computed with the full velocity field (as in Figures \ref{montage6_GGD3} and \ref{montage5_plots}) compared with the profiles computed with only the horizontal velocity component, $\Delta v_\varphi$ (dashes).  These profiles are nearly identical, showing that the radial component of the velocity field contributes very little towards the total line-profile variability.

\section{Conclusions and future directions}

	In this paper we have presented very high quality observations of line-profile variability. The combination of high spectral resolution with very high precision allows a much more detailed comparison of observation and theory. Though we do not have a continuous sequence of spectra, we presented some observations that show remarkable similarities after a full orbit while others show significant differences between orbits.

	We find that the Moreno et al. (2005) approach to computing line profile variability from tidal flows yields theoretical profiles that can be very similar to observational data.  Specifically, the theoretical profiles display two characteristics of Spica's line-profile variability: 1) the ``boxy" appearance of the line that is observed twice per orbital cycle and that is associated with the ``equilibrium tide"; and 2) the superposed narrow absorption-like features that travel from the blue wing to the red wing on timescales of $\sim$1 day and that may be associated with the ``dynamical tides". This is also the 'spike' reported in Smith 1985b. Both of these characteristics are caused primarily by the horizontal components of the surface velocity field and thus, we conclude that tidal flows are the dominant contributor to Spica's line-profile. Because the simple one-layer model cannot be expected to yield information on the internal structure of the star, the relatively good coincidence between the {\it ab initio} calculation and the observations indicates that in a close binary system such as Spica, the tidal forcing by the companion is the dominant factor determining the surface perturbations. The azimuthal velocity fields that are calculated from tidal interactions have the potential to explain much of the observed variability. 
	
	Given the fact that our computation is performed from first principles with no built-in assumptions regarding the surface velocity field and heavy physical constraints on our input parameters, we consider the coincidence between the theoretical and observational line profiles to be very encouraging. We do use a single layer fluid approximation but all the dynamics are controlled from the fundamental equations of motion. The structure of the fluid flow is not constrained to fit any specific mathematical form. The velocity field, subject to the single-layer approximation, is computed directly from the equations of motion.  This illustrates the importance of properly treating surface flows (the horizontal component of the surface motions) when analyzing observed line profile variability. 
	
	There are several aspects of this model that can be readily explored. The strength of the narrow spectral features (bumps) are not ideally reproduced. The feedback from shear energy dissipation into surface temperature and local line profile shape should be included to test the significance of dissipation in computing the predicted profiles. Ideally, model atmosphere line profiles and limb darkening should be used in the general scheme, although a proper treatment requires the use of a 3D model atmosphere calculation. In the meantime, however,  we have shown the shape of the local line profiles to be of little significance for this particular system.  Also, incorporating a  more realistic stellar structure into the calculation should allow exploration of the effects produced by the interplay between the tidal forcing and the normal modes of oscillation.

\begin{figure*} [!h, !t, !b]
\begin{center}
\includegraphics[width=0.35\linewidth, angle=90]{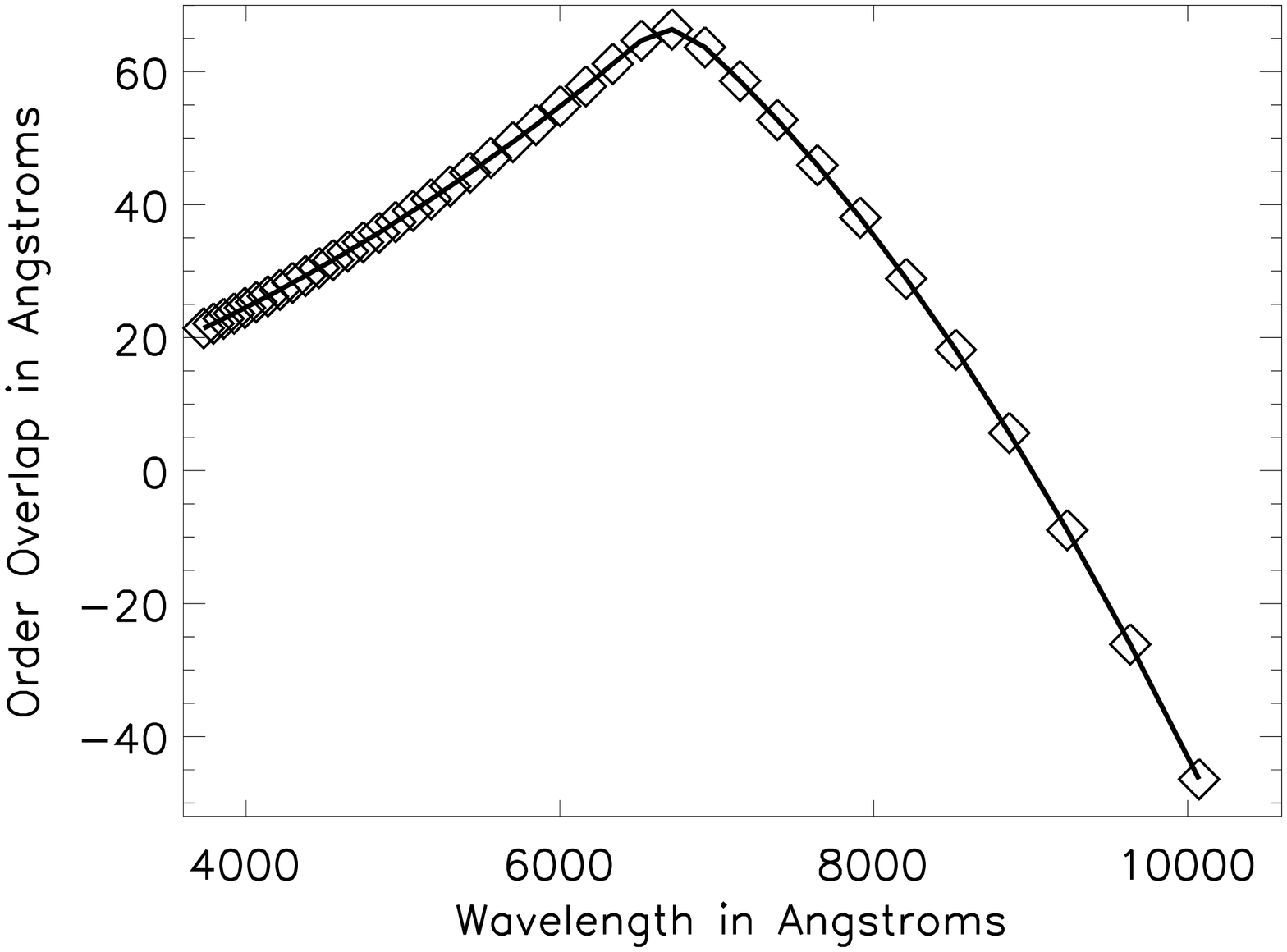}  
\includegraphics[width=0.35\linewidth, angle=90]{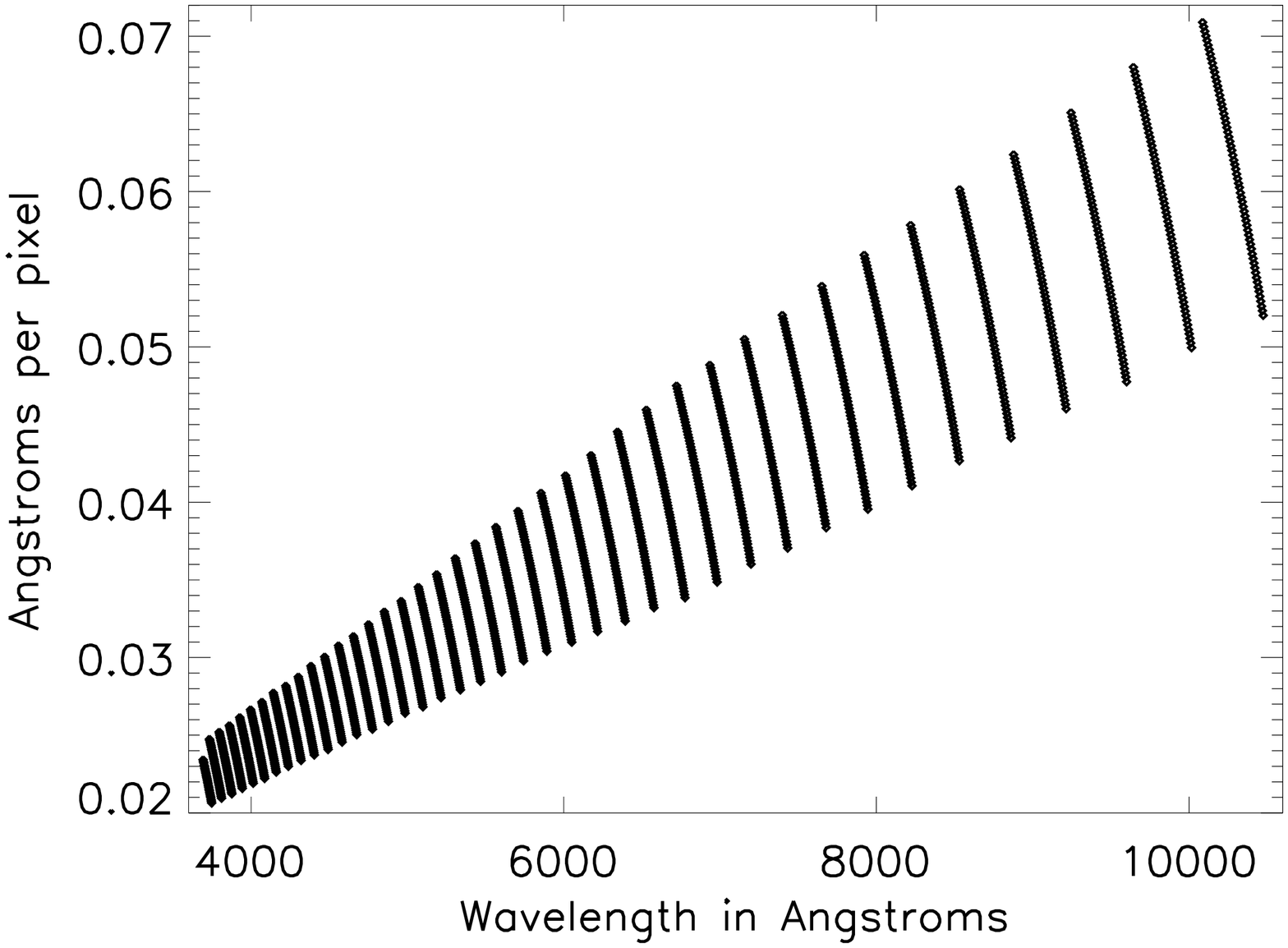}  \\ 
\includegraphics[width=0.35\linewidth, angle=90]{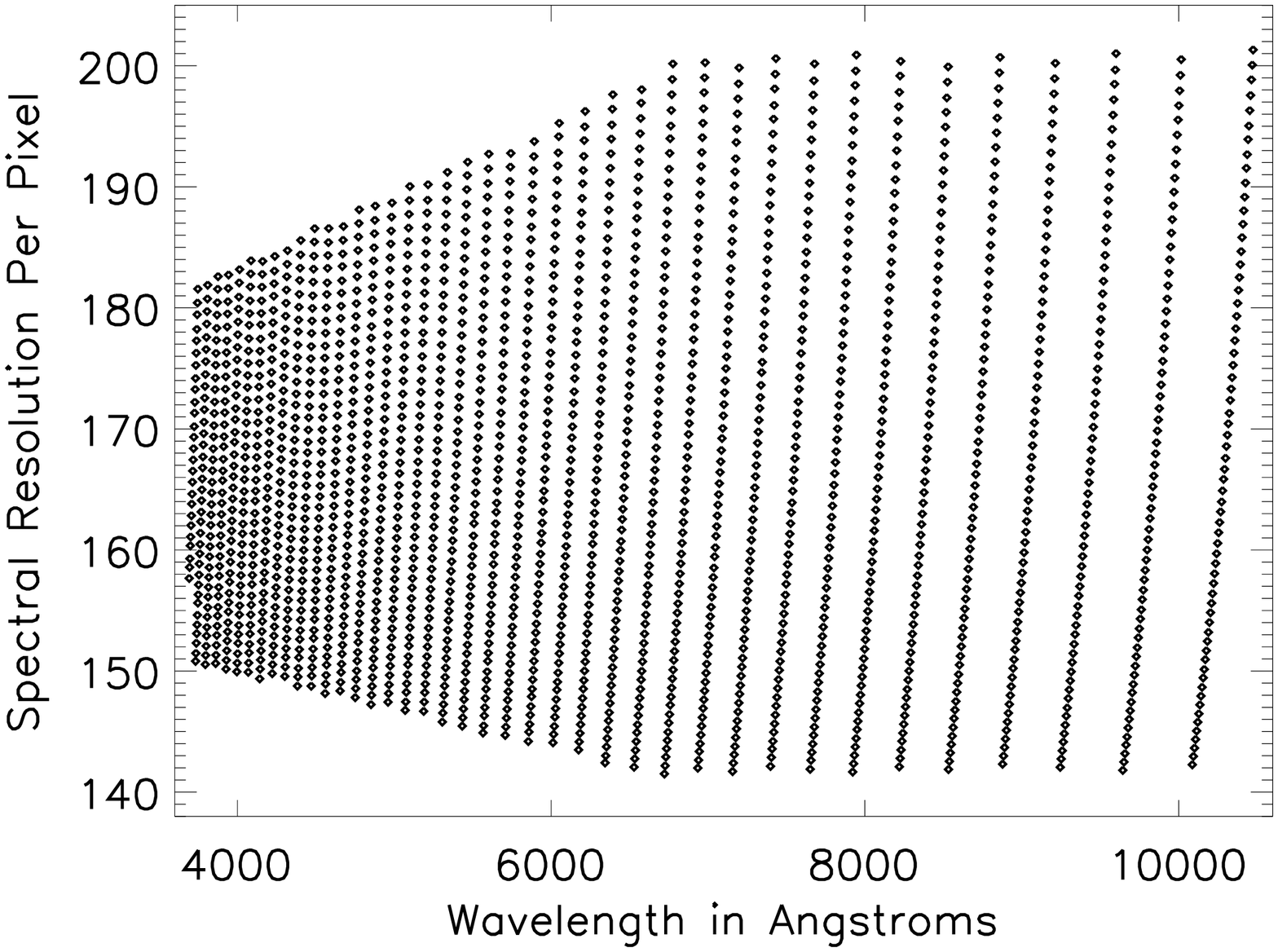}  
\includegraphics[width=0.35\linewidth, angle=90]{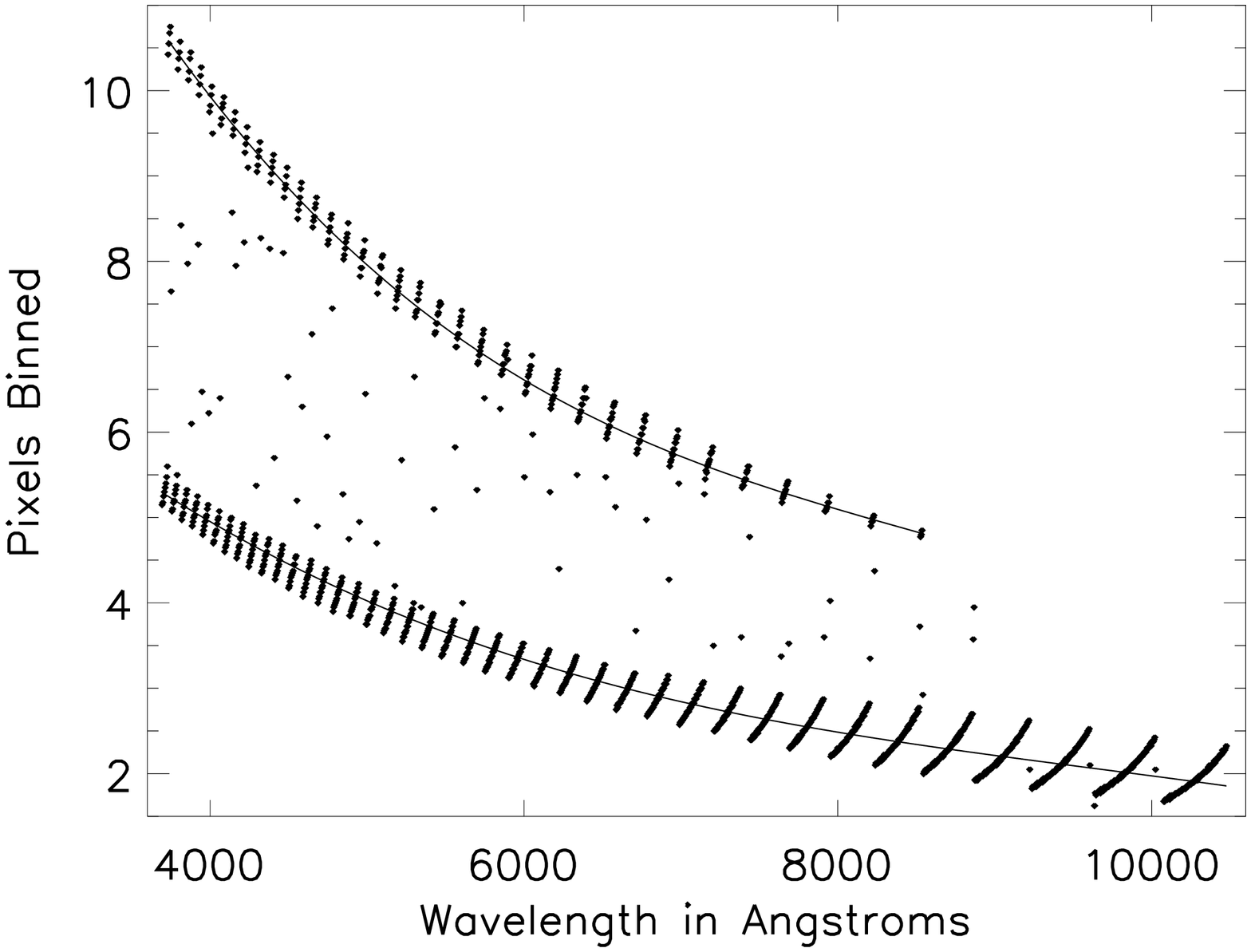} 
\caption{\label{espprop} The pre-processing properties of ESPaDOnS data as reduced by Libre-Esprit. Box a shows the order overlap as a function of wavelength. Box b shows the dispersion as a function of wavelength. Box c shows the corresponding spectral-resolution at single-pixel sampling. Box d shows the number of spectral-pixels used in the processing scripts to bin in wavelength to a constant 1{\AA}  per spectral-pixel coverage. Order-overlap and non-linear coverage leads to a bi-modal distribution of binned-pixels as well as a decrease with wavelength.}
\end{center}
\end{figure*}

\section{Acknowledgements }

	We express our gratitude to the referee, Myron Smith, for comments and suggestions that helped improve
the paper significantly, to John Hillier for providing the CMFGEN model spectra, Ben Brown for guidance in producing maps and Jason Aufdenberg for providing preliminary results on Spica's fundamental parameters. This program was partially supported by the NSF AST-0123390 grant, the University of Hawaii and the AirForce Research Labs (AFRL); as well as UNAM/PAPIIT IN-106708 and CONACYT 48929. This research used the facilities of the Canadian Astronomy Data Centre operated by the National Research Council of Canada with the support of the Canadian Space Agency. These observations were reduced with the dedicated software package Libre-Esprit made available by J. -F. Donati. This program used the Simbad data base operated by CDS, Strasbourg, France as well as the Markwardt and Coyote IDL Libraries available on the web. The NASA Astrophysics Data System (ADS) was frequently utilized.

\section{Appendix}

	The non-linear and irregular wavelength sampling of the ESPaDOnS spectropolarimeter led us to develop routines to linearize the output spectra. Libre-Esprit produces wavelength-calibrated spectra for every possible spectral pixel in the focal plane. As with all cross-dispersed echelle spectrographs, the dispersion, order-overlap and sampling are all functions of wavelength. An additional complication is that the ccd pixels correspond to 2.6km/s but the spectra are output by Libre-Esprit at 1.8km/s resolution, 30\% smaller than a pixel. These are called ccd and spectral pixels respectively. Figure \ref{espprop}a shows the order overlap as a function of wavelength for a typical ESPaDOnS exposure in our data-set. The order overlap increases from in overlap from 20{\AA}  at 4000{\AA}  to over 60{\AA}  around 7000{\AA}  before falling strongly to become a more than 40{\AA} gap in wavelength coverage by 10000{\AA} . The dispersion, shown in Figure \ref{espprop}b, runs from 0.02{\AA}  per pixel to 0.06{\AA}  per pixel in proportion to the wavelength increase. This leads to an essentially constant spectral resolution, shown in Figure \ref{espprop}c as roughly R$\sim$170,000. As the full-width-half-max of most calibration-lamp lines is over 2 pixels, the spectral resolution is around 70,000. 

	We desired a monotonic wavelength grid that assigns one intensity to one wavelength and combines all pixels covering the same wavelength to achieve higher signal-to-noise ratio. The processing scripts we wrote calculate the number of spectral-pixels to bin to achieve 0.12{\AA}  per pixel coverage for all wavelengths. This is shown in Figure \ref{espprop}d. Since the order overlap is significant, in some cases 30\% of an order, there is a bi-modal distribution in the number of pixels binned. For the part of an order with overlapping coverage, a linear wavelength coverage gives roughly twice the number of pixels binned. Beyond roughly 8000{\AA} , the order overlap disappears and the dispersion reaches 0.05{\AA}  per pixel resulting in a nearly regular 2-pixel binning factor.

\begin{figure*} [!h, !t, !b]
\begin{center}
\includegraphics[width=0.35\linewidth, angle=90]{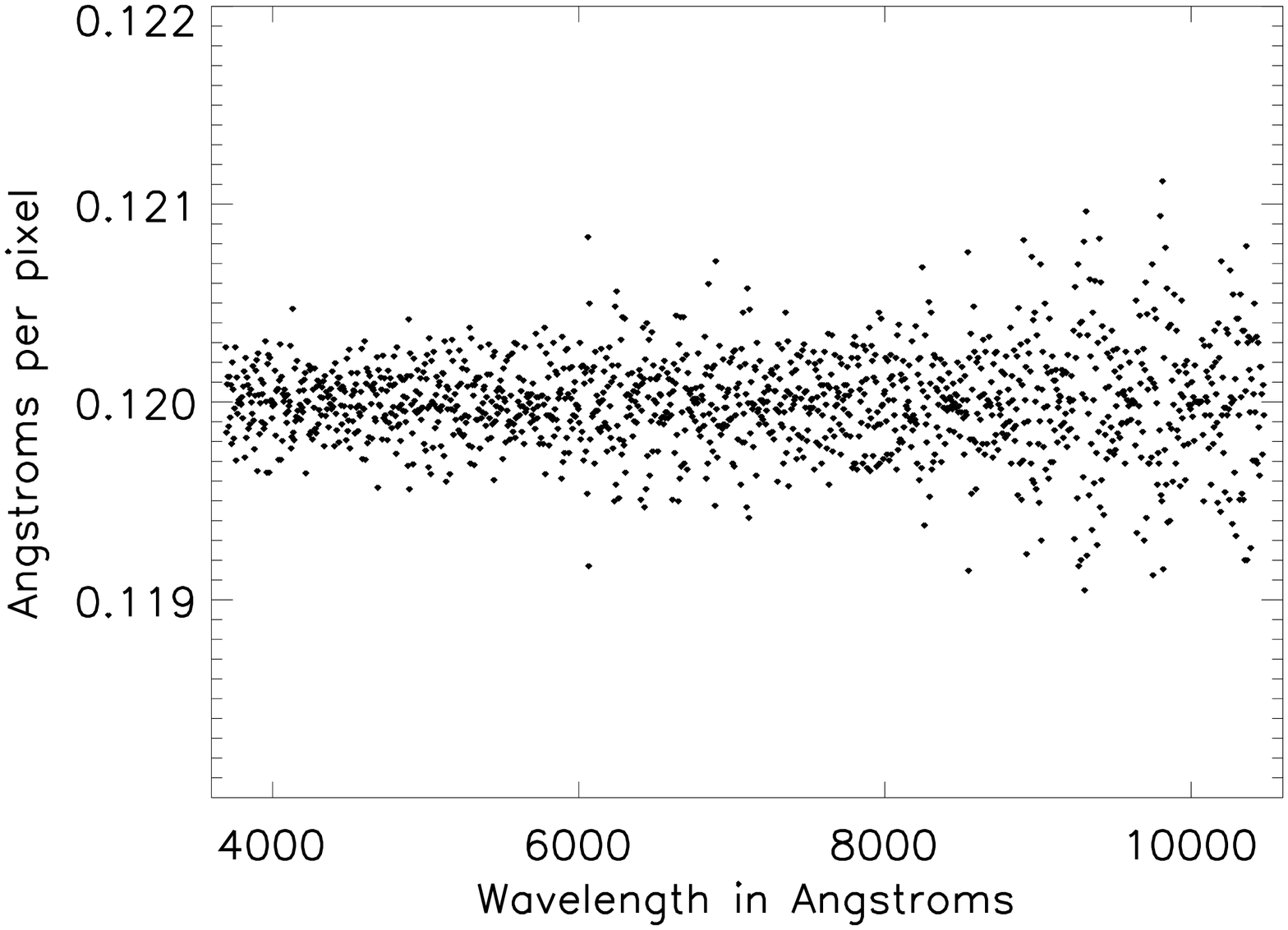}  
\includegraphics[width=0.35\linewidth, angle=90]{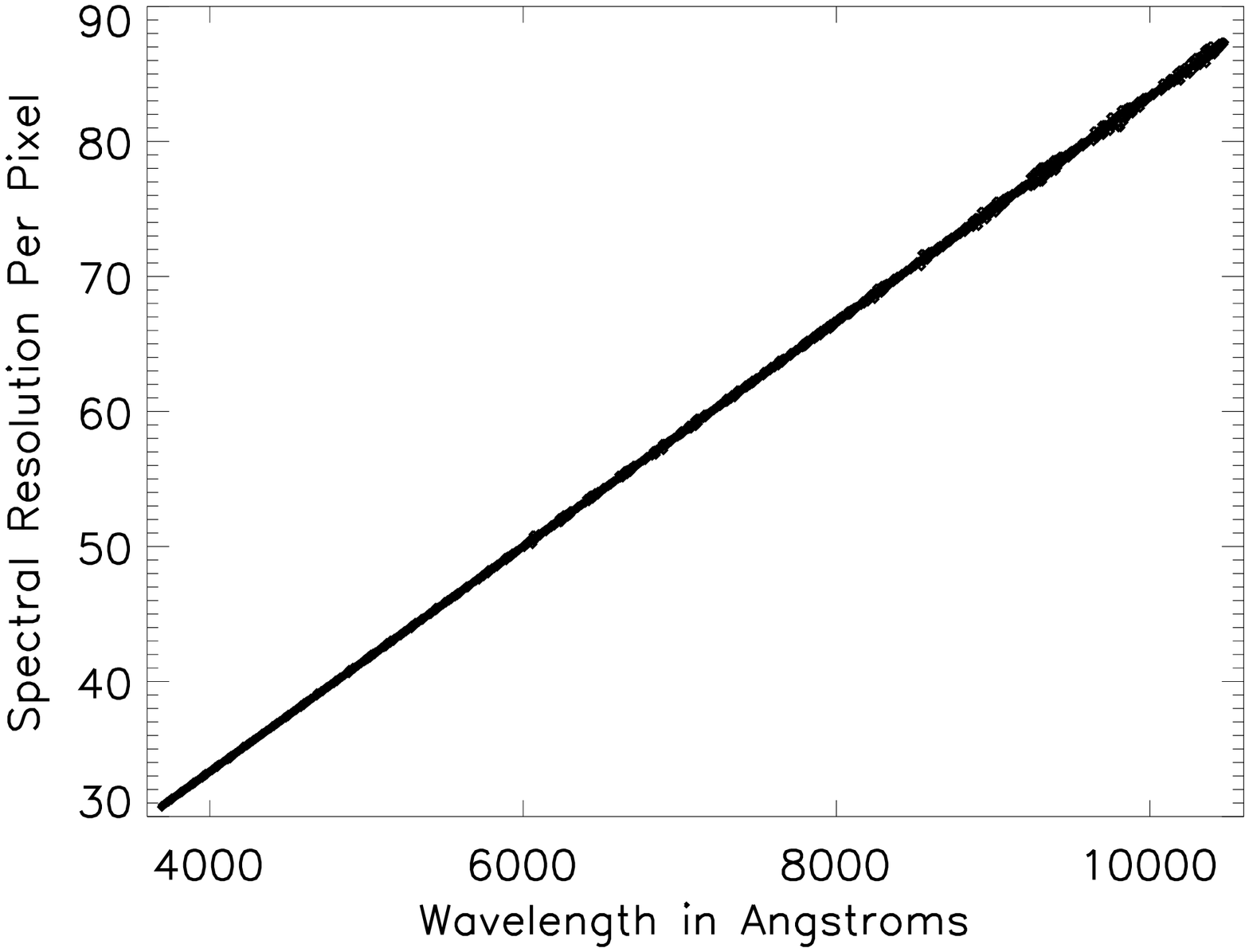}  \\ 
\includegraphics[width=0.35\linewidth, angle=90]{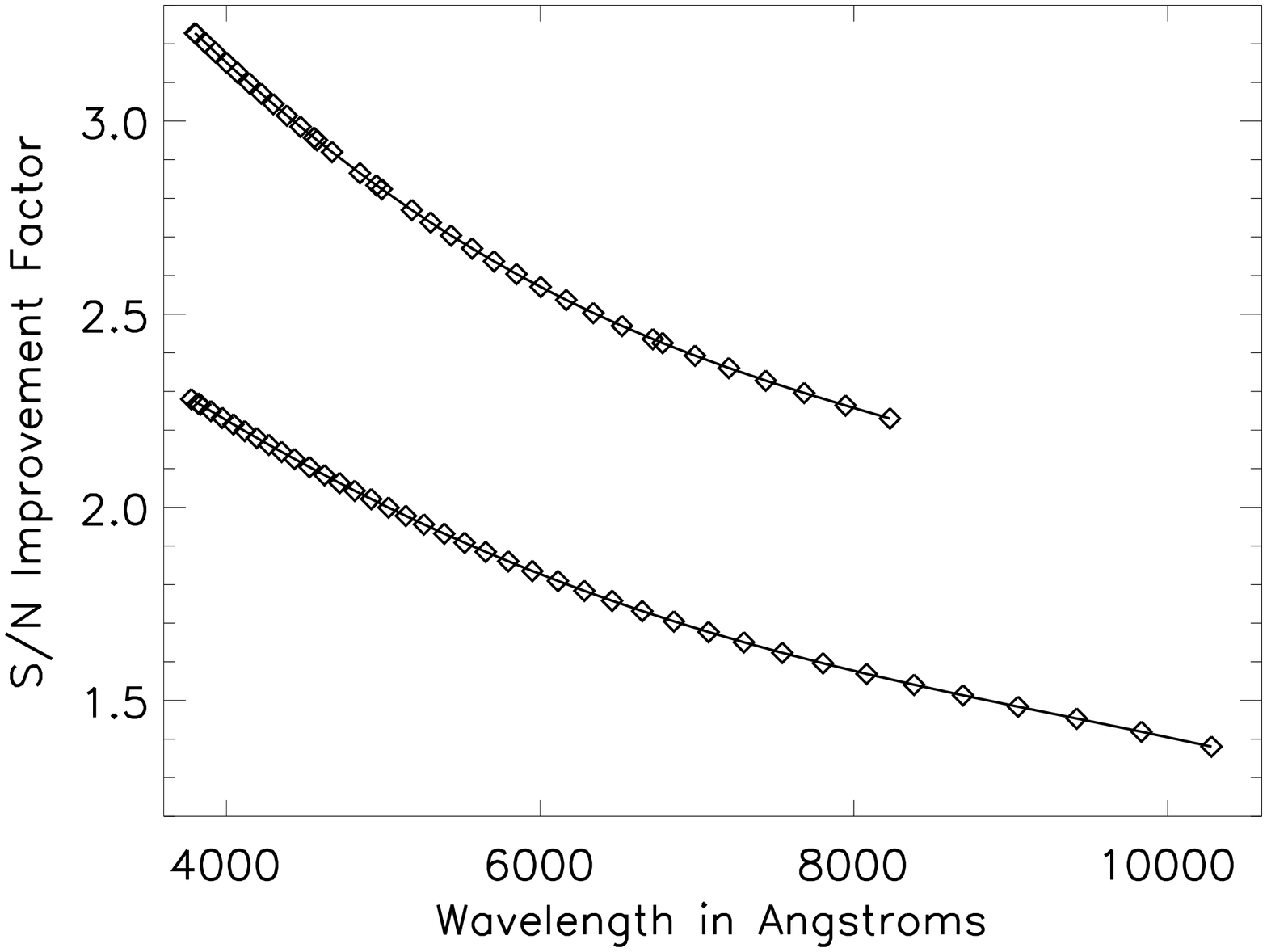}  
\includegraphics[width=0.35\linewidth, angle=90]{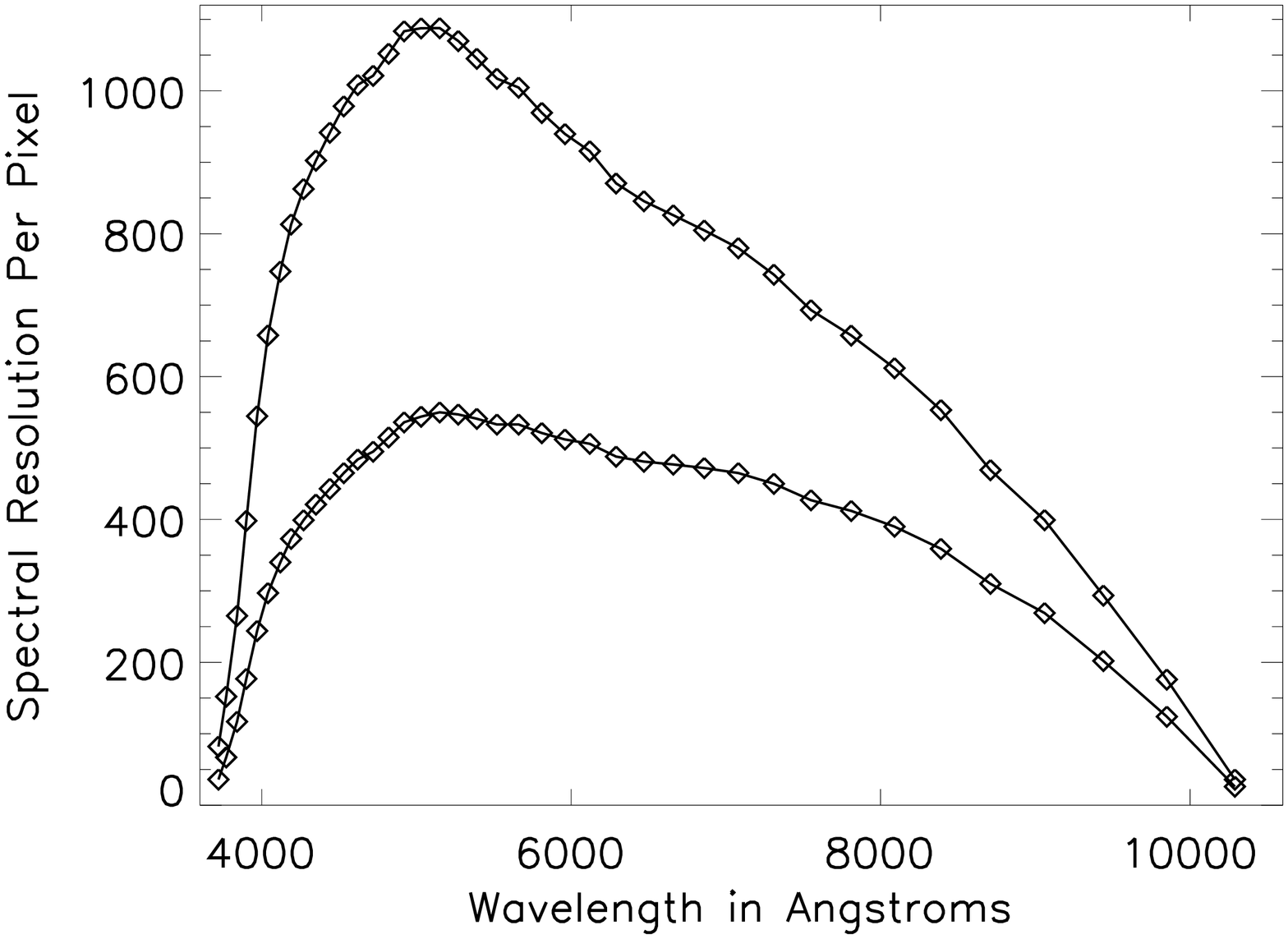}  
\caption{\label{espproprb} The post-processing properties of ESPaDOnS data. Box a shows the rebinned dispersion per pixel. Box b shows the resulting spectral resolution at single-pixel sampling. Box c shows the signal-to-noise improvement factor calculated as the square-root of the number of binned spectral-pixels. A polynomial fit is overplotted. Box d shows the signal-to-noise for each order calculated by the Libre-Esprit script (lower curve). The binning process results in an improved signal-to-noise (upper curve).}
\end{center}
\end{figure*}

	After applying the rebinning routines, the dispersion remains constant at 0.12{\AA} per pixel as seen in Figure \ref{espproprb}a. This results in a spectral resolution that is wavelength dependent, going from 30,000 to 80,000 per spectral-pixel at one pixel per resolution element. This is shown in Figure \ref{espproprb}b. As stated above, the physical spectral resolution of the raw output is roughly 70,000 with over 2 spectral pixels per resolution element. This new binned output undersamples 2:1 at 4000{\AA} increasing to 1:1 sampling at 7000{\AA} and rising with longer wavelengths. The signal-to-noise will be increased roughly as the square-root of the number of pixels averaged as all our observations are in the high signal-to-noise regime. The signal-to-noise improvement factors are shown in Figure \ref{espproprb}c for both order-overlap and no-overlap. Libre-Esprit outputs an estimate of the signal-to-noise for each order as part of the reduction routine. An example of this is seen in the lower curve of Figure \ref{espproprb}d. After the regularization of the wavelength array the signal-to-noise is increased. Using the modest improvement with no order-overlap, the resulting signal-to-noise is shown as the upper curve in Figure \ref{espproprb}d. Though these routines result in a wavelength-dependent resolution (with constant dispersion), the regular, linear, monotonic wavelength sampling allows for easier processing of large numbers of observations along with generally higher signal-to-noise.

\begin{table}[!h,!t,!b]
\begin{center}
\begin{footnotesize}
\caption{ESPaDOnS Orders \label{espord}}
\vspace{4mm}
\begin{tabular}{ccccrccc}
\hline
\hline
{\bf N} &{\bf $\lambda_0$} &{\bf $\lambda_c$} &{\bf $\lambda_1$} &{\bf    N} &{\bf $\lambda_0$} &{\bf $\lambda_c$} &{\bf $\lambda_1$}		 \\ 
\hline
\hline
61 &   369.12350  & 372 &  375.10460  &  41 &   542.71300  & 552  & 560.72960 \\
60 &   372.96250  & 377 &  381.41990  &	 40 &   556.02290  & 566  & 574.94500 \\
59 &   379.20690  & 384 &  387.95310  &	 39 &   570.00300  & 581  & 589.90050 \\
58 &   385.66620  & 390 &  394.71350  &	 38 &   584.70170  & 596  & 605.65430 \\
57 &   392.34930  & 397 &  401.71090  &	 37 &   600.17300  & 612  & 622.27430 \\
56 &   399.26270  & 404 &  408.96230  &	 36 &   616.49410  & 629  & 639.82890 \\
55 &   406.42750  & 412 &  416.47950  &	 35 &   633.71090  & 647  & 658.39830 \\
54 &   413.85750  & 419 &  424.27950  &	 34 &   651.92970  & 666  & 678.07840 \\
53 &   421.55920  & 427 &  432.37580  &	 33 &   671.44260  & 686  & 698.79780 \\
52 &   429.55120  & 435 &  440.78550  &	 32 &   692.43090  & 708  & 720.63430 \\
51 &   437.85630  & 444 &  449.52940  &	 31 &   714.77320  & 731  & 743.87970 \\
50 &   446.48130  & 453 &  458.62520  &	 30 &   738.60490  & 755  & 768.67450 \\
49 &   455.45800  & 462 &  468.09680  &	 29 &   764.07990  & 781  & 795.17950 \\
48 &   464.79900  & 472 &  477.97030  &	 28 &   791.37460  & 809  & 823.57770 \\
47 &   474.53390  & 482 &  488.26610  &	 27 &   820.69110  & 839  & 854.07930 \\
46 &   484.68840  & 492 &  499.01680  &	 26 &   852.26270  & 871  & 886.92720 \\
45 &   495.27700  & 503 &  510.24760  &	 25 &   886.36000  & 906  & 922.40310 \\
44 &   506.33940  & 515 &  521.99910  &	 24 &   923.29860  & 944  & 960.83560 \\
43 &   517.91600  & 527 &  534.30210  &	 23 &   963.44940  & 985  & 1002.6098 \\
42 &   530.02500  & 539 &  547.19600  &	 22 &   1007.2513  & 1029 & 1048.1787 \\
\hline
\hline
\end{tabular}
\end{footnotesize}
\end{center}
This Table shows order number, calculated beginning wavelength, Libre-Esprit log ``central'' wavelength and calculated ending wavelength. 
\end{table}

	We also performed some experiments on the accuracy of the signal-to-noise predictions given by Libre-Esprit. In the very high signal-to-noise regime, systematic effects caused by flat fielding issues, non-linearity or instrument instabilities can become significant. The Libre-Esprit reduction logs output two estimates of the signal-to-noise based on the flux. There is an estimate for each ``spectral pixel'' and ``ccd pixel''. The difference between these pixels is the resolution of 2.6km/s per physical ccd pixel compared to 1.8km/s per spectral pixel as resampled in Libre-Esprit.

\begin{figure*} [!h, !t, !b]
\begin{center}
\includegraphics[width=0.35\linewidth, angle=90]{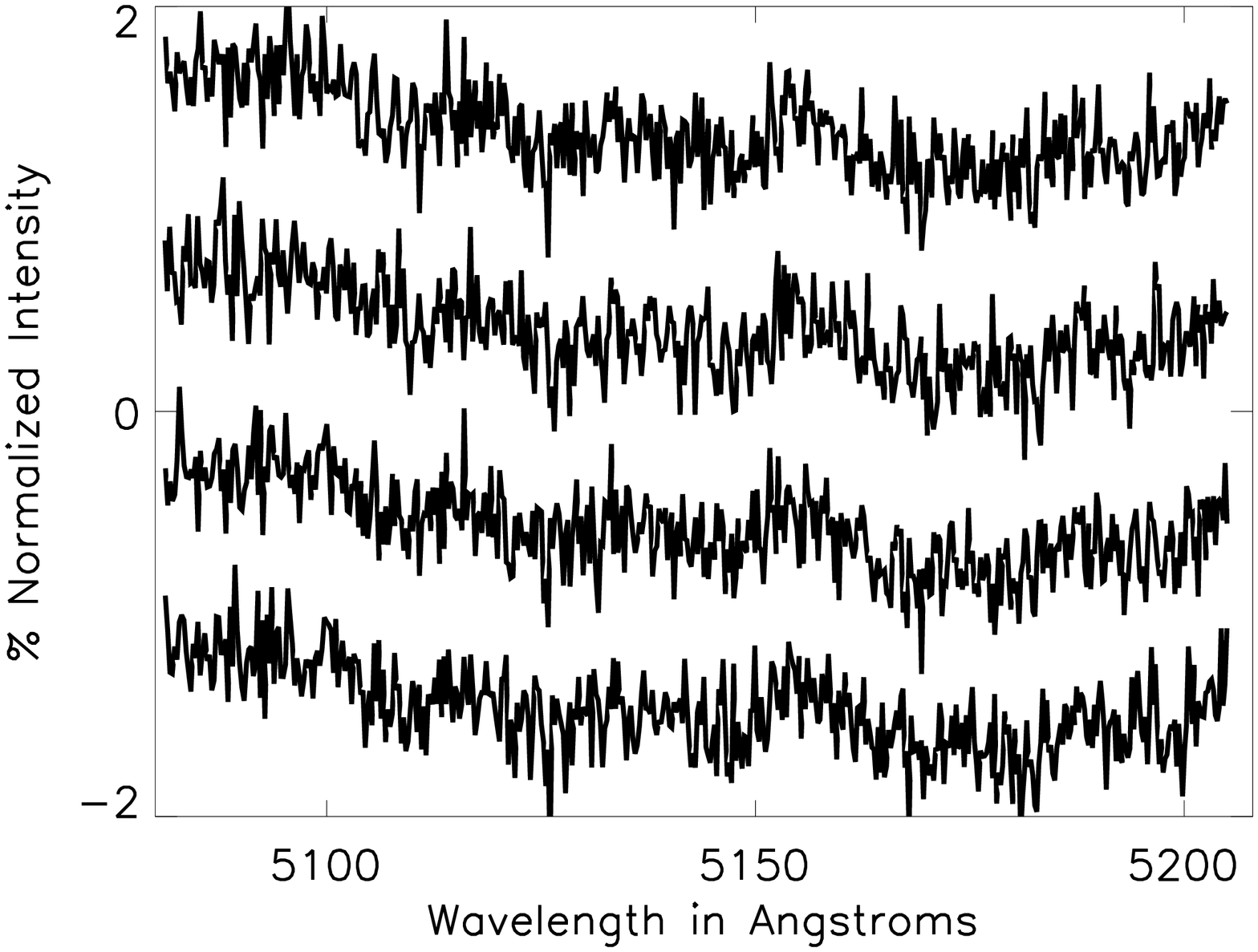}  
\includegraphics[width=0.35\linewidth, angle=90]{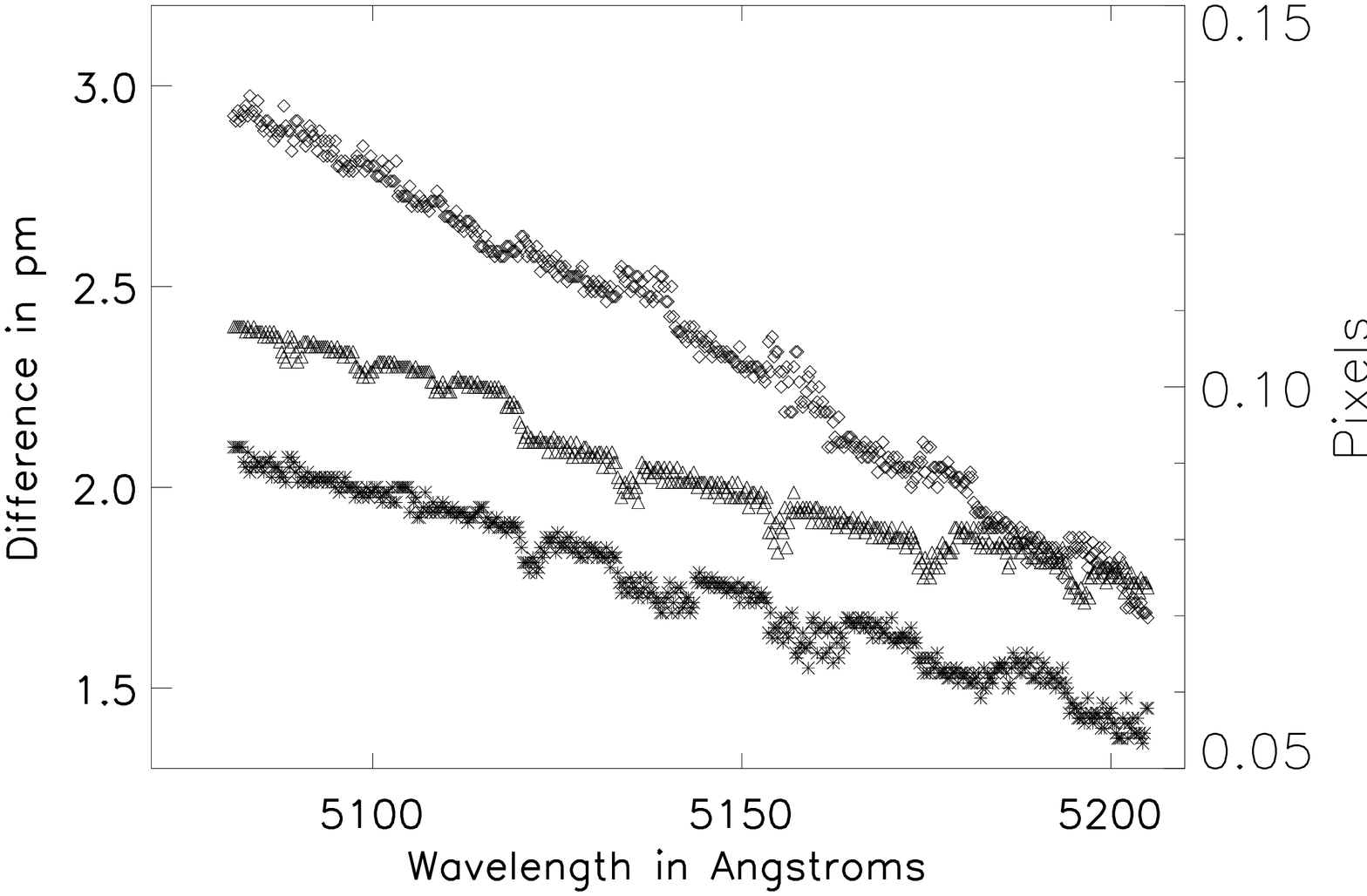}  \\ 
\includegraphics[width=0.35\linewidth, angle=90]{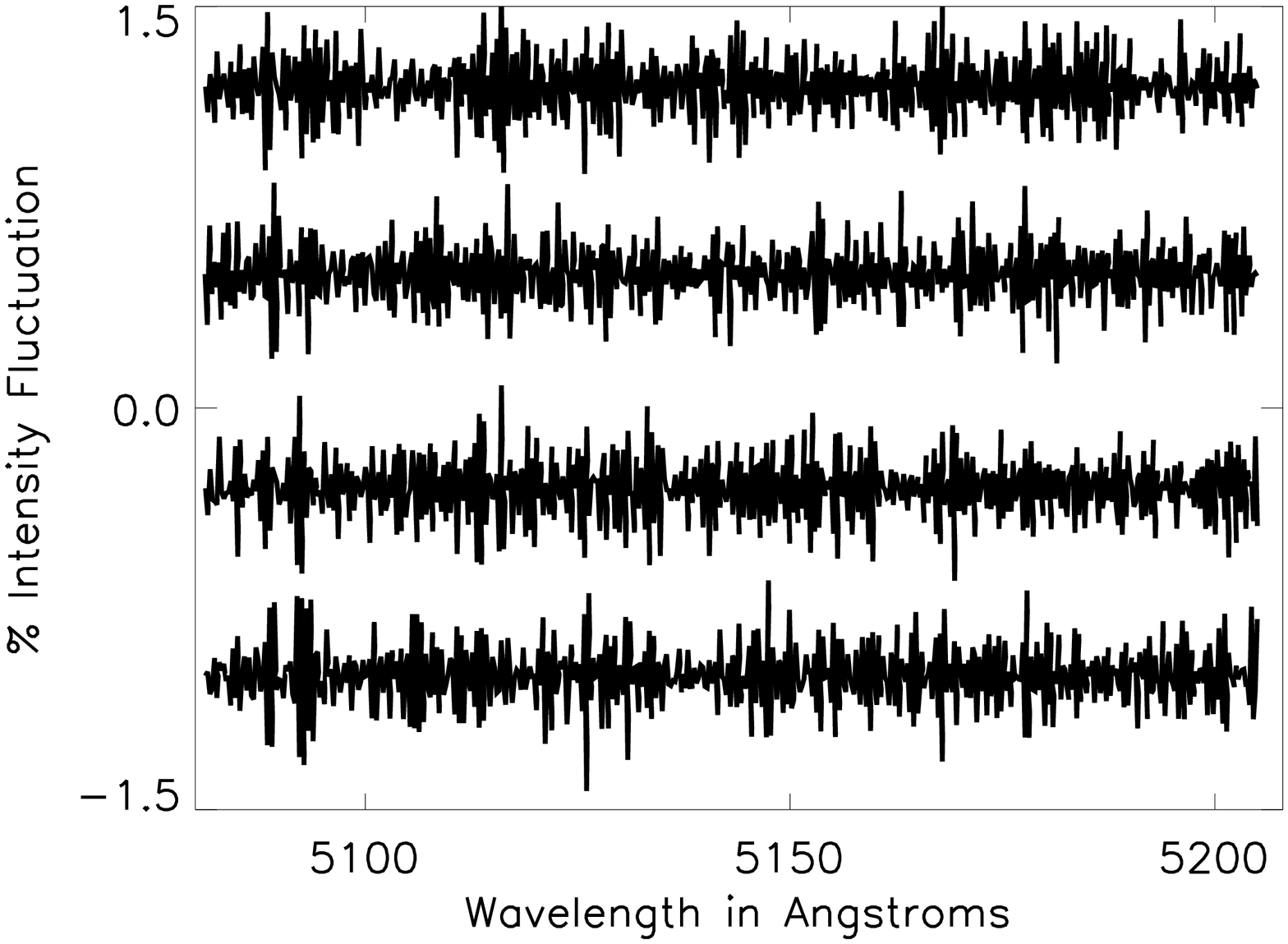}  
\includegraphics[width=0.35\linewidth, angle=90]{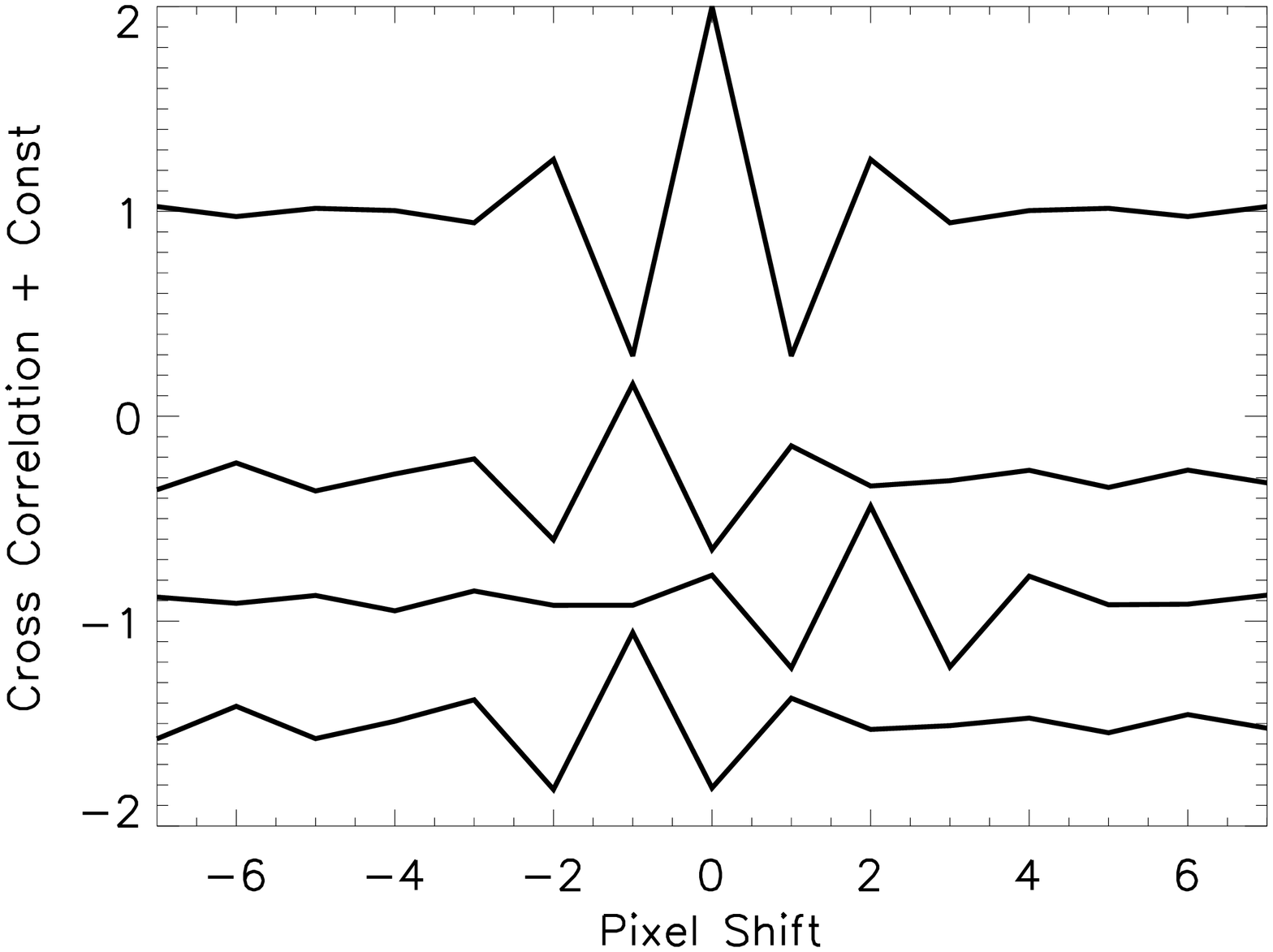} 
\caption{\label{noiseprop} The noise properties of the ESPaDOnS data in the highest signal-to-noise observation set: March 27th. Box a shows the spectrum of Spica in order 44 (5150{\AA}). We have two complete polarization sets each with two spectra from Libre-Esprit. Each individual spectrum here is calculated from two polarized orders in 4 separate exposures. The Libre-Esprit package gives signal-to-noise ratio's of 1350 / 1600 for each spectral / ccd pixel. Box a shows the four individual spectra after a simple linear continuum normalization offset for clarity. Box b shows the difference in wavelength solutions in pico-meters and pixel fractions after removing an offset of 1-2 pixels between individual solutions. The wavelength dependence of the shifts across an order are negligible. Box c shows the residual noise after subtracting a 3-pixel boxcar smooth from each intensity spectrum. This highlights the noise and removes spectral structure. Box d shows the cross correlation between each of the four noise spectra and the top noise spectra from box c. The top curve represents an autocorrelation. A clear signature from the boxcar smooth is seen. The lower three plots also show a correlation, but spectrally offset by 1 or 2 pixels. The wavelengths differ by less than 10\% of a pixel but a 1-2 pixel shift between exposures has been recorded by Libre-Esprit. This correlation between the noise spectra is likely caused by structure on the CCD.}
\end{center}
\end{figure*}

	The noise properties of the ESPaDOnS data in the high signal-to-noise regime can be best seen in our best observation set: March 27th. These properties are measured directly on the Libre-Esprit output data without applying our wavelength-regularizing scripts. Figure \ref{noiseprop} shows four separate spectra of Spica for order number 44 centered at 5150{\AA}. There are two complete polarization measurements which each consist of two intensity spectra output by Libre-Esprit. Each individual spectrum there is calculated from four individual exposures each with two orders. Thus, each spectrum in Figure \ref{noiseprop}a represents an average of 8 individual spectra. The Libre-Esprit package gives signal-to-noise ratio for each plotted spectrum of roughly 1350 / 1600 per spectral and ccd pixel. The wavelength solution is derived independently for each 4-exposure polarization measurement. The shift between the wavelength solutions is typically 0-2 pixels. In Figure \ref{noiseprop}b, the change in wavelength solution is shown to be a small fraction of a pixel across this spectral order. The illumination pattern on the CCD changes only slightly between individual exposure sets.
	
	 The noise in the spectra can be estimated by subtracting a boxcar-smoothed spectrum from a raw spectrum. As the full-width-half-maximum of spectral features is roughly 2.2 pixels, any wavelength variation present at the 1-spectral-pixel level is surely noise. Figure \ref{noiseprop}c shows residual noise from each spectrum after subtracting a 3-pixel boxcar smooth. This highlights the noise and removes spectral structure. This noise-spectrum can then be cross-correlated to determine the level of any systematic errors present. Figure \ref{noiseprop}d shows the cross correlation between each of the four noise spectra and the first noise spectra (the top spectrum in box c). The top curve represents an autocorrelation and a clear signature from the boxcar smooth is seen. The lower three cross-correlation curves also show this same signature, but spectrally offset by 1 or 2 pixels and at significantly lower amplitude. The wavelength solutions differ by less than 10\% of a pixel across an order and are generally stable. However, a 1-2 pixel shift between exposures has been recorded by Libre-Esprit between the four wavelength solutions . This correlation between the noise spectra is likely caused by structure on the CCD.

\begin{figure} [!h, !t, !b]
\begin{center}
\includegraphics[width=0.85\linewidth, angle=90]{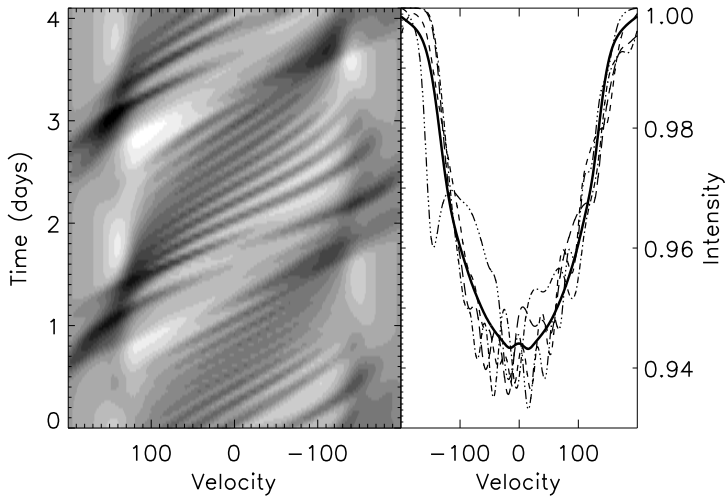}
\caption{\label{time-evolution} This illustrates the time evolution of the line profile over a single orbit. The left panel shows deviations from the average profile over an orbital cycle. The right hand panel shows the average line profile as well as some individual line profiles from various orbital phases. The velocity is corrected for $m_1$'s orbital motion.}
\end{center}
\end{figure}

\section{Traveling bumps}

	In this appendix section we will describe some general properties of the model calculations. Figure \ref{time-evolution} shows a typical model run.  The code is allowed to run for roughly 30 orbital cycles to damp out any perturbations from initial conditions. The perturbed velocity field is then calculated at each timestep, as is its projection along the line-of-sight to the observer, from which a total line profile is output. The right hand panel of Figure \ref{time-evolution} shows the average line profile over the orbital cycle as the dark solid line. Several individual spectra from throughout the orbit are also over-plotted as the dashed and dotted lines. Once the average line profile is subtracted from each individual measurement, a grey-scale plot of the deviations can be constructed to show how individual surface perturbations move across a line profile and preserve their identity over an orbital cycle. The left hand panel is dominated by the main tidal bulge - the light and dark patches that are most prominent on the line wings (largest velocities). However, the smaller-scale surface perturbations can be more easily seen near line center as the individual diagonal streaks.

\begin{figure} [!h, !t, !b]
\begin{center}
\includegraphics[width=0.7\linewidth, angle=90]{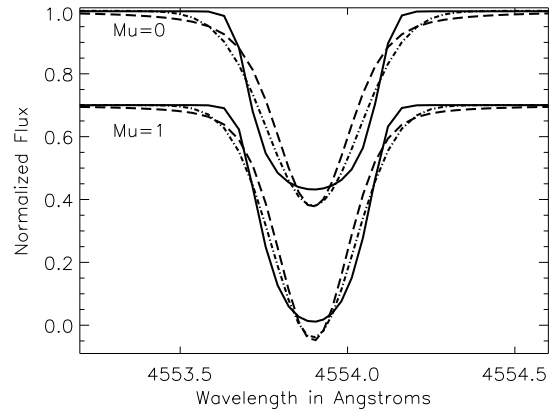}
\caption{\label{local-profs-mu} This illustrates the local line profiles used in computing perturbed average spectral line profiles. The solid line represents the CMFGEN local line profiles. The dashed line is a Voigt profile fit to the CMFGEN line profile. The dot-dash line is a Gaussian fit to the CMFGEN line profile. Fits are shown for both disk-center ($\mu$=1) and for the limb ($\mu$=0). The difference between Voigt and Gaussian fits lie primarily in the width and the wings of the profiles.}
\end{center}
\end{figure}

\section{Local Line Profile Effects}

	In order to test the accuracy and utility of using Gaussian local line profiles, we performed an investigation as to the effect of using different shape local-line profiles in computing the integrated rotationally broadened, tidally-perturbed profiles. We obtained CMFGEN model atmosphere profiles with T$_{eff}$=25000K and log(g)=3.75 provided by John Hillier. Since the CMFGEN model provides emergent flux (local line profiles) for a set of projected locations on the stellar disk ($\mu$) this database of profiles represents our starting point to simulate local line profile effects. We now have a model local line profile shape for any point on the stellar surface.
		
	These model atmosphere profiles are then fit with both Gaussian and Voigt functions at every $\mu$. These fits are straightforward using IDL's curve-fitting routines: Gaussfit and Curvefit. Once we have CMFGEN, Gaussian and Voigt local line profiles for every $\mu$ an accurate integrated line profile can be made to simulate an observation of a tidally perturbed star. The projected velocity is known for every surface element as is the value for $\mu$ and emergent flux at every surface element. An example of these local line profiles is shown in Figure \ref{local-profs-mu}
		
\begin{figure} [!h, !t, !b]
\begin{center}
\hspace{-15mm}
\includegraphics[width=1.0\linewidth]{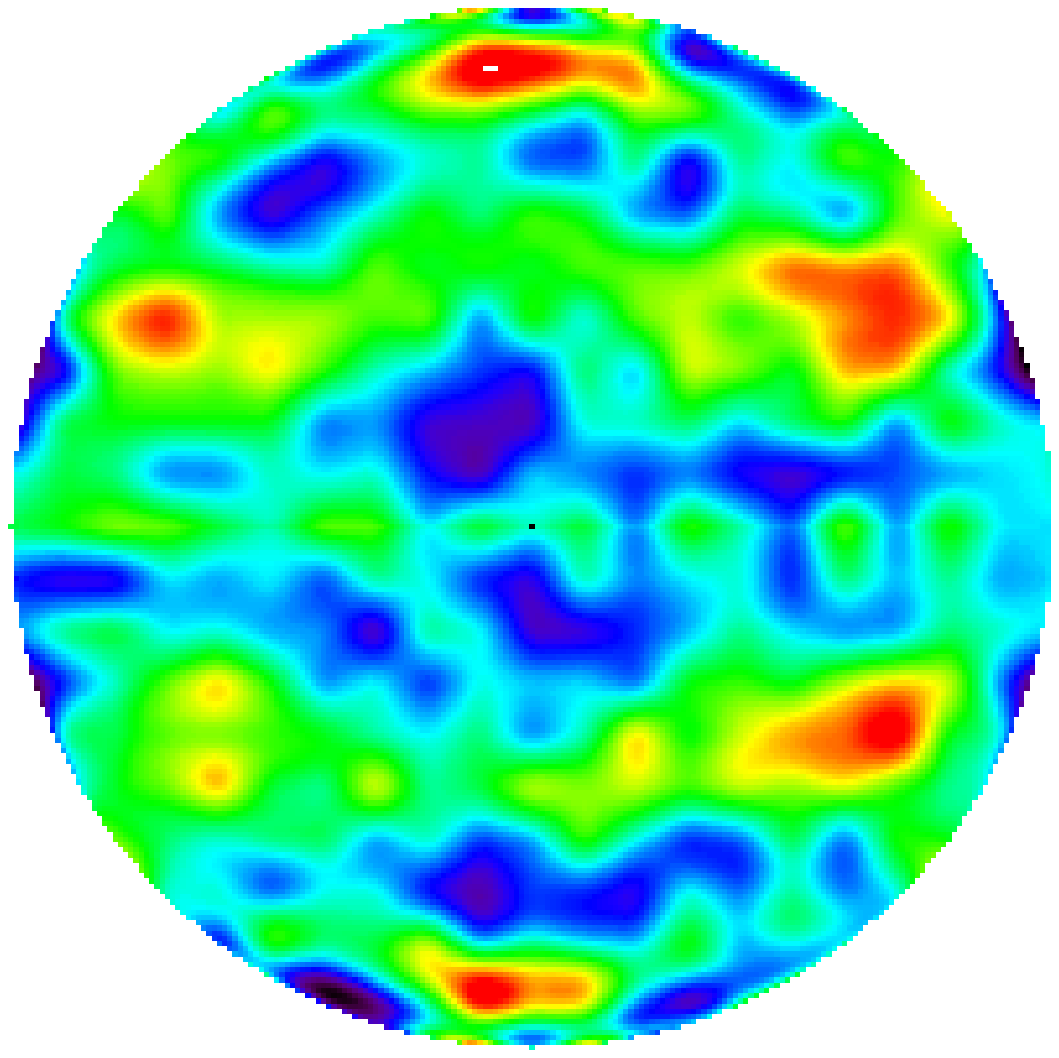} \\
\includegraphics[width=0.7\linewidth, angle=90]{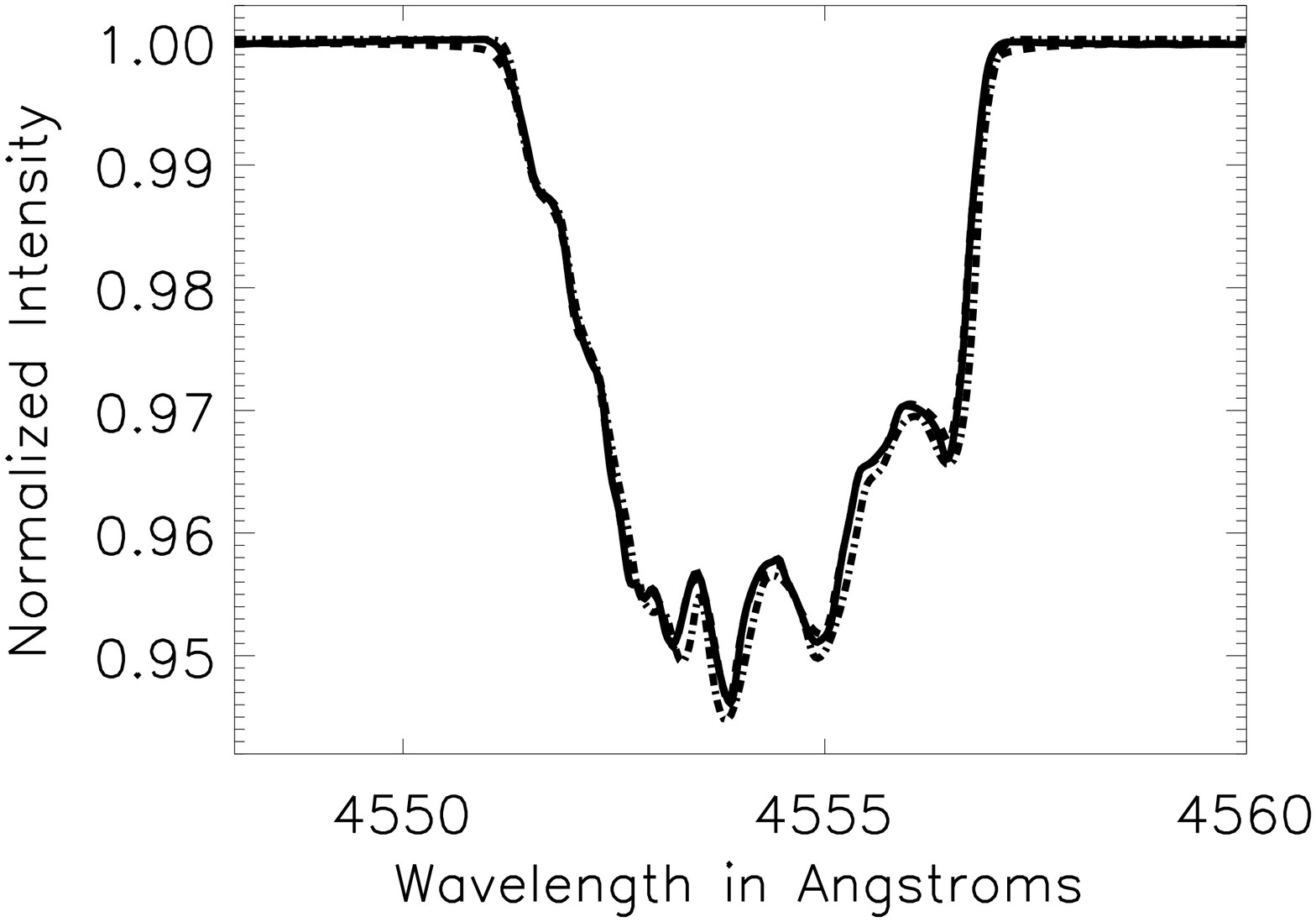}
\caption{\label{local-prof-nomatter} This illustrates the perturbed integrated profiles using different local profile shapes. The top panel shows a velocity perturbation on the stellar surface. The bottom panel shows the integrated line profile when calculated using CMFGEN (solid), Gaussian (dot-dash) and Voigt (dashed) local line profiles. The difference between local line profile morphology is significant - the Gaussian and Voigt local line profiles show broader wings and a sharper core while the CMFGEN spectra are more 'boxy'. However, these local differences have an entirely negligible effect on the integrated profile.}
\end{center}
\end{figure}

	To illustrate the effects of local line profile shapes on the perturbations caused by macroscopic surface flows, we computed a perturbed line profile for a velocity field using all three of these local line profile shapes. A simulated perturbing velocity field is created as a combination of spherical harmonic functions and random fluctuations with a maximum amplitude of $\pm$15km/s. This creates a highly perturbed line profile with known properties that can clearly demonstrate the effect of local line profile shapes. Figure \ref{local-prof-nomatter} shows a perturbed velocity field in the top panel as well as integrated line profiles in the bottom panel. The integrated line profiles are calculated using the perturbed stellar velocity field. The CMFGEN (solid), Gaussian (dot-dash) and Voigt (dash) local Si III line profiles are used to generate an integrated profiles seen in the bottom panel. There is essentially no difference between integrated line profiles even though significant differences in local line profile shapes are apparent. 

\begin{figure} [!h, !t, !b]
\includegraphics[width=0.95\linewidth]{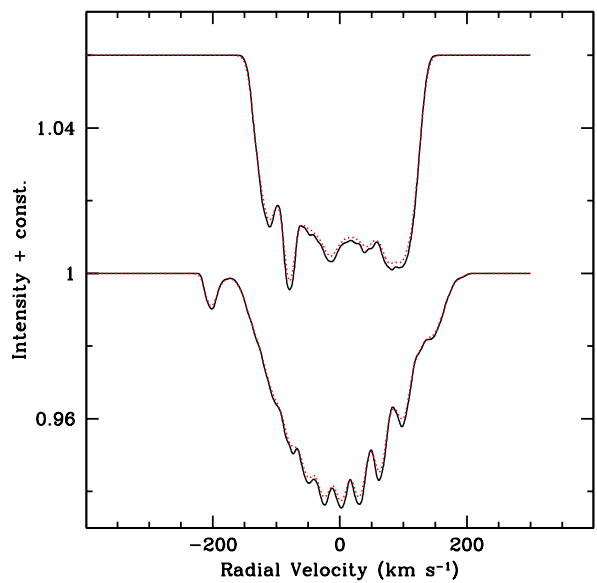}
\caption{\label{local-profs-tidal} Computed line profiles with Case GGD3 parameters showing that the intrinsic shape of the adopted line profile only affects the the depth of the sharp features, but does not modify the general characteristics of the perturbed line profile. The dotted profile was computed with a Gaussian intrinsic line-shape, while the continuous one with a more "boxy" ($\sim  e^{-x^4}$ ) intrinsic line-shape.}
\end{figure}

	As a third test of local line profile shape effects on integrated line profiles, we adapted our tidal-flow code to compute integrated line profiles using a more boxy local line profile shape - $\sim  e^{-x^4}$ instead of $\sim  e^{-x^2}$. As expected, the difference in local line profile shape was negligible. Figure \ref{local-profs-tidal} shows integrated line profiles computed for Case GGD3 parameters. The more boxy local line profiles are used for the solid curve and the Gaussian local line profiles are used for the dotted curve. There is a very small difference in the depth of the perturbations but the overall morphology is very similar.
	
	We consider the use of Gaussian local line profiles to be entirely sufficient to demonstrate this tidal-flow model for integrated line profiles. The integrated line profile intensities differ at the 0.05\% level for various local line profile shapes. This is considered to be entirely below any systematic errors or errors caused by uncertainties in the physical parameters used in calculating the properties of the binary systems.

\begin{figure*} [!h, !t, !b]
\begin{center}
\includegraphics[width=0.45\linewidth]{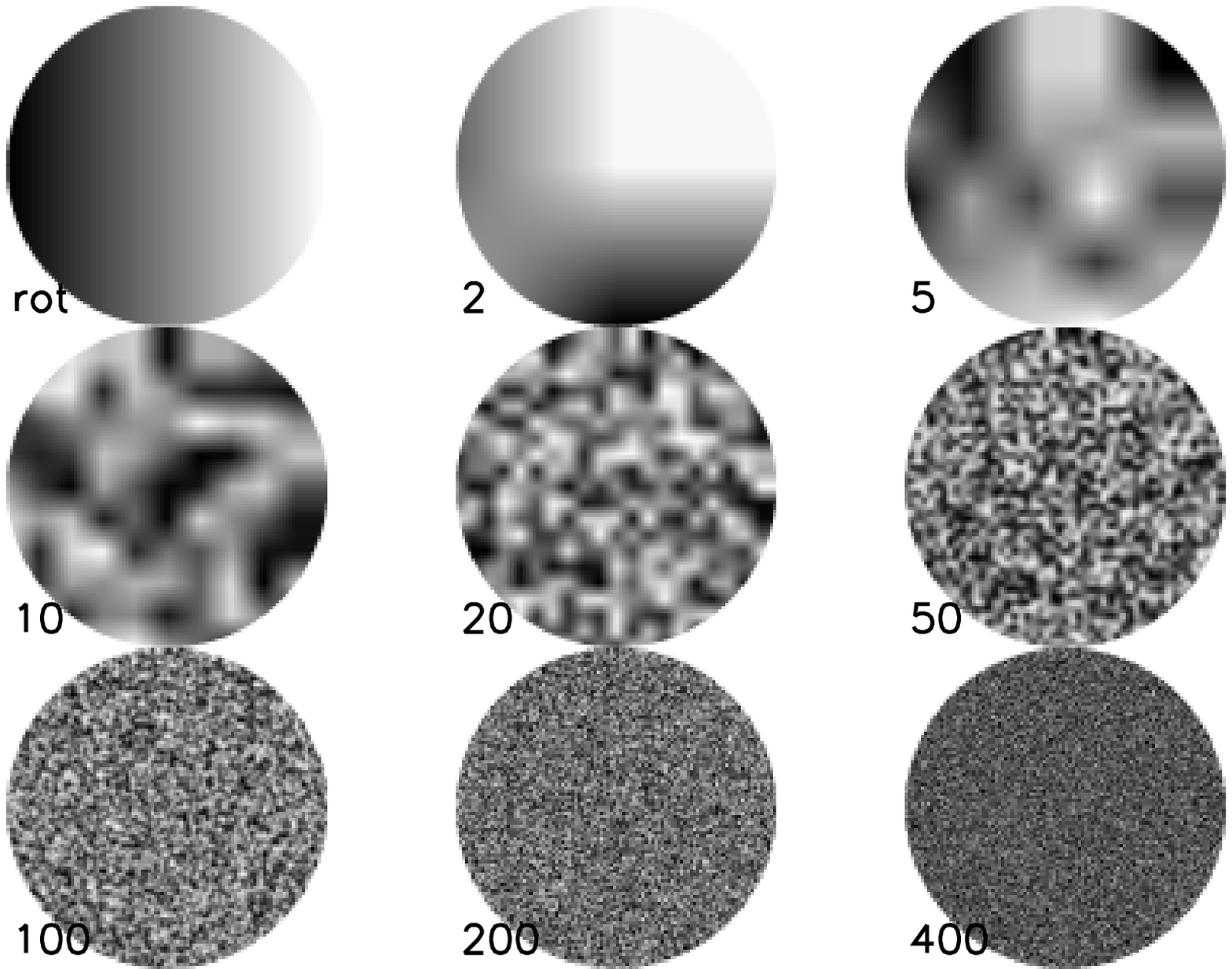}
\includegraphics[width=0.45\linewidth]{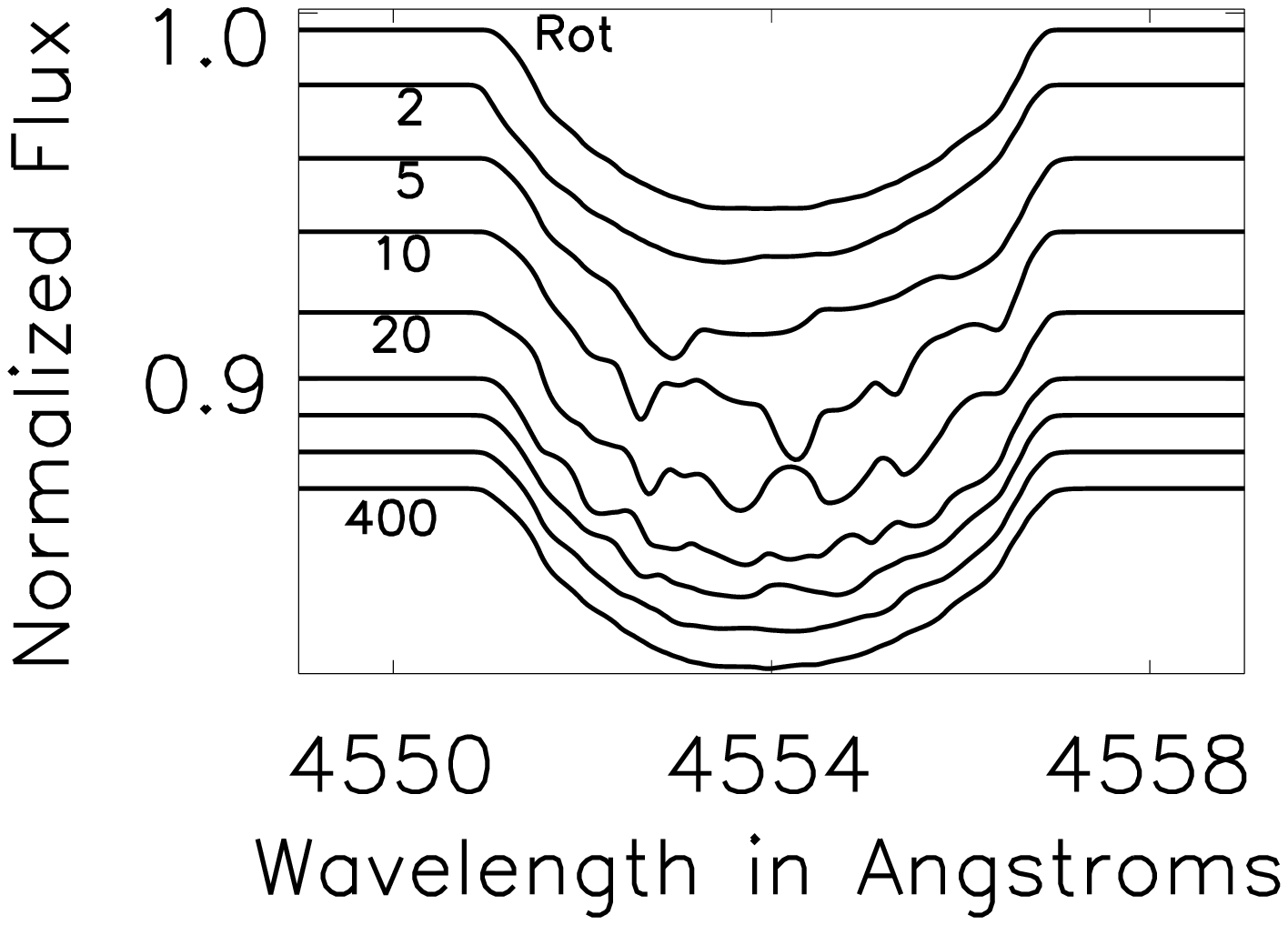}
\caption{\label{randomper} This illustrates how randomly generated velocity perturbations with various spatial distributions can give rise to line profiles similar to those observed. The top panel shows various velocity fields - the 160km/s rotational broadening in the top left and randomly generated fluctuations with 2, 5, 10, 20, 50, 100, 200 and 400 fluctuations across the equator. The line profiles corresponding to these velocity fields (scaled to $\pm$15km/s) are shown in the bottom panel. Lines with several 'bumps' of roughly the amplitude observed are seen in the examples with 10 to 20 equatorial bumps.}
\end{center}
\end{figure*}

\section{Velocity Perturbations}

	We feel it is worth pointing out that the amplitude and size of the 'bumps' fairly tightly constrain the size and amplitude of the corresponding velocity fluctuations on the stellar surface. In order to produce the bumps we observe, one simply needs a velocity perturbation of order 10-30km/s with a few 'patches' of velocity perturbation on the stellar surface for each observed bump. Figure \ref{randomper} illustrates this idea. 
	
	We created radially projected velocity perturbations on the stellar surface with a random number generator scaled to a peak amplitude of $\pm$15km/s. The surface of the star was perturbed with this random field scaled to have an increasing number of 'granules' on the stellar surface. This illustrates the effect of decreasing the size-scale of velocity perturbations on the integrated line profiles. Our experiments were run with spatial scales of 0, 2, 5, 10, 20, 50, 100, 200 and 400 individual perturbations across an equatorial slice. 
	
	The top panel of Figure \ref{randomper} shows first the rotational broadening of 160km/s in the top left followed by these perturbed velocity fields. Once these perturbations are incorporated with the rotational broadening, the corresponding line profiles are calculated. The bottom panel of Figure \ref{randomper} shows these perturbed line profiles. A fairly simple trend can be seen. A small number of perturbed surface patches (2-5) results in small bulk-shape deviations of the line profile. When the number of surface perturbation patches reaches 10-20, line profiles resembling those observed are seen. As the number of surface perturbations increase, more bumps appear at increasingly smaller amplitudes until (by 400 equatorial patches) there's essentially no visually apparent difference between perturbed and unperturbed line profiles. This example illustrates a simple constraint on the perturbed velocity field coming directly from the observed amplitude and number of bumps - relatively large-scale surface perturbations (a few to several patches in an equatorial slice) with an amplitude that is a fraction of the rotational velocity can cause the observed bumps. Since the structure of the observed perturbations (the line profile shape) is more or less similar from one orbit to the next, the corresponding velocity perturbations are clearly not random.


\begin{thebibliography}{}
\bibitem[]{}	Aufdenberg, J. et al., 2009, in prep.
\bibitem[]{}	Baade, D., 1984, A\&A, 134, 105.
\bibitem[]{} 	Dolginov, A.Z. \& Smel'Chakova, E.V. 1992, A\&A 257, 783.
\bibitem[]{}	Donati, J.F. et al., 1997, MNRAS, 291, 658.
\bibitem[]{} 	D'Alessio, P. et al, 1999, \apj, 527, 893.
\bibitem[]{} 	Eggleton, P.P., Kiseleva, L. G. \& Hut, P., 1998, ApJ, 499, 853.
\bibitem[]{}	Fullerton, A.W., Gies, D.R. \& Bolton, C.T., 1996, ApJS, 103, 475.
\bibitem[]{}	Gies, D.R. \& Kullavanija, A., 1988, ApJ, 326, 813.
\bibitem[]{} 	Hartmann, L., et al., 1998, ApJ, 495, 385. 
\bibitem[]{}  	Herbison-Evans, D. et al., 1971, MNRAS, 151, 161.
\bibitem[]{}	Hillier, D.J. \& Miller, D.L., 1998, ApJ, 496, 407.
\bibitem[]{} 	Lomb, N.R., 1978, MNRAS, 185, 325.
\bibitem[]{} 	Moreno, E. \& Koenigsberger, G. 1999, RMA\&A, 35, 157.
\bibitem[]{} 	Moreno, E., Koenigsberger, G. \& Toledano, O. 2005, A\&A, 437, 641.
\bibitem[]{}	Reid A.H.N. et al., 1993, ApJ, 417, 320.
\bibitem[]{} 	Riddle, R. L., 2000, Ph.D. thesis, Georgia State Universtiy.
\bibitem[]{}	Rivinius, T.H. et al., 2001, A\&A 369, 1058.
\bibitem[]{} 	Scharlemann, E.T., 1981, ApJ, 246, 292. 
\bibitem[]{} 	Shobbrook, R.R et al., 1969, MNRAS, 145, 131.
\bibitem[]{} 	Smak, J. 1970, Acta Astron. 20, 75.
\bibitem[]{} 	Smith, M.A. 1985a, ApJ, 297, 206.
\bibitem[]{} 	Smith, M.A. 1985b, ApJ, 297,224.
\bibitem[]{}	Sterken, C. Jerzykiewicz, M. \& Manfroied, J., 1986, A\&A, 169, 166.
\bibitem[]{} 	Tassoul, J.L. 1987, ApJ, 322, 856.
\bibitem[]{} 	Townsend, R.H.D. 1997a, MNRAS, 284, 839.
\bibitem[]{} 	Townsend, R.H.D. 1997b, PhD Thesis, University College London. 
\bibitem[]{}	Uytterhoeven K. et al., 2001, A\&A, 371, 1035.
\bibitem[]{} 	Vogt, S.S. \& Penrod, G.D., 1983, ApJ, 275, 661.
\bibitem[]{}	Walker, G.A.H., Yang, S. \& Fahlman, G.G., 1979, ApJ, 233, 199.
\end{thebibliography}
\end{document}